\documentclass[aps]{revtex4}
\usepackage{eurosym}
\usepackage{amsfonts}
\usepackage{amsmath}
\usepackage{amssymb,epsf}
\usepackage{color}
\usepackage{graphicx}
\usepackage{natbib}
\usepackage{float}
\usepackage{caption}
\usepackage{subfig}
\usepackage{epstopdf}

\begin{document}

 \title{ Critical behavior of charged AdS black holes surrounded by quintessence\\ via an alternative phase space}
 \author{S. H. Hendi$^{1,2,3}$ \footnote{email address: hendi@shirazu.ac.ir} and
 Kh. Jafarzade$^{4,5}$\footnote{email address: khadije.jafarzade@gmail.com}}

 \affiliation{
 $^1$Department of Physics, School of Science, Shiraz University, Shiraz 71454, Iran \\
 $^2$Biruni Observatory, School of Science, Shiraz University, Shiraz 71454, Iran \\
 $^{3}$Canadian Quantum Research Center 204-3002 32 Ave Vernon, BC V1T 2L7 Canada \\
 $^4$ Department of Theoretical Physics, Faculty of Basic Sciences, University of Mazandaran, P. O. Box 47416-95447, Babolsar, Iran\\
$^5$ ICRANet-Mazandaran, University of Mazandaran, P. O. Box 47416-95447, Babolsar, Iran }

\begin{abstract}
Considering the variable cosmological constant in the extended
phase space has a significant background in the black hole
physics. It was shown that the thermodynamic behavior of charged
AdS black hole surrounded by the quintessence in the extended
phase space is similar to the van der Waals fluid. In this paper,
we indicate that such a black hole admits the same criticality and
van der Waals like behavior in the non-extended phase space. In
other words, we keep the cosmological constant as a fixed
parameter, and instead, we consider the normalization factor as a
thermodynamic variable. We show that there is a first-order
small/large black hole phase transition which is analogous to the
liquid/gas phase transition in fluids. We introduce a new picture
of the equation of state and then we calculate the corresponding
critical quantities. Moreover, we obtain the critical exponents
and show that they are the same values as the van der Waals
system. Finally, we study the photon sphere and the shadow
observed by a distant observer and investigate how the shadow
radius may be affected by the variation of black hole parameters.
We also investigate the relations between shadow radius and phase
transitions and calculate the critical shadow radius where the
black hole undergoes a second-order phase transition.
\end{abstract}

\maketitle

\section{Introduction}

Undoubtedly, the black hole (BH) is one of the most fascinating
and mysterious subjects in the world of physics as well as
mathematics. BH was one of the interesting predictions of general
relativity which is confirmed by observational data \cite{LIGO}.
The observational evidence of massive objects and detection of the
gravitational waves open a new window in modern mathematical
physics and data analysis. On the other hand, the discovery of a
profound connection between the laws of BH mechanics with the
corresponding laws of ordinary thermodynamic systems has
been one of the remarkable achievements of theoretical physics \cite%
{Bardeen,Hawking}. In other words, the consideration of a BH as a
thermodynamic system with a physical temperature and an entropy
opened up new avenues in studying their microscopic structure.

In the past two decades, the study of BH thermodynamics in an
anti-de Sitter (AdS) space attracted significant attention.
Strictly speaking, the investigation of thermodynamic properties
of black holes in such a spacetime provides a deep insight to
understand the quantum nature of gravity \cite{Maldacena,Witten}.
In particular, the phase transition of AdS black holes has gained
a lot of attention due to the AdS/CFT correspondence in recent
years. The pioneering work in this regard was realized by Hawking
and Page who proved the existence of a certain phase transition
(so called Hawking-Page) between thermal radiation and
Schwarzschild-AdS BH \cite{Hawking1}. Afterward, our understanding
of phase transition has been extended by studying in more
complicated backgrounds \cite{Cvetic1,Cvetic2}. Among conducted
efforts, thermodynamics of charged black holes in the background
of an asymptotically AdS spacetime is of particular interest, due
to a complete analogy between them and the van der Waals
liquid-gas system. Such an analogy will be more precise by
considering the cosmological constant as a dynamical pressure and
its conjugate quantity as a thermodynamic volume in the extended
phase space \cite{Kubiznak1}.

It is worthwhile to mention that the conducted investigations in
this regard are based on the first law and Smarr relation by
comparing BH mechanics with ordinary thermodynamic systems not CFT
point of view. Despite the interesting achievements of AdS/CFT
correspondence such as describing the Hawking radiation mechanism
and dual interpretation of Hawking-Page phase transition and so on
with fixed $ \Lambda $, this method is not yet able to provide a
suitable picture in the extended thermodynamics. In the context of
AdS/CFT correspondence, the cosmological constant is set by $N$,
related to the number of coincident branes ($M$ branes or $D$
branes) on the gravity side. On the field theory side, $N$ is
typically the rank of a gauge group of the theory, and as such, it
also determines the maximum number of available degrees of
freedom. The thermodynamic volume in the bulk gravity theory
corresponds to the chemical potential in the boundary field theory
which is the conjugate variable of the number of colors
\cite{Kubiznak3,Johnson}. Variation of the pressure of the bulk,
or equivalently variation of the AdS radius ($l$) leads to
variation of the boundary quantities: i) the number of colors $N$.
ii) the volume of the space on which the field theory is
formulated (since $V \propto l^{d-2}$). iii) the CFT charge $Q$
which is related to the bulk charge $Q_{b}$ according to $Q = l
Q_{b}$. So, considering the cosmological constant as a new
thermodynamical parameter may give the phase diagram an extra
dimension. An alternative interpretation is suggested in
\cite{Karch}, where the variation of the pressure in the bulk
theory corresponds to vary the volume of the boundary field
theory. In this approach, the number of colors is kept fixed,
which requires the variation of Newton's constant to compensate
the variation of the volume of the boundary field theory. This
shows that one cannot employ such an approach (AdS/CFT
correspondence) to investigate the extended thermodynamics of
black holes.

Considering the defined thermodynamic pressure and volume, one can
study thermodynamics of black holes in a new framework, sometimes
referred to as BH Chemistry \cite{Mann}. This change of
perspective has led to a different concept of known processes and
the discovery of a broad range of new phenomena associated with
black holes such as van der Waals behavior
\cite{Kubiznak1,Kubiznak2}, solid/liquid phase transitions
\cite{Altamirano}, triple points \cite{WWei}, reentrant phase
transitions \cite{Kubiznak4} and heat engines \cite{Johnson}.
Also, using BH volume, one can study the BH adiabatic
compressibility which has attracted attention in connection with
BH stability \cite{Dolan1}. This new perspective has also been
successful in describing the thermodynamic structure of black
holes in other gravitational theories such as Lovelock gravity
\cite{Dolan2}, nonlinear electrodynamics \cite{Hendi},
Einstein-Yang-Mills gravity \cite{Zhang1}, black holes with scalar
hair \cite{Hristov1}, dyonic black holes \cite{Dutta}, f(R)
gravity \cite{Chen}, STU black holes \cite{Caceres}, quasi
topological gravity \cite{Hennigar1}, conformal gravity
\cite{Zhao1} Poincare gauge gravity \cite{ZLiu}, Lifshitz gravity
\cite{Brenna,Jafarzade,Zeng} and massive gravity \cite{Hendi1}.

Since the thermodynamic behavior of BH is highly affected by the
variation of electric charge, an alternative approach for
investigating van der Waals like phase transition is proposed
\cite{Chamblin1,Chamblin2} by considering the electric charge as a
thermodynamic variable and fixing the cosmological constant. In
this regard, a phase transition between the large and small black
holes in the $Q-\Phi $ plane is seen. In addition, it is shown
that \cite{Dehyadegari} such a suggestion is mathematically
problematic and physically unconventional and it is logical to
consider the square of the electric charge, $Q^{2}$, as a
thermodynamic variable. In Ref. \cite{Dehyadegari}, the van der
Waals like behavior of charged AdS BH in $Q^{2}-\Psi$ plane with a
fixed cosmological constant is studied. It is also obtained the
critical exponents which were exactly coincident with those
obtained for van der Waals liquid. As a keynote, we should
emphasize that unlike the van der Waals liquid the mentioned phase
transition occurred for temperature higher than critical
temperature ($T>T_{c}$) in $Q^{2}-\Psi$ plane. However,
$Q^{2}-\Psi$ criticality has been investigated in alternative
theories of gravity \cite{Yazdikarimi}.

Alternative theories of gravity are proposed to overcome different
shortcomings of Einstein's general relativity. In recent years,
modern observational evidence has shown that the universe is
expanding with
acceleration, demanding the existence of dark energy \cite%
{Bachall,Sahni}. Among the various dark energy candidates,
consideration of the cosmological constant or the quintessence is
more common. The quintessence is proposed as the canonical scalar
field with state parameter $-1< \omega <1 $. However, in order to
explain the late-time cosmic acceleration, one has to restrict
such a parameter to $-1< \omega <-\frac{1}{3} $. The cases of
$\omega=\frac{1}{3} $ and $\omega=0 $ are, respectively,
representing the radiation and dust around the black hole, while
the quintessential dark energy likes cosmological constant for
$\omega \longrightarrow -1 $. As the first attempt, Kiselev
obtained the solutions of Einstein's field equations for the
quintessence matter around a charged (an uncharged) BH
\cite{Kiselev}. These solutions are described in terms of the
state parameter $\omega$ and the normalization factor $a$. The
normalization factor indicates the intensity of the quintessence
field and the state parameter refers to its nature and behavior.
The rotating generalization of Kiselev solutions and their AdS
modifications have been reported in Refs.
\cite{Ghosh,Oteev,Toshmatov,Xu}. In addition to the exact Kiselev
solutions and their generalizations, considerable efforts were
conducted in context of thermodynamic and phase transition of such
black holes \cite{Majeed,Wang2}. The obtained results showed that
variation of the quintessence field affects the thermodynamic
behavior of quintessential black holes and consequently, it can
lead to interesting critical behavior. Such a criticality is
studied in $Q^{2}-\Psi$ and $P-V$ planes, as common methods. It
will be interesting to probe the van der Waals like phase
transition and critical behavior by treating the normalization
factor as a thermodynamic variable and keeping both the
cosmological constant and electric charge as fixed parameters. To
achieve this goal, we would like to investigate the critical
behavior of Reissner-Nordstr\"{o}m-AdS (RN-AdS) black holes
surrounded by quintessence via this alternative viewpoint.

The organization of this paper is as follows. In the next section,
we would like to give a brief review of RN-AdS black holes
surrounded by quintessence and their criticality via the common
methods. In Sec. \ref{Criticality}, we study the critical behavior
of the corresponding BH solution using the alternative phase
space. We will calculate the critical exponents and compare the
obtained results with those obtained for van der Waals fluid. Sec.
\ref{PS-S} is devoted to exploring the photon orbits near the
black hole and formation of shadow. In Sec. \ref{shadow-critical},
we investigate the connection between BH shadow and phase
transition. Finally, we finish the paper with concluding remarks.

\section{ Thermodynamics of charged AdS BH surrounded by
quintessence: A brief review}\label{Review}

In this section, we first introduce the thermodynamics charged AdS
BH surrounded by quintessence by reviewing Refs.
\cite{Li,Liu,Hong,Guo}. The line element of such a BH is expressed
as
\begin{eqnarray}
ds^{2}=-f(r)dt^{2}+\frac{1}{f(r)}dr^{2}+r^{2} d\theta^{2}+r^{2}
sin^2\theta\; d \phi^{2},
\end{eqnarray}
where
\begin{eqnarray}
f(r)=1-\frac{2M}{r}+\frac{Q^{2}}{r^{2}}-\frac{a}{r^{3\omega +1}}+\frac{r^{2}%
}{l^{2}},  \label{EqFr}
\end{eqnarray}
where $M $ and $Q$ are the mass and electric charge of the black hole,
respectively and $\ell=\sqrt{-\frac{3}{\Lambda}} $ is the AdS radius which
is related to the cosmological constant. The state parameter $\omega $
describes the equation of state $p = \omega\rho$ where $p $ and $\rho $ are
the pressure and energy density of the quintessence, respectively. The
normalization factor $a $ is related to the density of quintessence $\rho $
as
\begin{eqnarray}
\rho =-\frac{a}{2}\frac{3\omega}{r^{3(1+\omega)}},  \label{Eqrho}
\end{eqnarray}
with $[length]^{-2}$ dimensions. Solving the equation $(f(r =
r_{+}) = 0)$, one can obtain the total mass of the BH ($M $) as
\begin{eqnarray}
M=\frac{r_{+}}{2}+\frac{Q^{2}}{2r_{+}}+\frac{r_{+}^{3}}{2l^{2}}-\frac{a}{2}%
r_{+}^{-3\omega}.  \label{Eqmass}
\end{eqnarray}

In addition, one can use the Hawking and Bekenstein area law to obtain the
entropy as
\begin{eqnarray}
S=\frac{A}{4}=\pi r_{+}^{2}.  \label{Eqentropy}
\end{eqnarray}

Working in the extended phase space, the cosmological constant and
thermodynamic pressure are related to each other with the
following relation
\begin{eqnarray}
P=-\frac{\Lambda}{8\pi}=\frac{3}{8\pi l^{2}},  \label{Eqpressure}
\end{eqnarray}
where the variability of the cosmological constant is associated to the
dynamical vacuum energy.

It it easy to rewrite Eq. (\ref{Eqmass}) in terms of pressure and
entropy as
\begin{eqnarray}
M(S,Q,P,a)=\frac{1}{6\sqrt{\pi S}}\left( 3\pi Q^{2}+3S+8PS^{2}-3a\pi^{\frac{%
3\omega +1}{2}}S^{\frac{1-3\omega }{2}}\right) .  \label{EqMS}
\end{eqnarray}

It is obvious that the total mass of BH plays the role of enthalpy
instead of internal energy in the extended phase space. Therefore,
regarding the enthalpy representation of the first law of BH
thermodynamics, the intensive parameters conjugate to $S$, $Q$,
$P$ and $a$ are, respectively, calculated as
\begin{eqnarray}
T\equiv \left(\frac{\partial M}{\partial S}\right)_{P,Q,a}&=&\frac{1}{4\pi
r_{+}}\left(1-\frac{Q^{2}}{r_{+}^{2}}+8\pi Pr_{+}^{2}+3\omega a
r_{+}^{-3\omega -1} \right),  \label{Temp1} \\
\Phi \equiv \left( \frac{\partial M}{\partial Q}\right)_{S,P,a}&=&\frac{Q}{%
r_{+}},  \label{Phi1} \\
V\equiv \left( \frac{\partial M}{\partial
P}\right)_{S,Q,a}&=&\frac{4}{3}\pi
r_{+}^{3},  \label{Vol1} \\
y \equiv \left( \frac{\partial M}{\partial a}\right)_{S,Q,P}&=&-%
\frac{1}{2}r_{+}^{-3\omega},  \label{Eqintensive1}
\end{eqnarray}
where $T$, $\Phi $ and $V $ are the temperature, electric potential and
thermodynamic volume, respectively and $y $ is the quantity
conjugate to the dimensionful factor $a$. Considering the dimensional
analysis, we can obtain the following Smarr relation
\begin{eqnarray}
M=2TS+\Phi Q-2VP+(1+3\omega)y a,  \label{EqSmarr}
\end{eqnarray}
where confirms that we have to regard $a$ as a thermodynamic
quantity. Now, it is straightforward to find that the first law of
the BH is obtained as
\begin{eqnarray}
dM = TdS + \Phi dQ + VdP + y da.  \label{EqFlaw}
\end{eqnarray}

\subsection{$P-V$ criticality and van der Waals phase transition: Usual
method}

In this section, we briefly review the critical behavior of such
black holes in the usual way. Considering the temperature
relation, Eq. (\ref{Temp1}) with the definition of pressure, Eq.
(\ref{Eqpressure}), one can easily derive the equation of state of
the BH as
\begin{eqnarray}
P=\frac{T}{2r_{+}}-\frac{1}{8\pi r_{+}^{2}}+\frac{Q^{2}}{8\pi r_{+}^{4}}-\frac{3a\omega}{8\pi r_{+}^{3(1+\omega)}}.  \label{EqPr}
\end{eqnarray}

Since the event horizon radius $r_{+}$ is associated with
the van der Waals fluid specific volume ($\upsilon$)
\cite{Kubiznak1,Kubiznak2, Li} as $\upsilon = {2\ell
_{\rm{P}}^{{2}}{r_ + }}$ ( $\ell _{\rm{P}}$ is the Planck length
that we set $\ell _{\rm{P}} =1$ since we work in the geometric
units), Eq. (\ref{EqPr}) can be rewritten as
\begin{eqnarray}
P=\frac{T}{\upsilon}-\frac{1}{2\pi \upsilon^{2}}+\frac{2Q^{2}}{\pi
\upsilon^{4}}-\frac{8^{\omega}\times 3a\omega}{\pi
\upsilon^{3(1+\omega)}}.  \label{EqPnu}
\end{eqnarray}

\begin{figure}[!htb]
\centering
\subfloat[]{
        \includegraphics[width=0.3\textwidth]{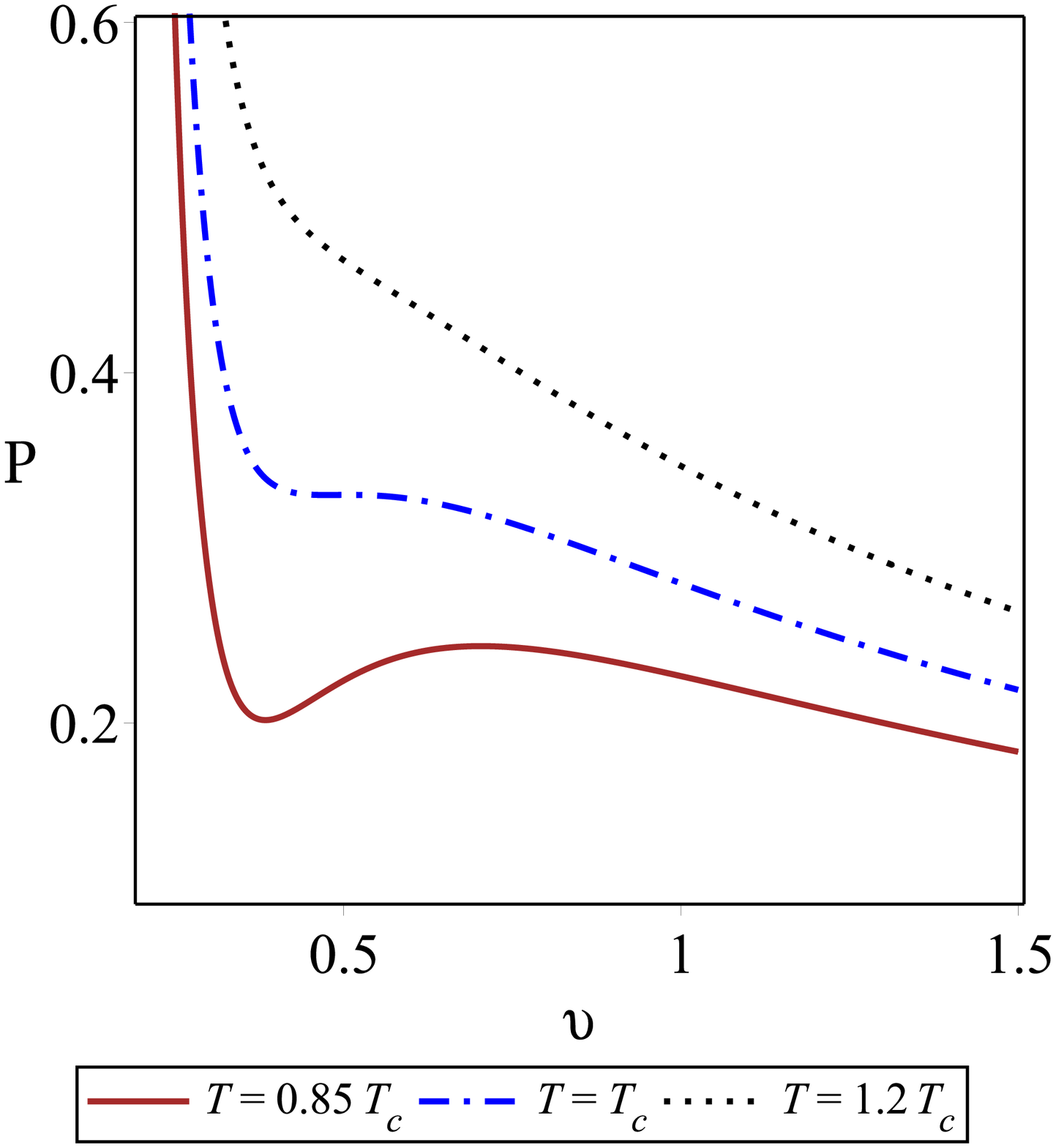}}
\subfloat[]{
        \includegraphics[width=0.31\textwidth]{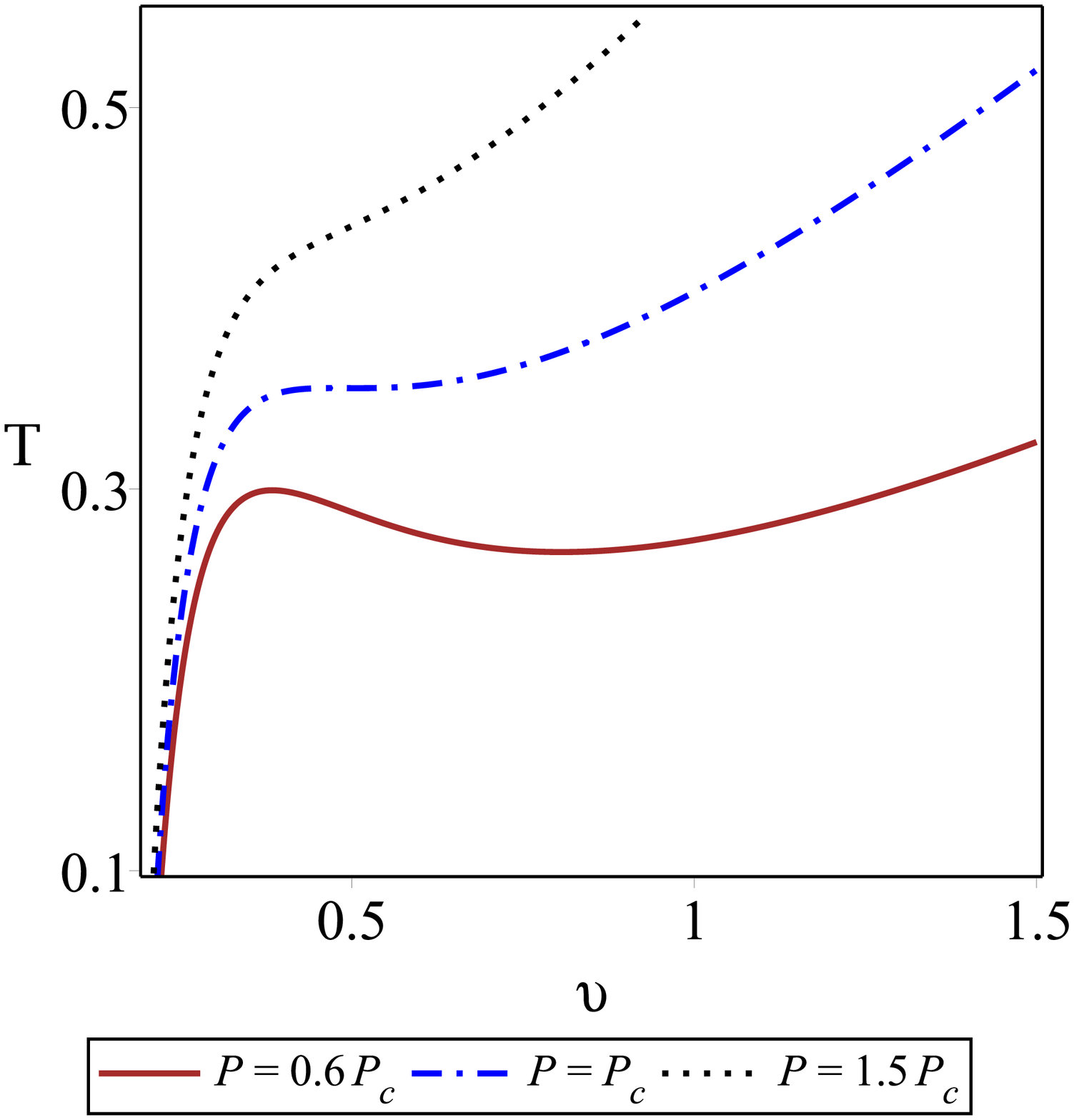}}\newline
\caption{The van der Waals like phase diagrams for $Q=0.1 $, $\protect\omega %
=-\frac{2}{3} $ and $a=0.5$.}
\label{FigPT}
\end{figure}

The corresponding $P-V$ and $T-V$ diagrams are depicted in Fig.
\ref{FigPT}. Evidently, the behavior is reminiscent of the van der
Waals fluid which confirms the first-order small-large BH
transition for temperatures smaller than the critical temperature.
The critical point can be extracted from
\begin{equation}
\frac{\partial p}{\partial \upsilon }\bigg|_{\upsilon =\upsilon _{c},T=T_{c}}=0~~~\&~~~%
\frac{\partial ^{2}p}{\partial \upsilon ^{2}}\bigg|_{\upsilon =\upsilon _{c},T=T_{c}}=0.
\label{EqCpoint11}
\end{equation}
which results into the following equation for calculating critical
volume
\begin{equation}
\upsilon_{c}^{3\omega +1}-24Q^{2}\upsilon_{c}^{3\omega -1}+8^{\omega}\times 9 a \omega (2+5\omega +3\omega^{2})=0.
\label{Eqnuc}
\end{equation}

The critical temperature and pressure is calculated as
\begin{eqnarray}
T_{c} &=&\frac{1}{\pi \upsilon_{c}}-\frac{8Q^{2}}{\pi
\upsilon_{c}^{3}}+ \frac{8^{\omega}\times 9a \omega (\omega
+1)}{\pi \upsilon_{c}^{3\omega +2}},  \nonumber \\
&&  \nonumber \\
P_{c}&=& \frac{1}{2\pi \upsilon_{c}^{2}}-\frac{6Q^{2}}{\pi
\upsilon_{c}^{4}}+\frac{8^{\omega}\times 9a \omega
(2+3\omega)}{\pi \upsilon_{c}^{3\omega +3}}. \label{EqPTc}
\end{eqnarray}

Equation (\ref{Eqnuc}) can be analytically solved for $\omega =
-\frac{2}{3}$, resulting into  the following critical quantities
\begin{equation}
\upsilon_{c}=2\sqrt{6}Q;~~~~T_{c}=\frac{\sqrt{6}}{18\pi Q}-\frac{a}{2\pi};~~~~P_{c}=\frac{1}{96\pi Q^{2}}.
\label{EqCqu}
\end{equation}

It is worthwhile to mention that the critical volume and pressure
are exactly the same as those presented for  RN-AdS BH
\cite{Kubiznak1}, and only the critical temperature is affected by
quintessence dark energy. For $ a=0 $, all these critical
quantities reduce to those of the RN-AdS black hole.

\section{ Critical Behavior of the charged AdS BH surrounded by
quintessence: An alternative approach} \label{Criticality}

Kubiznak and Mann in Ref. \cite{Kubiznak1} showed that charged AdS
black holes have a critical behavior similar to van der Waals
fluid in the extended phase space. Li also employed such an idea
for investigating the critical behavior of the charged AdS BH
surrounded by quintessence \cite{Li}. Although the idea of
considering the variable cosmological constant has attracted a lot
of attention in BH thermodynamics, it was shown that by keeping
the cosmological constant as a fixed parameter and instead
considering the square of electric charge as a thermodynamic
variable, one
can observe such a critical behavior in $Q^{2}-\Psi $ plane \cite%
{Dehyadegari}. The study of phase transition via this alternative
approach was made for the charged AdS BH in the presence of
quintessence field in Ref. \cite{Chabab}. Now, we are interested
in studying the critical behavior of charged AdS black holes
surrounded by quintessence via a new approach by considering both
the cosmological constant (Eq. \ref{Eqpressure}) and electric
charge as fixed external parameters and allow the normalization
factor to vary. Here, we investigate critical behavior of the
system for two different cases to find a proper alternative
approach.

\subsection{ Critical behavior of the BH via approach
I}\label{AppI}

\begin{figure}[!htb]
\centering \subfloat[ $Q=0.1$ and $ \omega=-0.5 $]{
   \label{ay1}     \includegraphics[width=0.31\textwidth]{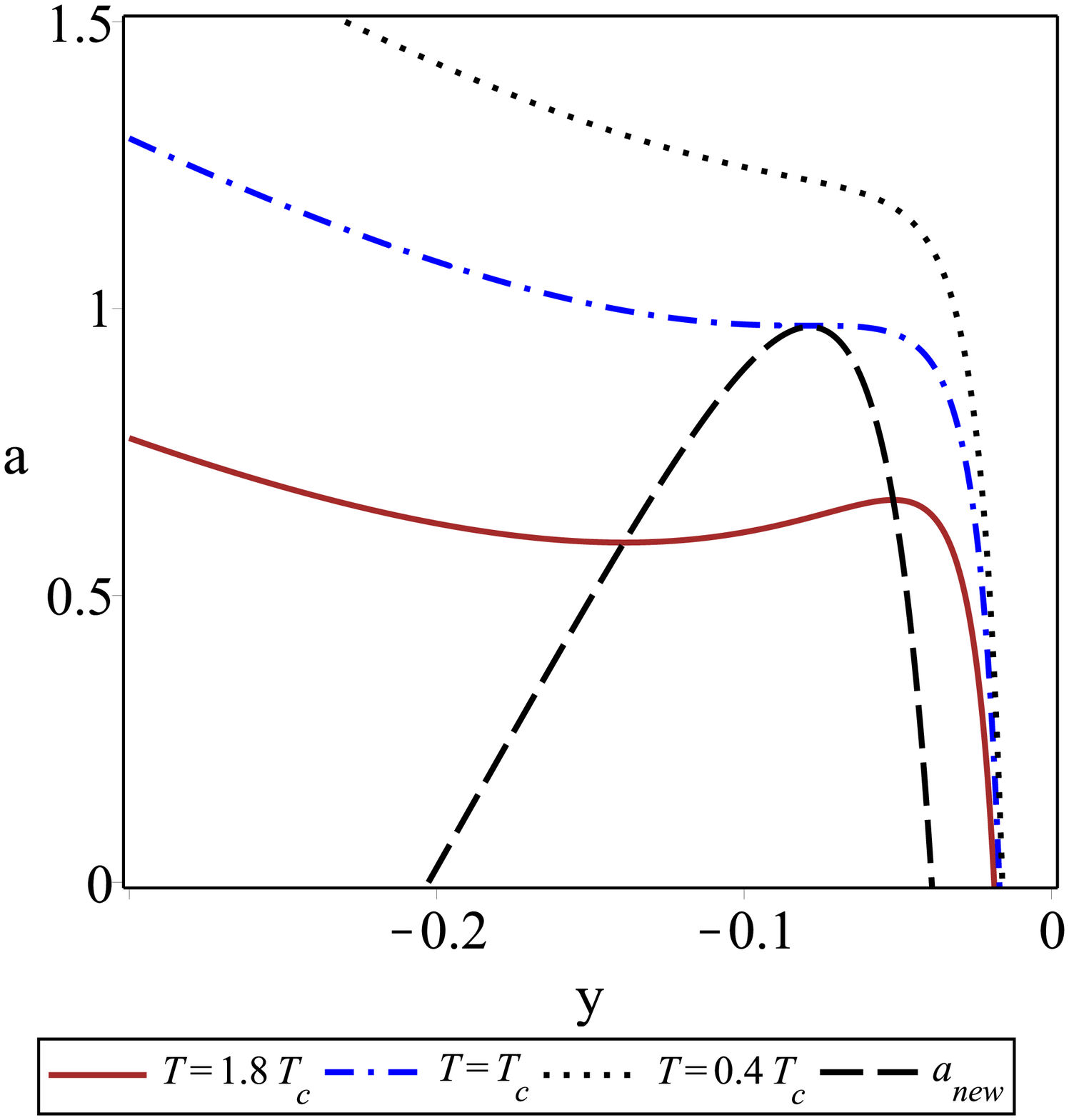}}
\subfloat[$ Q=0.1 $ and $ \omega=-0.5 $]{
     \label{Ga1}   \includegraphics[width=0.31\textwidth]{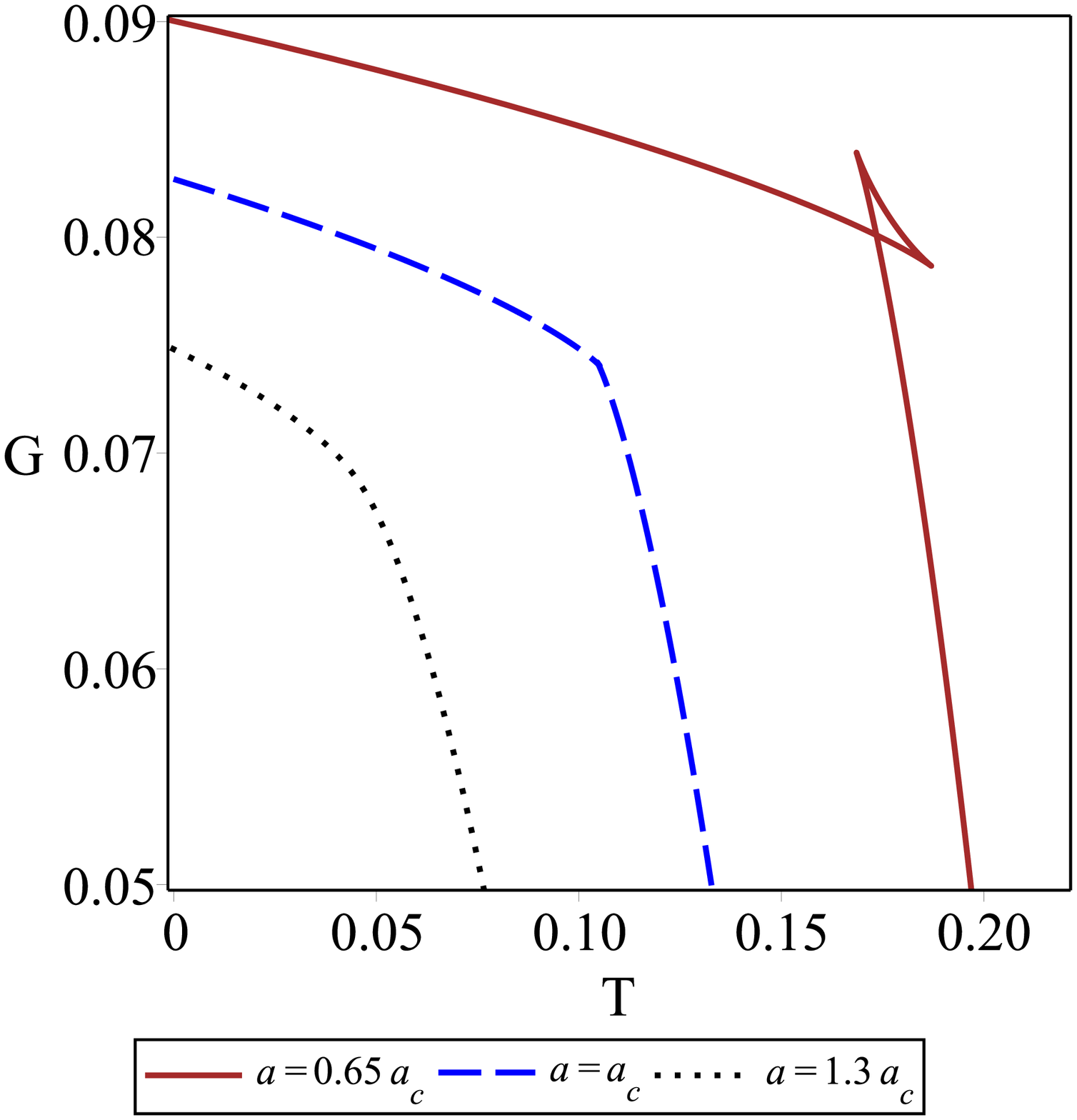}}
 \subfloat[$ Q=0.1 $ and $ \omega=-0.5 $]{
    \label{aT1}    \includegraphics[width=0.33\textwidth]{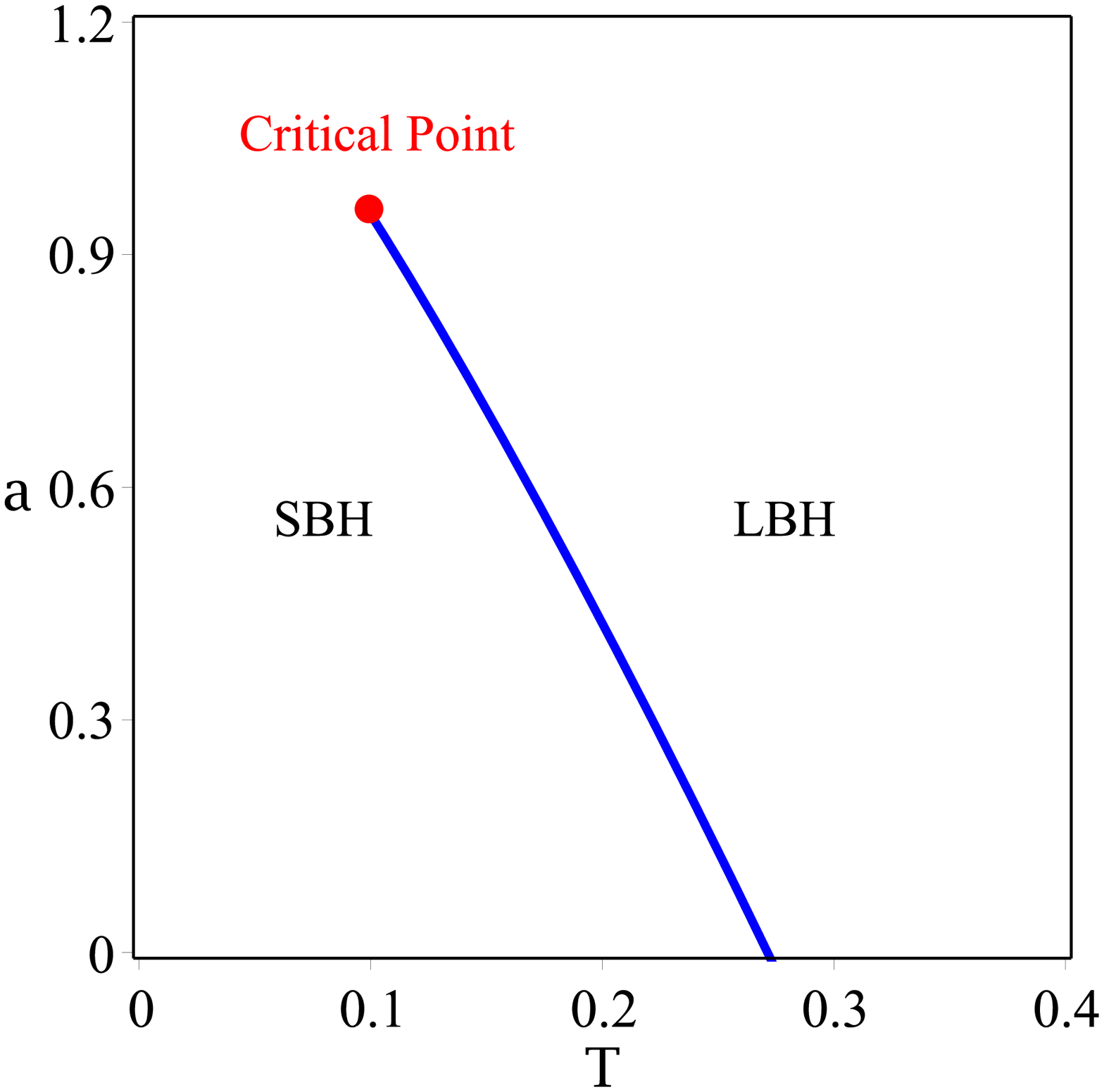}}
\newline
\subfloat[ $Q=0.2$ and $ \omega=-0.8 $]{
   \label{ay2}     \includegraphics[width=0.31\textwidth]{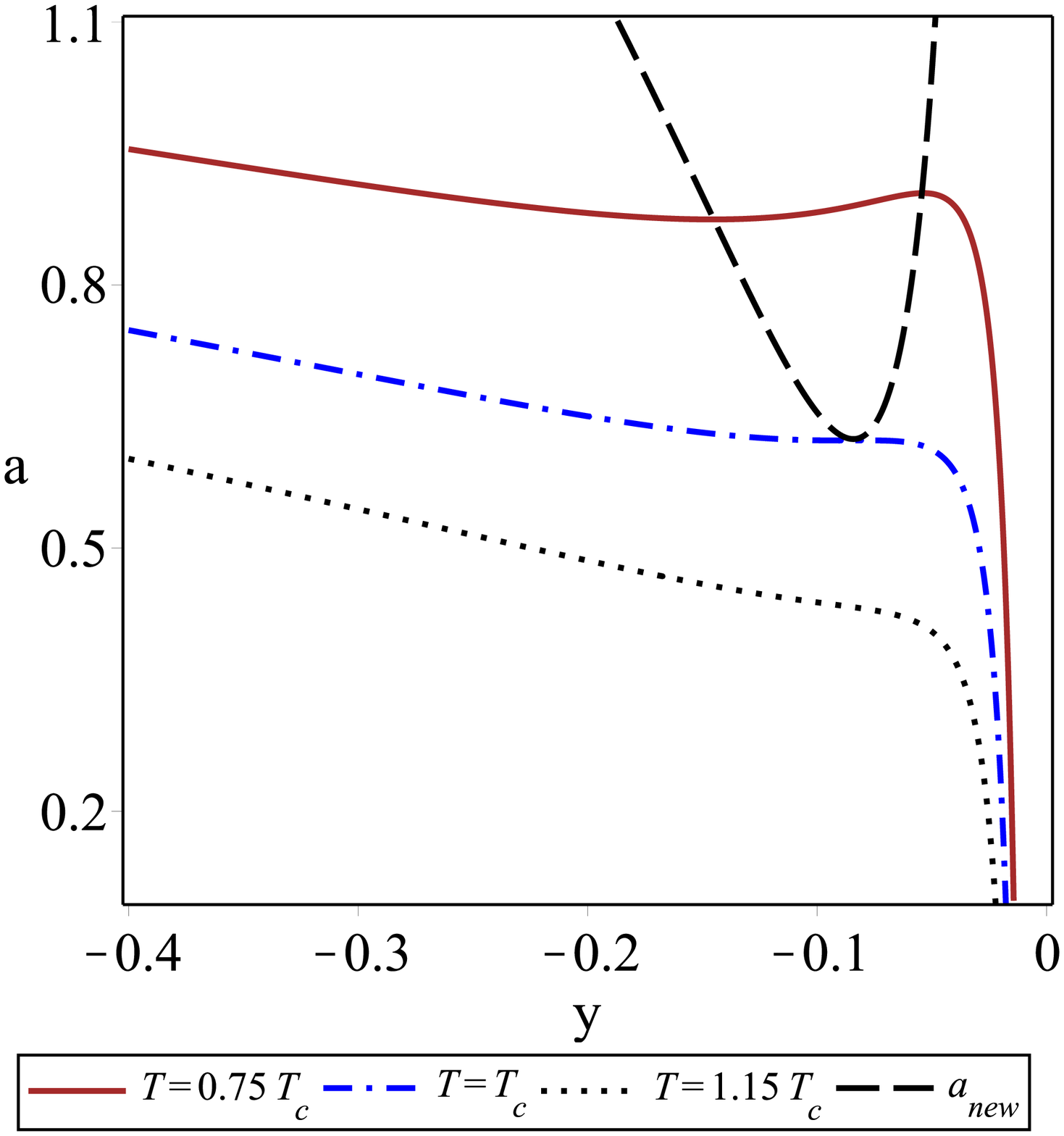}}
\subfloat[$Q=0.2$ and $ \omega=-0.8 $]{
    \label{Ga2}    \includegraphics[width=0.31\textwidth]{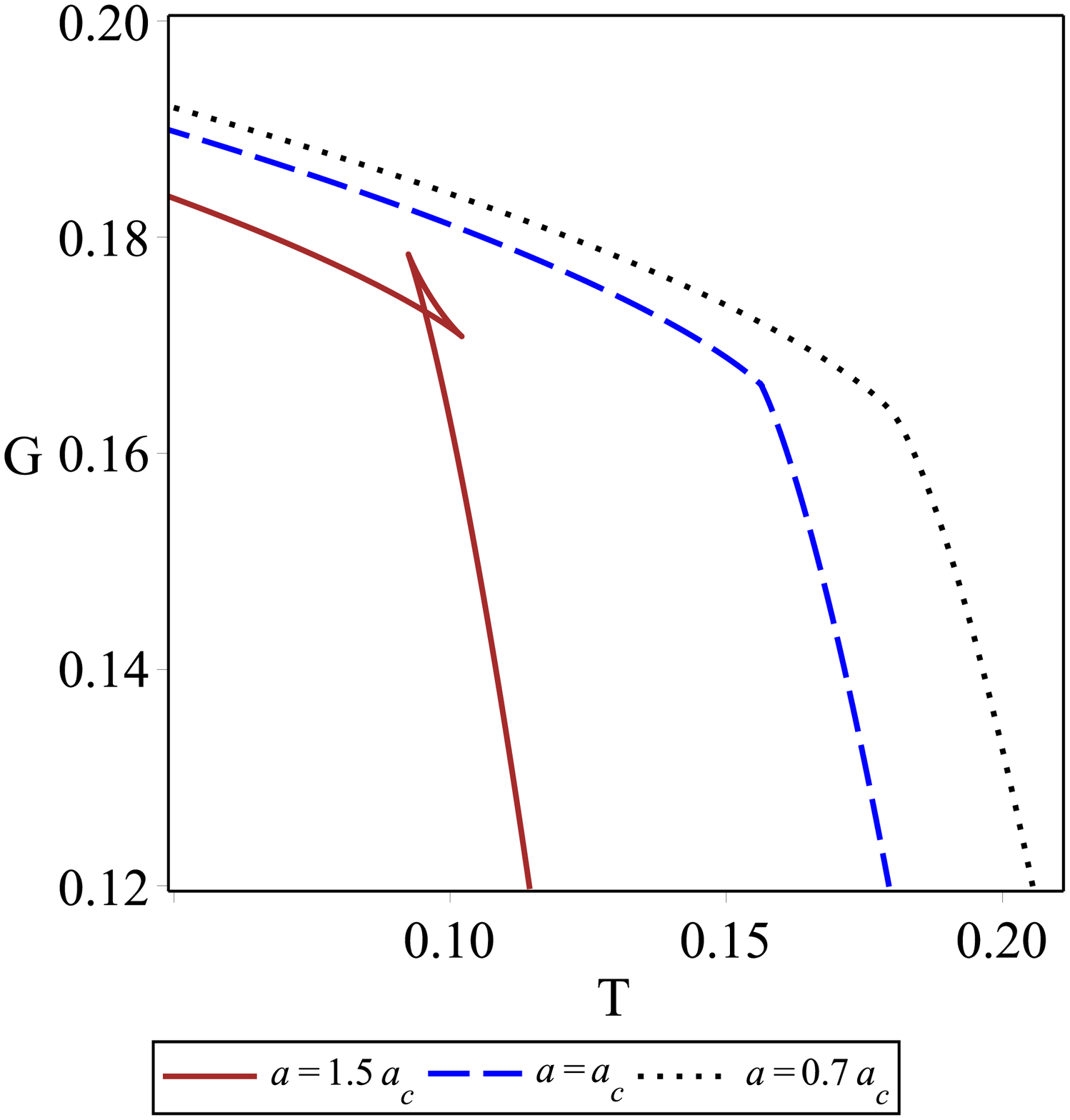}}
 \subfloat[$Q=0.2$ and $ \omega=-0.8 $]{
    \label{aT2}    \includegraphics[width=0.33\textwidth]{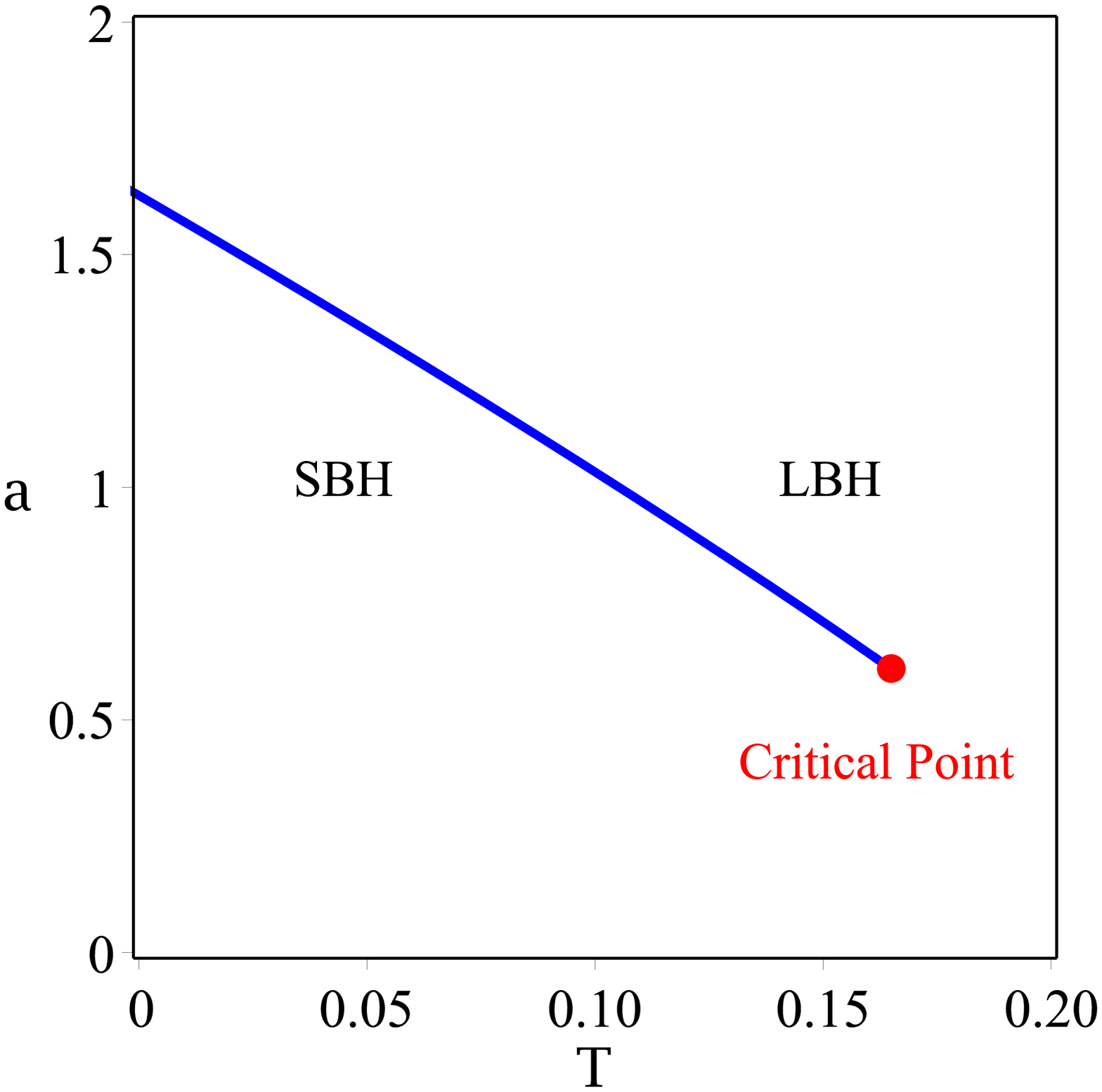}}
\caption{Left panels:  The behavior of isothermal $ a - y$. Middle
panels: the Gibbs free energy diagram $ G - T$. Right panels:
Coexistence curve of small-large BH phase transition in the $a - T
$ plane.} \label{Figay}
\end{figure}

In this subsection, we consider the normalization factor $
a $ as a thermodynamic variable and study the critical behavior of
the system under its variation. We start by writing the equation
of state in the form $a(T,y)$ by using Eq. (\ref{Temp1}).
Inserting Eq. (\ref{Eqintensive1}) into the relation of
temperature, the equation of state is obtained as
\begin{eqnarray}
a(T,y)=\frac{4\pi T}{3\omega (-2y)^{\frac{3\omega +2}{3\omega}}}+\frac{Q^{2}}{3\omega (-2y)^{\frac{3\omega -1}{3\omega}}}-\frac{1}{3\omega (-2y)^{\frac{3\omega +1}{3\omega}}}-\frac{1}{\omega l^{2} (-2y)^{\frac{3\omega +3}{3\omega}}} .  \label{EqaTy}
\end{eqnarray}
In order to investigate the critical behavior of the
system and compare with the van der Waals fluid, we should plot
isotherm diagrams. The corresponding $a - y$ diagram is
illustrated in Figs. \ref{Figay}(a) and  \ref{Figay}(d).  The
diagrams show that, for  constant $ Q $ and $ l $, there is an
inflection point which may be interpreted as the critical point
where two phases of small and large black holes are in
equilibrium. Using the equation of state (\ref{EqaTy}) and the
concept of the inflection point, the critical point can be
characterized by
\begin{equation}
\frac{\partial a}{\partial y }\bigg|_{y =y _{c},T=T_{c}}=0~~~\&~~~%
\frac{\partial ^{2}a}{\partial y ^{2}}\bigg|_{y =y _{c},T=T_{c}}=0,
\label{EqCriticalpoint}
\end{equation}
which leads to
\begin{eqnarray}
y_{c} &=&-\frac{1}{2}\left( \frac{\sqrt{2l}}{6}\sqrt{\frac{l(1+3\omega)+\sqrt{108Q^{2}(1+\omega)(1-3\omega)+l^{2}(1+3\omega)^{2}}}{1+\omega}}\right) ^{-3\omega},  \nonumber \\
&&  \nonumber \\
T_{c}&=& \frac{Q^{2}l^{2}(1-3\omega)(-2y_{c})^{\frac{1}{\omega}}+l^{2}(1+3\omega)(-2y_{c})^{\frac{1}{3\omega}}+9(1+\omega)(-2y_{c})^{-\frac{1}{3\omega}}}{4\pi l^{2}(2+3\omega)},  \nonumber \\
&&  \nonumber \\
a_{c}&=&\frac{3+3Q^{2}l^{2}(-2y_{c})^{\frac{4}{3\omega}}-l^{2}(-2y_{c})^{\frac{2}{3\omega}}}{\omega l^{2}(3\omega +2)(-2y_{c})^{\frac{\omega +1}{\omega}}}. \label{EqaTc}
\end{eqnarray}

Studying the heat capacity, one can confirm the
criticality behavior mentioned above via the method of reported in
Refs. \cite{HendiA,HendiB}. After some manipulations, we find
\begin{eqnarray}
C=\frac{2\pi \left( l^{2}+3(-2y)^{-\frac{2}{3\omega}}-Q^{2}l^{2}(-2y)^{\frac{2}{3\omega}}+3a \omega l^{2}(-2y)^{\frac{1+3\omega}{3\omega}}\right) }{3-l^{2}(-2y)^{\frac{2}{3\omega}}+3Q^{2}l^{2}(-2y)^{\frac{4}{3\omega}}-3(2+3\omega)a \omega l^{2}(-2y)^{\frac{3+3\omega}{3\omega}}}
. \label{EqHCa}
\end{eqnarray}

Solving denominator of the heat capacity with respect to $a $, a new
relation for normalization factor $(a_{new})$ is obtained which is different
from what was obtained in Eq. (\ref{EqaTy}). $ a_{new} $ is obtained as
follows
\begin{eqnarray}
a_{new}=\frac{3+3Q^{2}l^{2}(-2y)^{\frac{4}{3\omega}}
-l^{2}(-2y)^{\frac{2}{3\omega}}}{\omega l^{2}(3\omega
+2)(-2y)^{\frac{\omega +1}{\omega}}}. \label{Eqamax}
\end{eqnarray}

Evidently, the above relation diverges at $ \omega
=-\frac{2}{3} $ . This new relation for normalization factor $ a $
has an extremum which exactly coincides with the inflection point
of $a - y $ diagram (see dashed lines in Figs. \ref{Figay}(a) and
\ref{Figay}(d)). In other words, its extremum is the same critical
normalization factor  and its proportional $ y $ (in which
$a_{new}$ is maximum) is $ y_{c} $. By deriving the new
normalization factor with respect to $ y $, one can obtain $
y_{max} $ as
\begin{eqnarray}
y_{max} &=&-\frac{1}{2}\left( \frac{\sqrt{2l}}{6}
\sqrt{\frac{l(1+3\omega)+\sqrt{108Q^{2}(1+\omega)
(1-3\omega)+l^{2}(1+3\omega)^{2}}}{1+\omega}} \right) ^{-3\omega},
\label{Eqymax}
\end{eqnarray}
which is the same $ y_{c} $ in Eq. (\ref{EqaTc}). It is evident
that by inserting Eq. (\ref{Eqymax}) into Eq. (\ref{Eqamax}), one
can reach $ a_{c} $ (compare $a_{new}  $ to $a_{c}  $ in Eq.
(\ref{EqaTc})). As we see from Fig. \ref{Figay}(a), for $ \omega >
-\frac{2}{3} $, the new normalization factor has a maximum which
matches to the inflection point of $a - y$ diagram. Whereas an
opposite behavior can be observed for $ \omega < -\frac{2}{3} $
(see Fig. \ref{Figay}(d)). The phase structure of a thermodynamic
system can also be characterized by the Gibbs free energy, $
G=M-ST $.  The behavior of the Gibbs free energy in term of $ T $
is depicted in  Figs. \ref{Figay}(b) and \ref{Figay}(e). The
existence of  swallow-tail shape in $G - T$ diagram indicates that
the system has a first order phase transition from small BH to
large black hole. For $ \omega > -\frac{2}{3} $, system undergoes
a first order phase transition for   $ a < a_{c} $ (see Fig.
\ref{Figay}(b)) and $ T> T_{c} $ (see Fig. \ref{Figay}(a)).
Whereas for $ \omega < -\frac{2}{3} $, such a phase transition is
observed for $ a > a_{c} $ and $ T< T_{c} $ (see Figs.
\ref{Figay}(d) and \ref{Figay}(e)). The coexistence line of two
phases of small and large black holes, along which these two
phases are in equilibrium, is obtained  from Maxwell's equal area
law. Figures \ref{Figay}(c) and \ref{Figay}(f) displays the
coexistence line of small-large BH phase transition. The critical
point is highlighted by a small circle at the end of the
coexistence line.

\subsection*{Critical exponents}\label{Crit-Exp}

Critical exponents describe the behavior of physical quantities
near the critical point. For fixed dimensionality and range of
interactions, the critical exponents are independent of the
details of a physical system, and therefore, one may regard them
quasi-universal. Now, we aim to calculate the critical exponents
in this new approach. To do so, we first introduce the following
useful relations
\begin{equation}
C_{y}=\mid t\mid^{-\alpha},~~~\eta=\mid t\mid^{\lambda},~~~\kappa_{T}=\mid
t\mid^{-\gamma},~~~\mid a-a_{c}\mid=\mid y-y_{c}\mid^{\delta}.
\label{EqCexponent}
\end{equation}
where the critical exponents $\alpha$ , $\lambda $, $\gamma$ and $\delta$
describe the behavior of specific heat $C_{y} $, the order parameter $%
\eta $, the isothermal compressibility $\kappa_{T} $ and behavior on the
critical isotherm $T=T_{c} $, respectively. To find the critical exponent,
we define the below dimensionless quantities
\begin{equation}
\xi=\frac{a}{a_{c}},~~~\zeta=\frac{y}{y_{c}},~~~\tau=\frac{T}{T_{c}}.
\label{Eqreduced1}
\end{equation}

Since the critical exponents are studied near the critical point, we can
write the reduced variables in the following form
\begin{equation}
\zeta=1+\nu,~~~\tau=1+t.  \label{EqNereduced1}
\end{equation}

First, we rewrite the entropy in terms of $T $ and $y $ as,
\begin{equation}
S(T,y)=\pi (-2y)^{\frac{2}{3\omega}},  \label{EqNS}
\end{equation}
which is independent of temperature. So, we find that
\begin{equation}
C_{y}=T\frac{\partial S}{\partial T}\bigg|_{y}=0,
\end{equation}
and hence $\alpha =0 $. By using Eq. (\ref{EqNereduced1}),
one can expand Eq. (\ref{EqaTy}) near the critical point as
\begin{equation}
\xi = \mathcal{A}+A_{1}t+\mathcal{A}_{1}\nu t
+\mathcal{A}_{2}\nu+\mathcal{A}_{3}\nu^{2}+\mathcal{A}_{4}\nu^{3}+O(t\nu^{2},
\nu^{4}), \label{Eqnu}
\end{equation}
where
\begin{eqnarray}
\mathcal{A}_{1}&=&-\frac{1}{3\omega}A_{1}(2+3\omega),  \nonumber \\
&&  \nonumber \\
\mathcal{A}_{2}&=&-\frac{1}{3\omega}\left( 3\mathcal{A}\omega +2A_{1}-A_{2}+A_{3}+3A_{4}\right) ,  \nonumber \\
&&  \nonumber \\
\mathcal{A}_{3}&=&\frac{1}{18\omega^{2}}\left( 9\omega (2A_{1}-A_{2}+A_{3}+3A_{4})+3A_{1}+8A_{4}+\mathcal{A}+18\mathcal{A}\omega^{2}\right),  \nonumber \\
&&  \nonumber \\
\mathcal{A}_{4}&=& -\frac{1}{162\omega^{3}}\left( 99\omega^{2} (2A_{1}-A_{2}+A_{3}+3A_{4})+18\omega (\mathcal{A}+3A_{1}+8A_{4})+8A_{1}-A_{2}+A_{3}+27A_{4}+162\mathcal{A}\omega^{3} \right)
,\label{EqaTc1}
\end{eqnarray}
and
\begin{eqnarray}
\mathcal{A}&=& A_{1}+A_{2}+A_{3}+A_{4},  ~~~A_{1}=\frac{4\pi T_{c}}{3\omega a_{c}(-2y_{c})^{\frac{3\omega +2}{3\omega}}},~~~
A_{2}=\frac{Q^{2}}{3\omega a_{c}(-2y_{c})^{\frac{3\omega -1}{3\omega}}},\nonumber \\
&&  \nonumber \\
A_{3}&=& -\frac{1}{3\omega a_{c}(-2y_{c})^{\frac{3\omega +1}{3\omega}}}, ~~~
A_{4}= -\frac{1}{\omega a_{c}l^{2}(-2y_{c})^{\frac{3\omega +3}{3\omega}}}
. \label{EqaTc11}
\end{eqnarray}

\begin{table}[tbp]
\begin{center}
\begin{tabular}{||c| c| c| c| c| c| c| c| c|c|c||}
\hline\hline $Q $ & $\omega$ & $y_{c}$ & $T_{c}$ &
$a_{c}$&$\mathcal{A}$& $A_{1}$ & $\mathcal{A}_{1}$&
$\mathcal{A}_{2}$& $|\mathcal{A}_{3}|$& $\mathcal{A}_{4}$ \\ \hline\hline
$~0.1~$ &$~-0.4~$ & $~-0.12~$ & $~0.17~$ & $~0.52~$ &
$~0.99~$  & $~-1.35~$ & $~-0.9~$  & $~< 10^{-10}~$ & $~<
10^{-10}~$& $~0.46~$ \\ \hline
$~0.1~$ & $~-0.5~$ & $~-0.07~$ &
$~0.09~$  & $~0.96~$  & $~1.0~$ & $~0.45~$  & $~-0.15~$& $~<
10^{-10}~$& $~< 10^{-10}~$ & $~0.15~$ \\ \hline
$~0.15~$ & $~-0.6~$ & $~-0.09~$ & $~0.16~$  & $~0.59~$  & $~1.0~$
& $~-1.57~$  & $~-0.17~$& $~<10^{-9}~$& $~< 10^{-10}~$ & $~0.21~$ \\ \hline
 $~0.18~$ & $~-0.7~$ & $~-0.08~$ & $~0.11~$  & $~0.96~$  & $~1.0~$
& $~-0.7~$  & $~0.03~$& $~<10^{-9}~$& $~< 10^{-10}~$ & $~0.09~$ \\ \hline
$~0.2~$ & $~-0.8~$ & $~-0.08~$ & $~0.16~$  & $~0.61~$  &
$~0.99~$ & $~-1.87~$  & $~0.31~$& $~<10^{-9}~$& $~< 10^{-9}~$ &
$~0.10~$ \\ \hline
$~0.2~$ & $~-0.9~$ & $~-0.06~$ & $~0.20~$  & $~0.39~$  &
$~1.0~$ & $~-3.98~$  & $~1.03~$& $~<10^{-9}~$& $~< 10^{-9}~$ &
$~0.11~$ \\ \hline \hline
\end{tabular}
\end{center}
\caption{Numerical solution of the coefficients $  \mathcal{A}_{i}$ for $ l=1 $.}
\label{tab1}
\end{table}

Considering table \ref{tab1}, it is evident that the
coefficients $  \mathcal{A}_{2}$ and $  \mathcal{A}_{3}$ can be
ignorable. So, Eq. (\ref{Eqnu}) reduces to
\begin{equation}
\xi = \mathcal{A}+A_{1}t+\mathcal{A}_{1}\nu t+\mathcal{A}_{4}\nu^{3}
,  \label{Eqnu2}
\end{equation}

Differentiating Eq. (\ref{Eqnu2}) with respect to $ \nu $
for a fixed $ t $, we get
\begin{equation}
da= a_{c}(\mathcal{A}_{1} t+3\mathcal{A}_{4}\nu^{2})d\nu
,  \label{Eqda}
\end{equation}

Now, using the fact that the normalization factor remains
constant during the phase transition and employing the Maxwell's
area law, we have the following two equations:
\begin{eqnarray}
\xi &=&  A_{1}t+\mathcal{A}_{1}\nu_{l} t+\mathcal{A}_{4}\nu_{l}^{3}=A_{1}t+\mathcal{A}_{1}\nu_{s} t+\mathcal{A}_{4}\nu_{s}^{3},  \nonumber \\
&&  \nonumber \\
0&=&\int_{\nu_{l}}^{\nu_{s}}\nu (\mathcal{A}_{1} t+3\mathcal{A}_{4}\nu^{2})d\nu ,
\label{EqMaxwell2}
\end{eqnarray}
where $\nu_{s} $ and $\nu_{l} $ denote the event horizon
of small and large black holes, respectively. Equation
(\ref{EqMaxwell2}) has a unique non-trivial solution given by
\begin{equation}
\nu_{s}=-\nu_{l}=\sqrt{-\frac{\mathcal{A}_{1}}{\mathcal{A}_{4}}t}.
\label{Eqnusl}
\end{equation}

According to the table \ref{tab1}, the argument under the square
root function is always positive. From Eq. (\ref{Eqnusl}), one can
find that
\begin{equation}
\eta=y_{c}(\nu_{l}-\nu_{s})=2y_{c}\nu_{l}=2\sqrt{-\frac{\mathcal{A}_{1}}{\mathcal{A}_{4}}t}
~~\Longrightarrow~~\lambda=\frac{1}{2}.
\end{equation}

Now, we can differentiate Eq. (\ref{Eqnu2}) to calculate the
critical exponent $\gamma $ as
\begin{equation}
\kappa_{T}=-\frac{1}{y}\frac{\partial y}{\partial a}\bigg|_{T}\propto
\frac{y_{c}}{ \mathcal{A}_{1}  a_{c}t}~~\Longrightarrow~~\gamma=1.
\end{equation}

Finally, the shape of the critical isotherm $t = 0$ is given by
\begin{equation}
\xi - \mathcal{A}=\mathcal{A}_{4}
\nu^{3}~~\Longrightarrow~~\delta=3.
\end{equation}

The obtained results show that the critical exponents in this new
approach (with fixed $\Lambda $ and variable $a $) are the same as
those obtained in \cite{Li} (with $\Lambda $ variable and fixed $a
$) and coincide with the van der Waals fluid system
\cite{Kubiznak1}.

\subsection{ Critical behavior of the BH via approach II}

One of the issues in previous section \ref{AppI}
(considering the positive normalization factor as a thermodynamic
variable) is that we could not interpret (or adapt) the positive
$a$ and its negative definite conjugate $y=\left(\frac{\partial
M}{\partial a}\right)_{P,Q,S}=-\frac{1}{2}r_{+}^{-3\omega}$
($[y]=[length]^{-3\omega}$) with the known thermodynamical
quantities. Besides, considering Fig. \ref{FigPT}a with Fig.
\ref{Figay}d, one has to apply a rotation of $\pi$ radian for an
appropriate comparison.

Here, we define a new parameter depending on the
normalization factor with $[length]^{-2}$ dimensions to adapt it
(its conjugate) as an ad hoc pressure (volume). Here, we consider
negative values of normalization factor (with negative $\omega$)
to define a positive variable $\beta$ as
\begin{equation}
\beta =\frac{3a \omega}{8\pi l^{3\omega +3}}.
\label{beta}
\end{equation}

Inserting Eq. (\ref{beta}) into Eq. (\ref{EqFr}), the metric
function can be rewritten as
\begin{equation}
f(r)=1-\frac{2M}{r}+\frac{Q^{2}}{r^{2}}-\frac{8\pi}{3}\frac{\beta l^{3\omega +3}}{\omega r^{3\omega +1}}+\frac{r^{2}}{l^{2}}.
\label{Fr2}
\end{equation}

Before discussing van der Waals phase transition, we determine the
admissible space of the parameters to ensure a well-posed
thermodynamics. To do so,  we require to investigate  the
existence of a BH in the bulk. We can do it by studying the
extremal BH criteria
\begin{equation}
f(r_{e})=0=f^{\prime}(r_{e}),
\label{Extrem}
\end{equation}

Equation (\ref{Extrem}) shows that the function $f(r)$ has one
degenerate horizon at $r_{e}$ which corresponds to the coincidence
of the inner and outer BH horizons. Solving these two equations,
simultaneously, leads to
\begin{eqnarray}
M &=& \frac{4\pi \beta (3\omega -1) l^{3\omega +5} +3\omega
r_{e}^{3\omega +1}(l^{2}+2r_{e}^{2})}{3\omega
l^{2}r_{e}^{3\omega}}, \label{EeExt1M} \\  Q &=&
\frac{\sqrt{\omega^{2} r_{e}^{3\omega +3}\left( 8\pi \beta
l^{3\omega +5}+l^{2}r_{e}^{3\omega +1} +3r_{e}^{3\omega
+3}\right)}}{l \vert \omega \vert r_{e}^{3\omega +1}}.
\label{EeExt1Q}
\end{eqnarray}

\begin{figure}[!htb]
\centering \subfloat[$ \beta=0.1 $ and $ \omega=-0.5 $]{
        \includegraphics[width=0.32\textwidth]{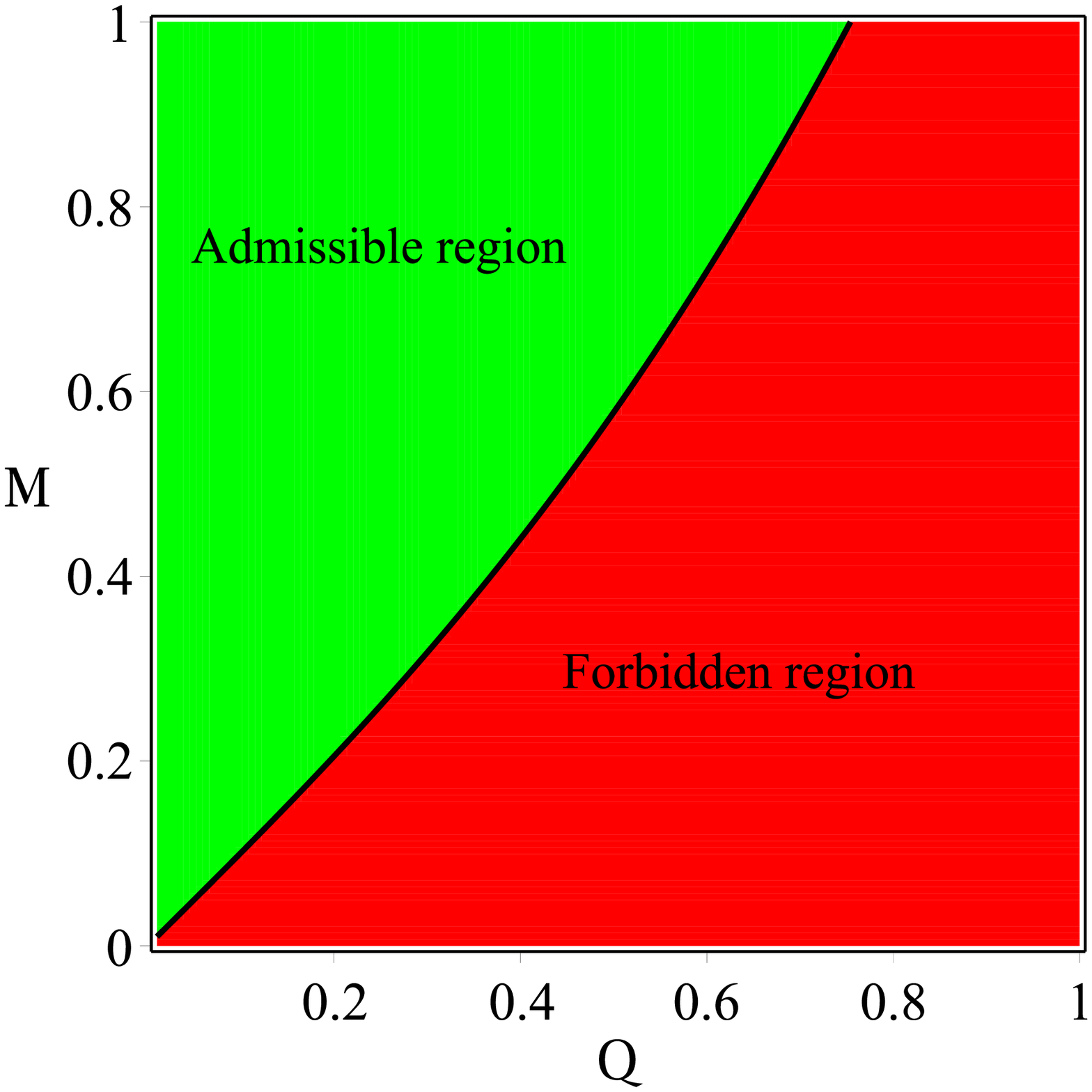}}
\subfloat[  $ Q=0.2 $ and $ \omega=-0.5 $]{
        \includegraphics[width=0.32\textwidth]{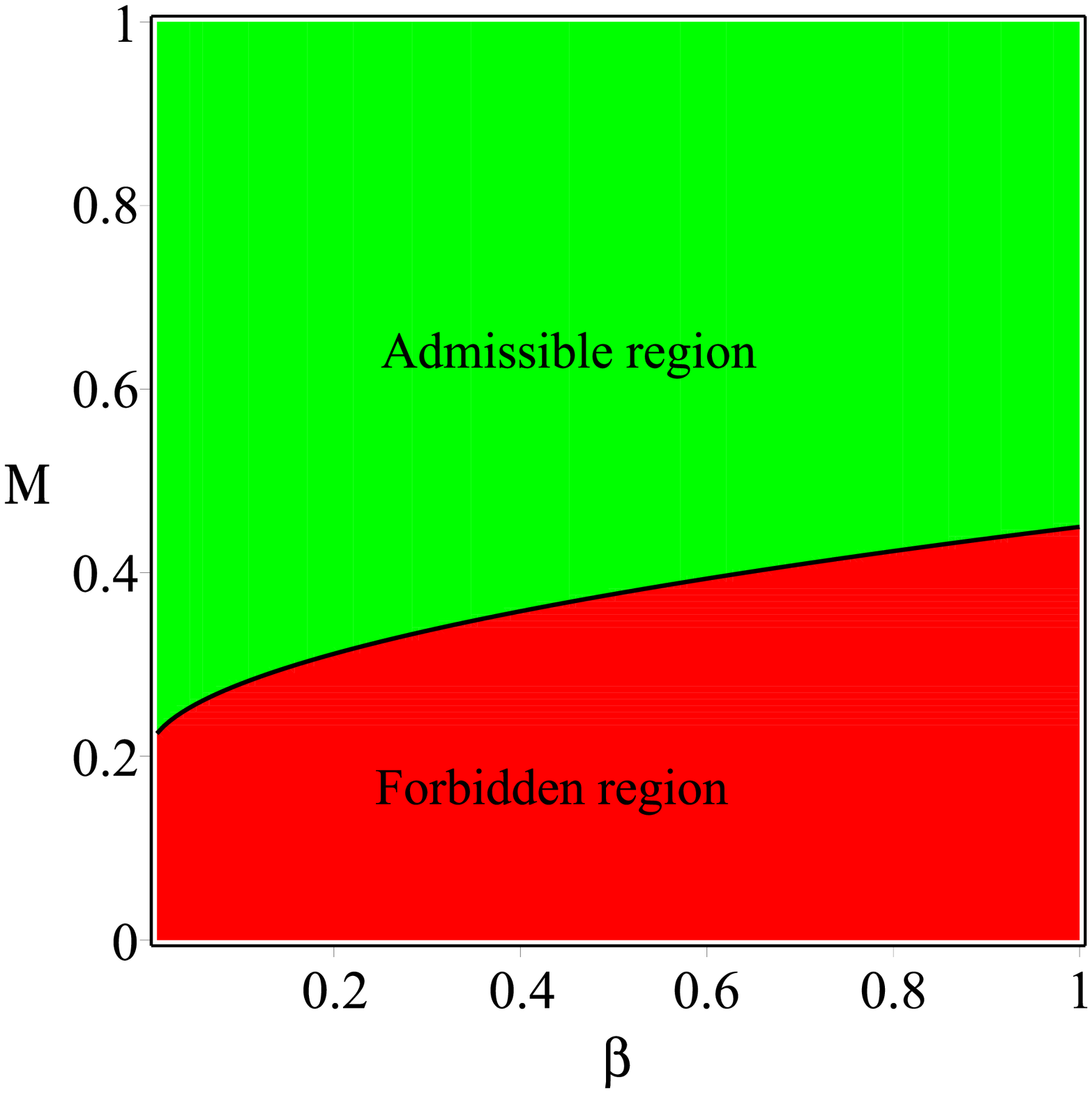}}
\subfloat[ $ M=1 $ and $ \omega=-0.5 $]{
        \includegraphics[width=0.32\textwidth]{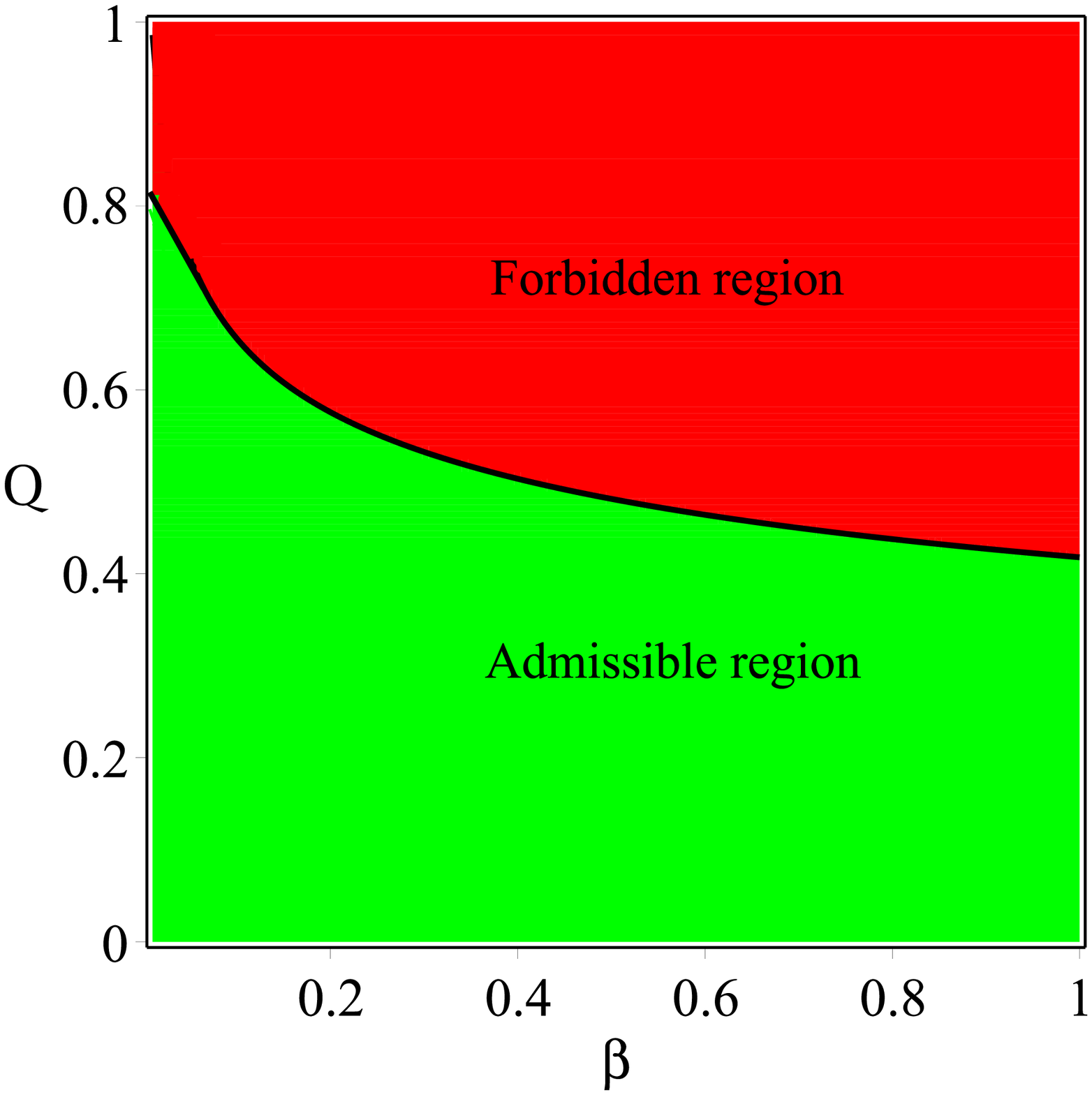}}
\newline
\caption{ The admissible parameter space for $ l=1 $. The black curve is the boundary for the existence of
black holes in the bulk, with extremal black holes sitting on the curve. } \label{Figadmissible}
\end{figure}

According to Eqs.  (\ref{EeExt1M}) and (\ref{EeExt1Q}), we can
plot the admissible parameter space.  The resultant curve is
depicted in Fig. \ref{Figadmissible}a by the black line. This
curve, denoting the extremal limit, provides a lower bound for the
existence of the black hole. Above this line, a BH (with two
horizons) is present, whereas no BH exists below it. In the same
way, one can determine the admissible parameter space in the
$M-\beta$ plane (see Fig. \ref{Figadmissible}b). We use the
following equations for the mentioned plane
\begin{eqnarray}
M &=& \frac{(3\omega -1) Q^{2}l^{2}+(3\omega +1)l^{2}r_{e}^{2}+3(\omega +1)r_{e}^{4}}{6\omega l^{2}r_{e}}, \\ \nonumber
\beta &=& \frac{(Q^{2}l^{2}-l^{2}r_{e}^{2} -3r_{e}^{4})r_{e}^{3\omega -1}}{8\pi l^{3\omega +5}}.
\label{EeExt2}
\end{eqnarray}

We can also use the corresponding equations to plot the $Q-\beta$
plane (Fig. \ref{Figadmissible}c) as
\begin{eqnarray}
Q &=& \frac{\sqrt{\omega^{2}r_{e}^{6\omega +3}(6M\omega l^{2}-(3\omega +1)l^{2}r_{e} -(3\omega +3)r_{e}^{3})}}{l \vert \omega \vert \sqrt{(3\omega -1)} r_{e}^{3\omega +1}}, \\ \nonumber
\beta &=& \frac{3\omega r_{e}^{3\omega}(Ml^{2}-r_{e}l^{2} -2r_{e}^{3})}{4\pi (3\omega -1) l^{3\omega +5}}.
\label{EeExt3}
\end{eqnarray}


Now, we study van der Waals phase transition by the use of definition (\ref{beta}). Inserting this relation into Eq. (\ref{Temp1}), one finds
\begin{equation}
T = \frac{1}{4\pi}\left( \frac{1}{r_{+}}-\frac{Q^{2}}{r_{+}^{3}}+\frac{3r_{+}}{l^{2}}+\frac{8\pi \beta l^{3\omega +3}}{r_{+}^{3\omega +2}}\right) ,
\label{TH1}
\end{equation}

From Eq. (\ref{TH1}), the equation of state $ \beta (T,r_{+})$ is given by
\begin{equation}
\beta = \frac{Tr_{+}^{3\omega +2}}{2l^{3\omega +3}}+\frac{Q^{2}r_{+}^{3\omega -1}}{8\pi l^{3\omega +3}}-\frac{r_{+}^{3\omega +1}}{8\pi l^{3\omega +3}}-\frac{3r_{+}^{3\omega +3}}{8\pi l^{3\omega +5}},
\label{beta1}
\end{equation}
where for $\omega =-1 $, Eq. (\ref{beta1}) reduces to
\begin{equation}
\beta = \frac{T}{2r_{+}}+\frac{Q^{2}}{8\pi r_{+}^{4}}-\frac{1}{8\pi r_{+}^{2}}-\frac{3}{8\pi l^{2}},
\label{beta2}
\end{equation}

Substituting Eq. (\ref{beta}) in Eq. (\ref{Eqmass}) and
differentiating with respect to $ \beta $, one can obtain
conjugate quantity of $\beta $ as
\begin{equation}
\chi =\frac{\partial M}{\partial \beta}=\frac{4\pi l^{3\omega +3}}{3\vert\omega\vert r_{+}^{3\omega}}.
\label{chi}
\end{equation}

\begin{figure}[!htb]
\centering \subfloat[ $Q=0.1$ and $ \omega=-0.8 $]{
        \includegraphics[width=0.32\textwidth]{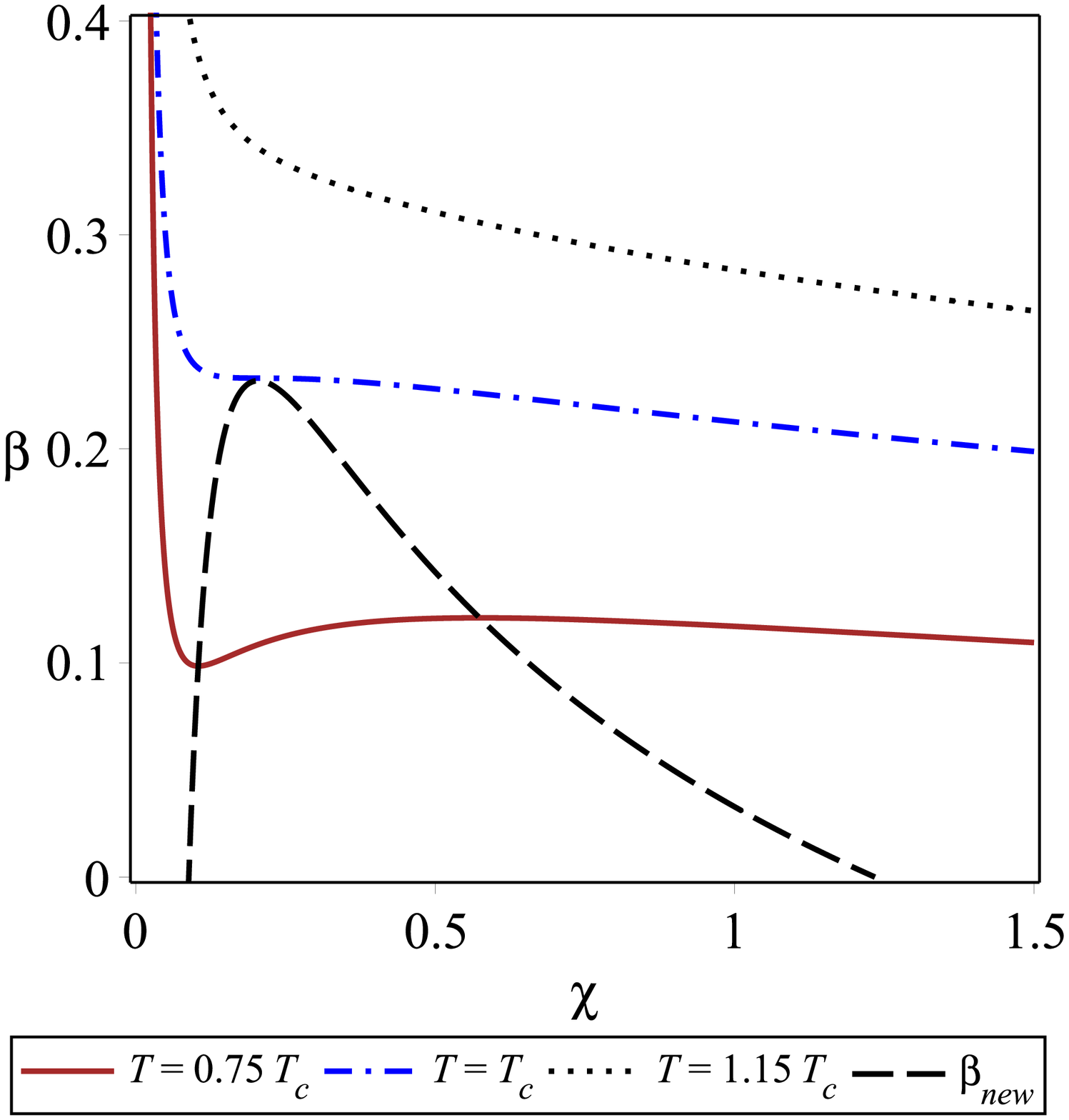}}
\subfloat[ $Q=0.1$ and $ \omega=-0.8 $]{
        \includegraphics[width=0.31\textwidth]{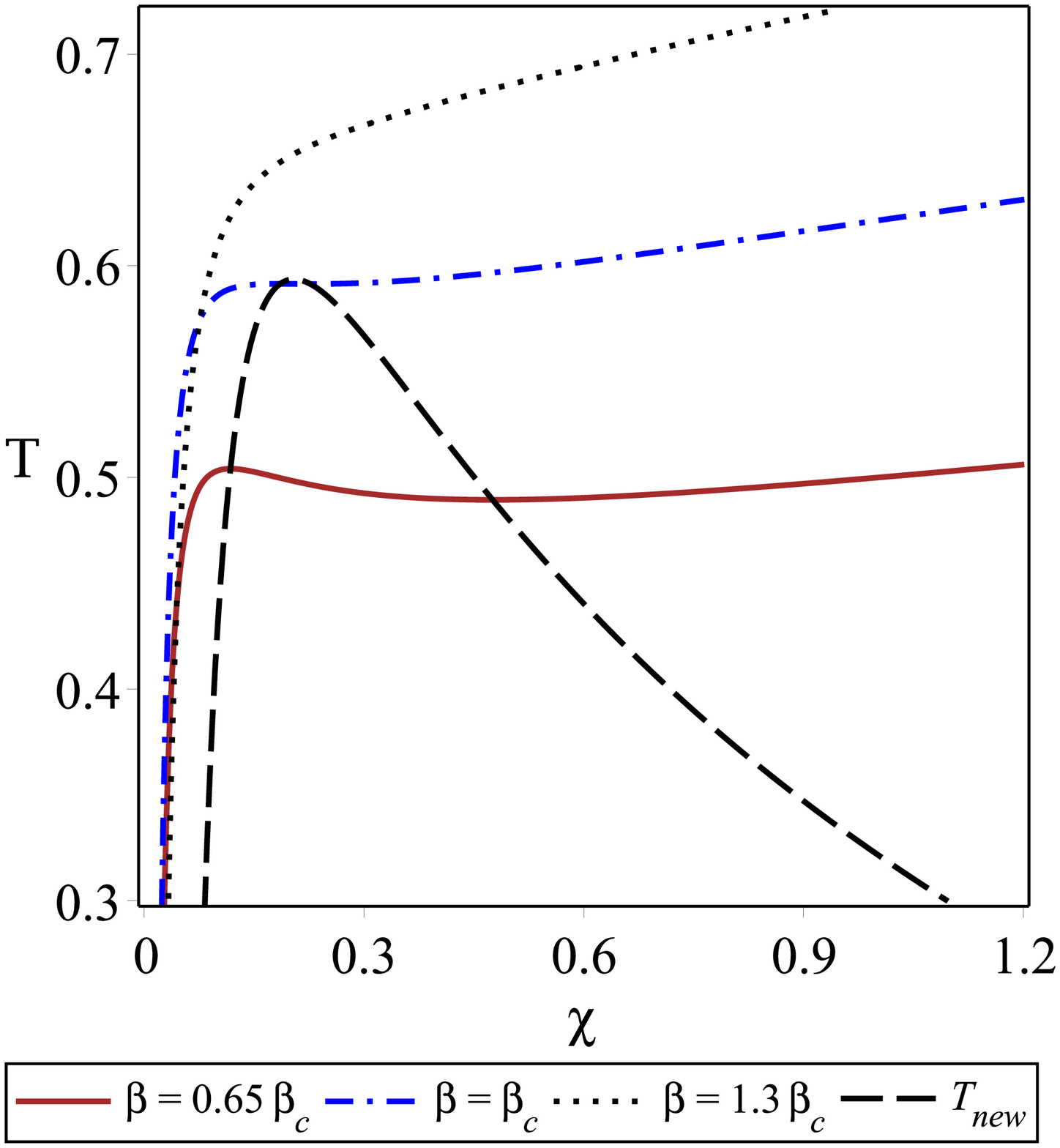}}\newline
\subfloat[ $Q=0.2$ and $ \omega=-0.5 $]{
        \includegraphics[width=0.32\textwidth]{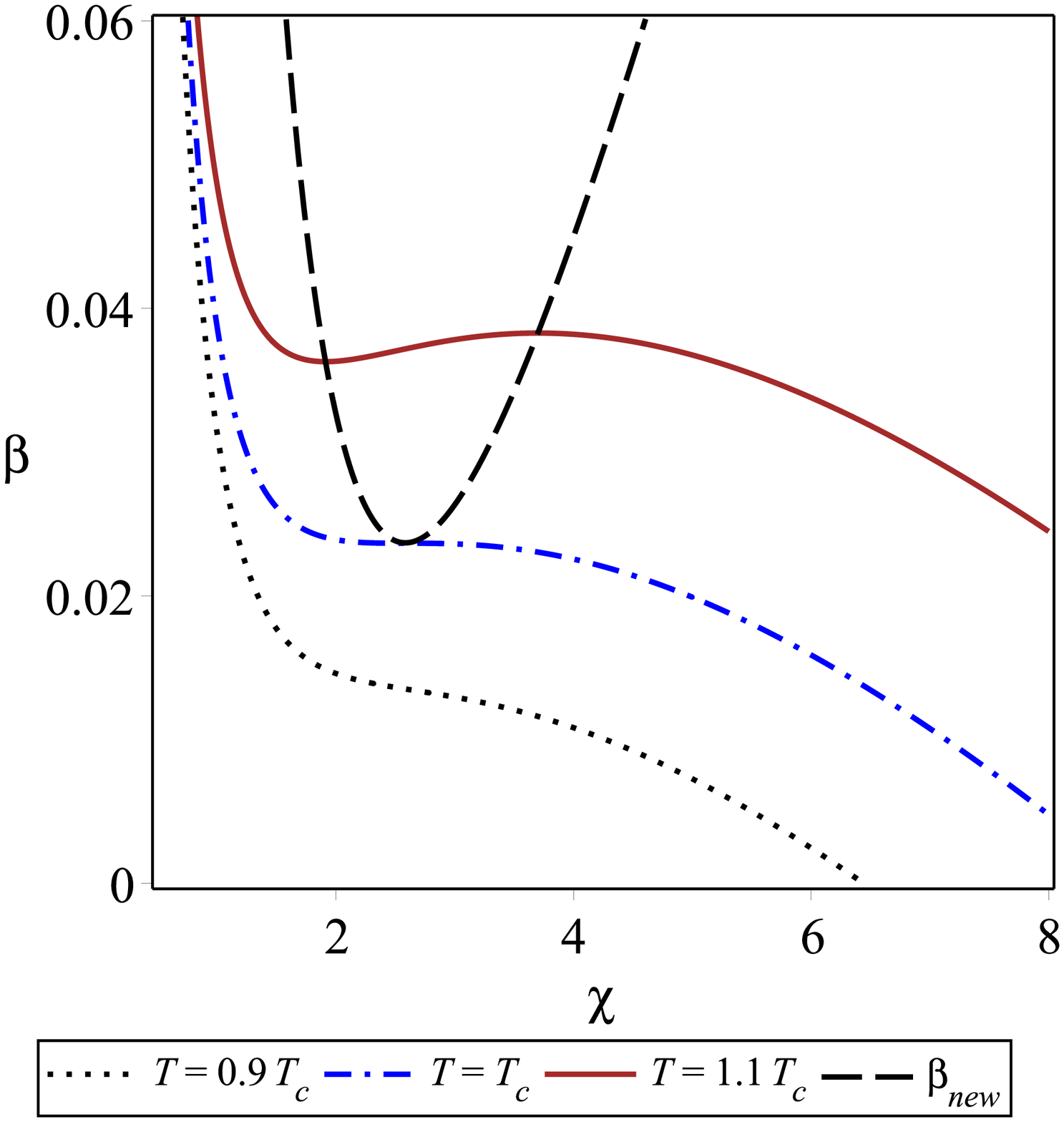}}
\subfloat[ $Q=0.2$ and $ \omega=-0.5 $]{
        \includegraphics[width=0.305\textwidth]{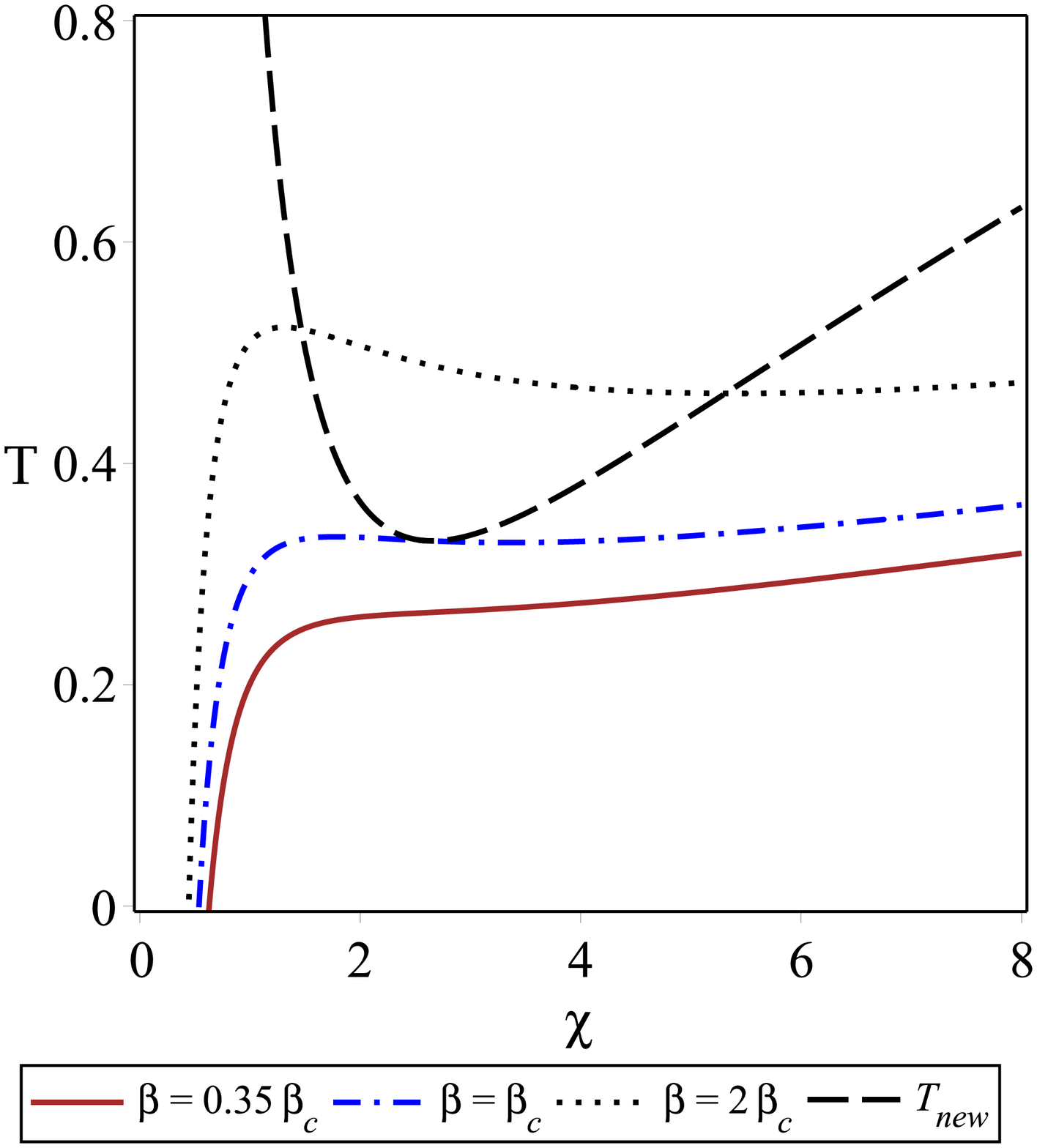}}\newline
\caption{ van der Waals like phase diagrams for $l=1$. Left panels: $\beta$ (continuous, dash-dotted and dotted line)
for different temperatures and $\beta_{new} $ (dashed line) versus $\protect\chi$. Right panels: $T$ (continuous, dash-dotted and dotted line) for different
$ \beta $ and $T_{new} $ (dashed line) versus $\protect\chi$.} \label{Figax}
\end{figure}

According to dimensional analysis, it is worth mentioning
that we can match $\beta$ and $\chi$ to ad hoc pressure
($[\beta]=[length]^{-2}$) and volume ($[\chi]=[length]^{3}$),
respectively.  Equation (\ref{beta1}) can be rewritten in terms
of $ \chi $ as
\begin{equation}
\beta = \frac{\mathcal{B}^{3\omega -1}\chi^{\frac{1-3\omega}{3\omega}}\left(Q^{2}l^{2}+4\pi T l^{2}\mathcal{B}^{3}\chi^{-\frac{1}{\omega}}
-\mathcal{B}^{2}l^{2}\chi^{-\frac{2}{3\omega}}-3\mathcal{B}^{4}\chi^{-\frac{4}{3\omega}}\right) }{8\pi l^{3\omega +5}},
\label{beta3}
\end{equation}
where $\mathcal{B}=\left(\frac{4\pi l^{3\omega
+3}}{3\vert\omega\vert }\right) ^{\frac{1}{3\omega}}$. The
behavior  $ \beta $ and $ T $ as a function of $ \chi $ is
depicted in Fig. \ref{Figax} which shows that the type of phase
transition is van der Waals like. Taking a look at Figs.
\ref{Figax}a and \ref{Figax}b, one can find that for $ \omega
<-\frac{2}{3} $, the van der Waals like phase transition occurs
for $ T<T_{c} $ and $ \beta<\beta_{c} $, whereas for $ \omega
>-\frac{2}{3} $, such a behavior is possible for $ T>T_{c} $ and $
\beta>\beta_{c} $ (see Figs. \ref{Figax}c and \ref{Figax}d).

\begin{figure}[!htb]
\centering
\subfloat[$T=0.2$ and $\omega =-0.8$]{
        \includegraphics[width=0.32\textwidth]{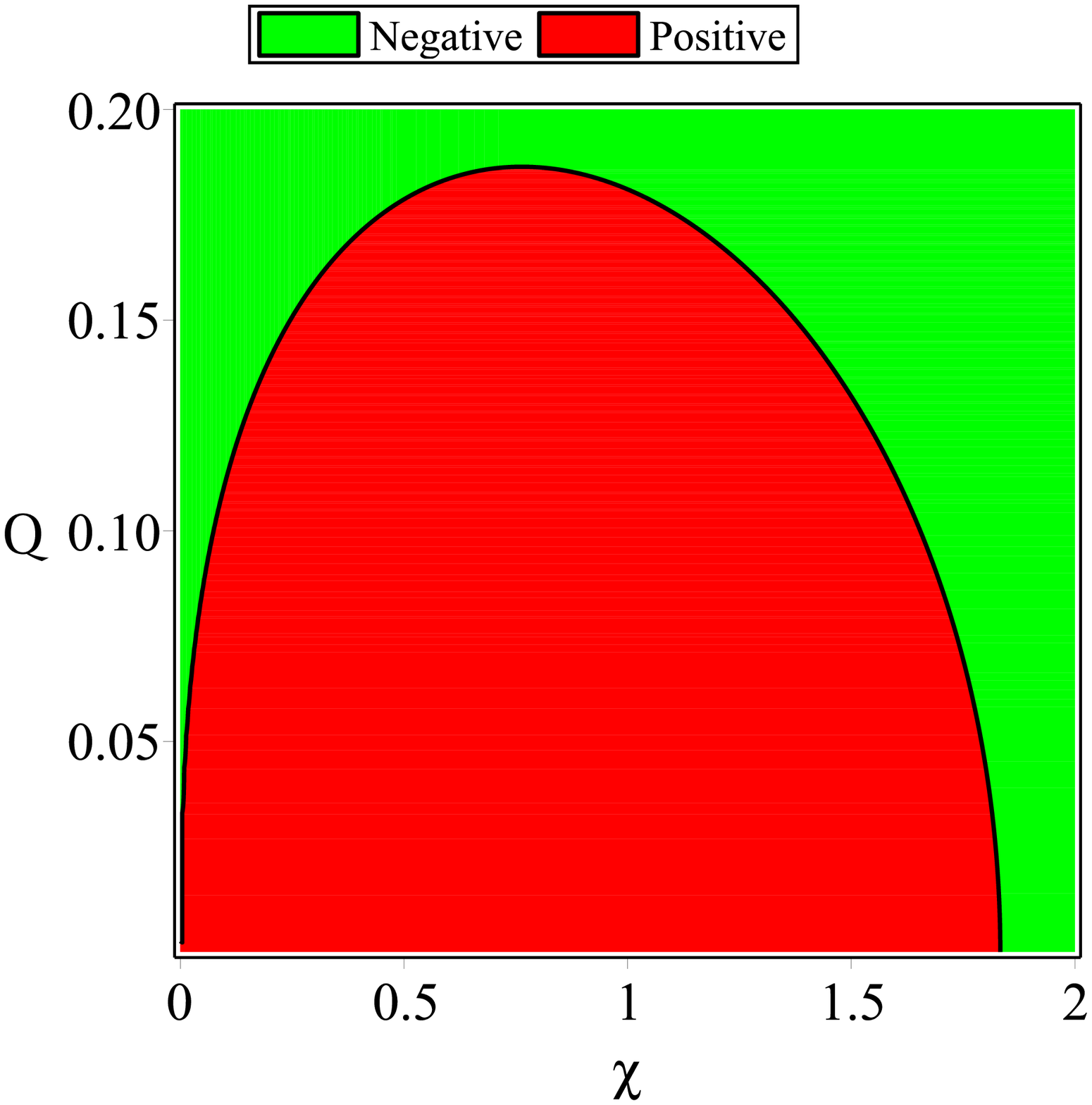}}
\subfloat[$Q=0.1$ and $\omega =-0.8$]{
        \includegraphics[width=0.32\textwidth]{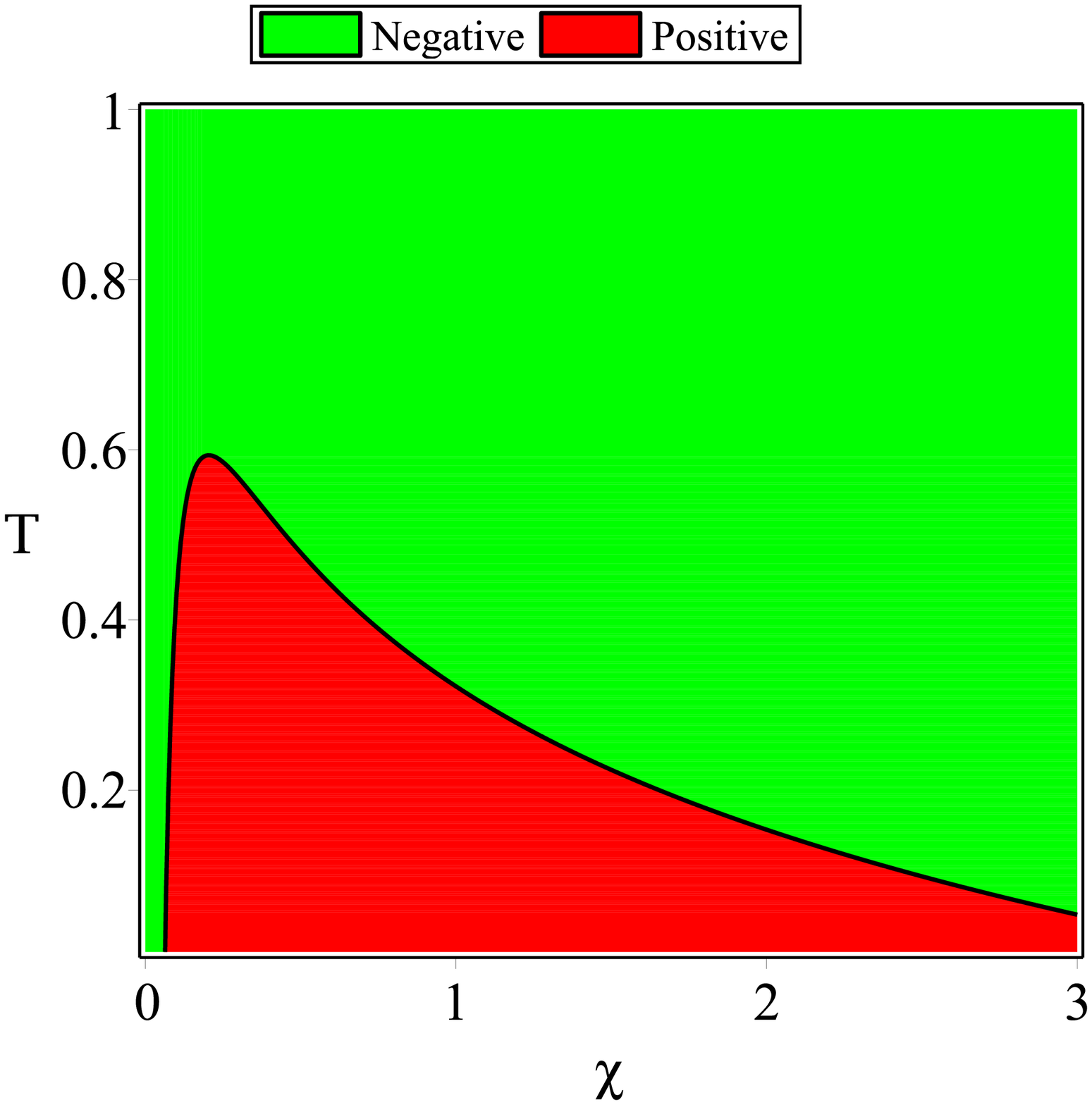}}
\subfloat[$T=0.2$ and $Q=0.1$]{
        \includegraphics[width=0.315\textwidth]{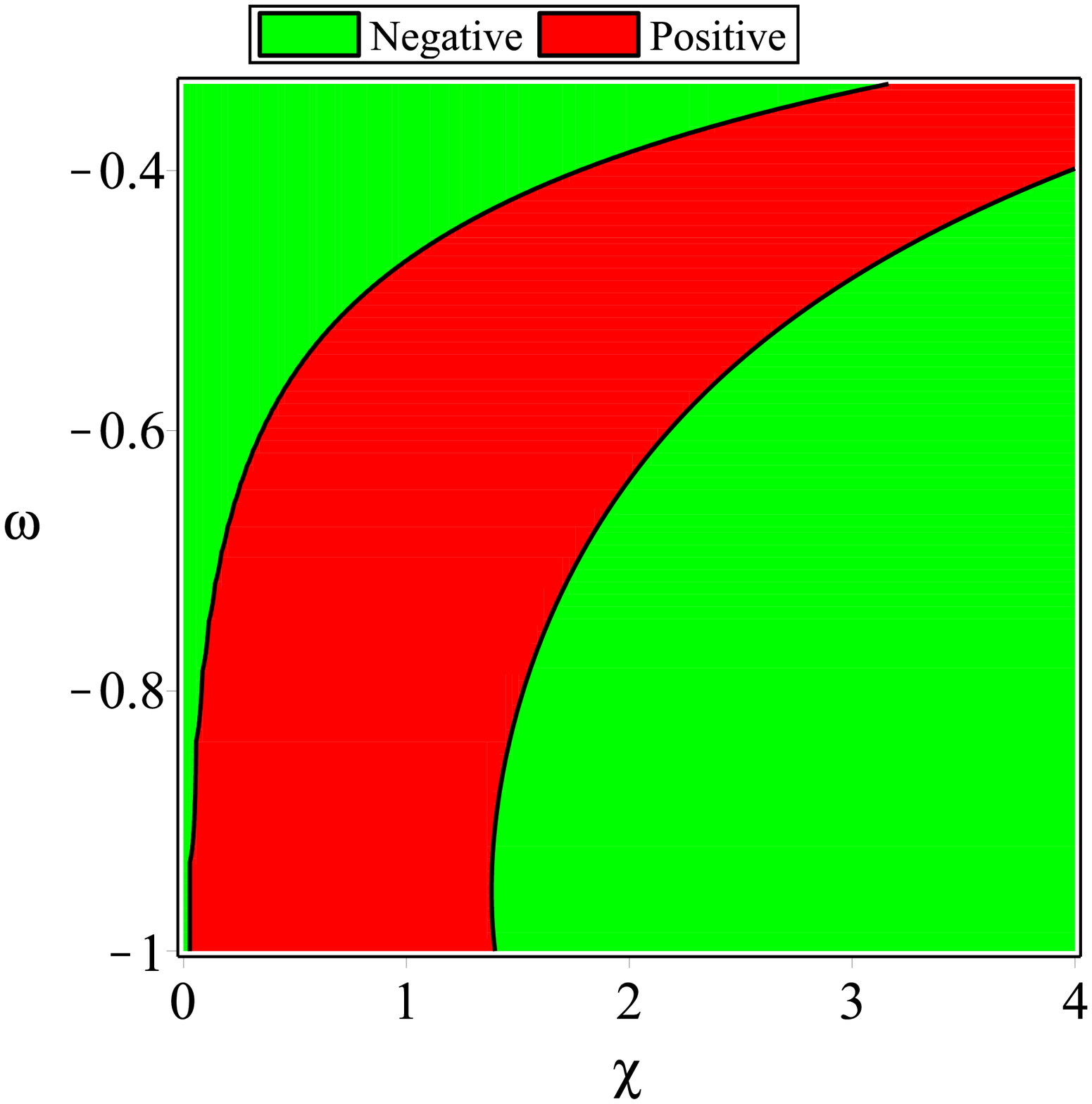}}\newline
\subfloat[$T=0.2$ and $\omega =-0.5$]{
        \includegraphics[width=0.32\textwidth]{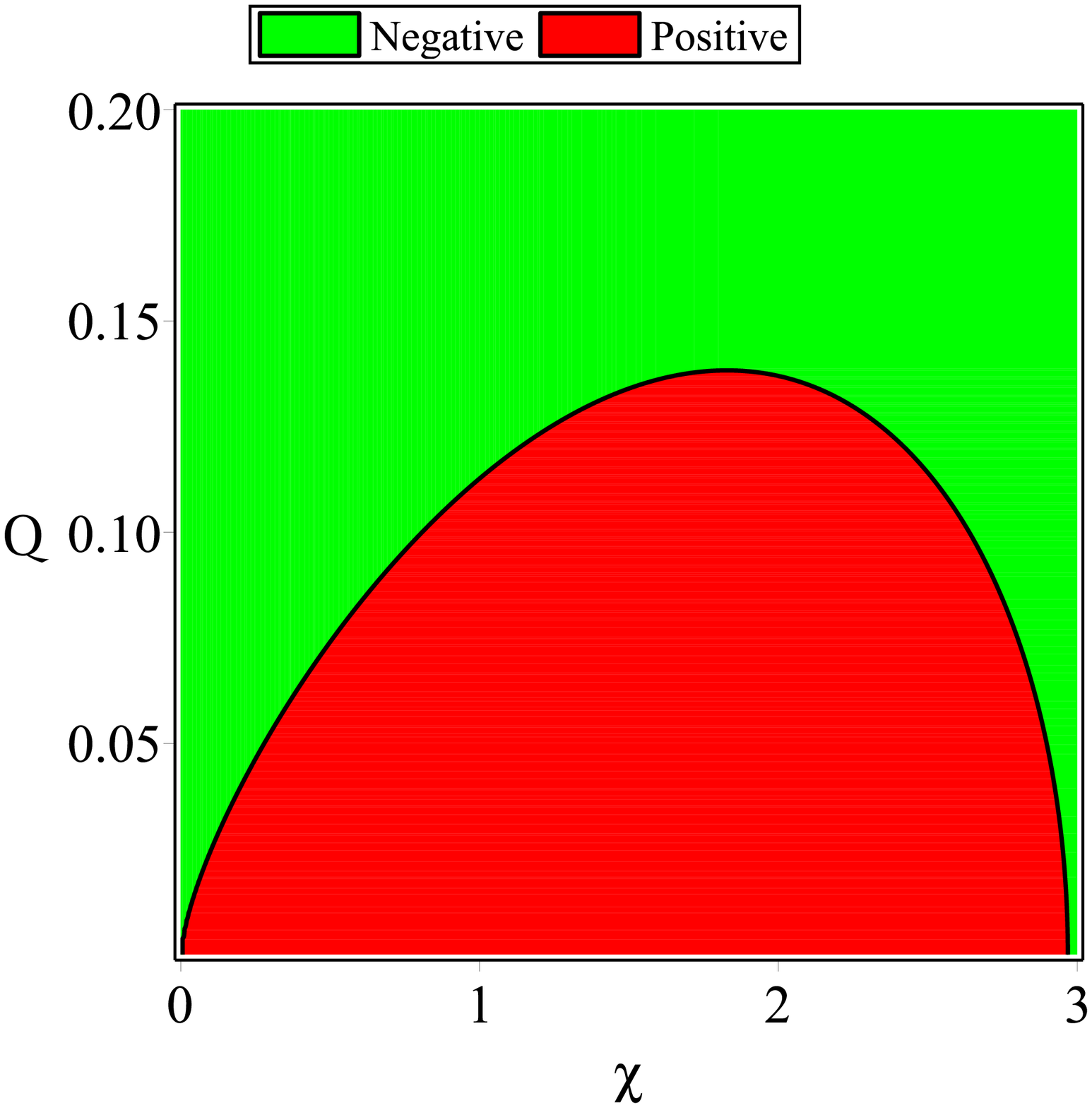}}
\subfloat[$Q=0.2$ and $\omega =-0.5$]{
        \includegraphics[width=0.32\textwidth]{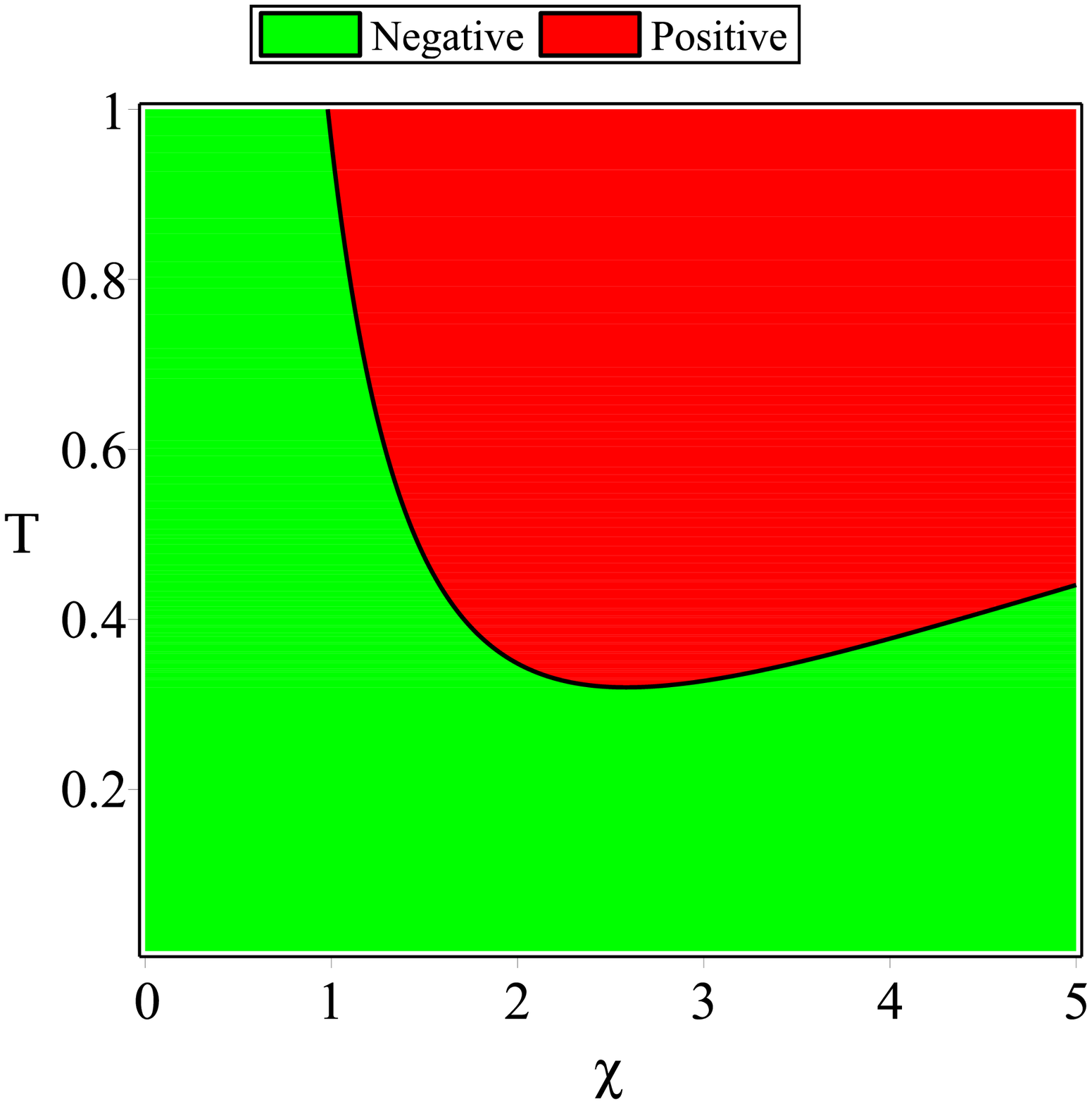}}
\subfloat[$T=0.2$ and $Q=0.2$]{
        \includegraphics[width=0.315\textwidth]{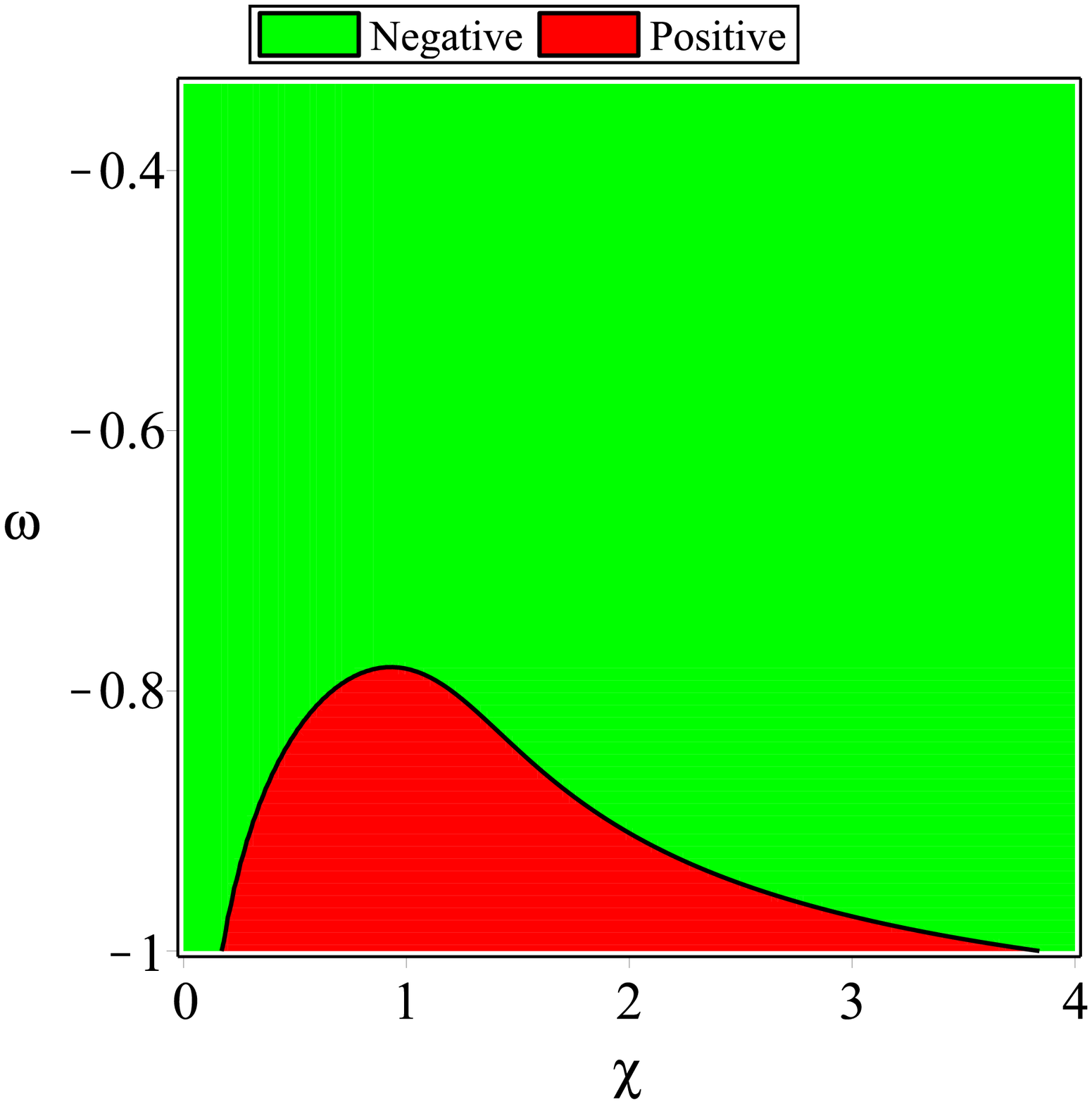}}
\caption{Variation of the $(\frac{d\beta}{d\protect\chi})$ as a function of different parameters for $l=1 $.}
\label{Fig2}
\end{figure}

As we see from Figs. \ref{Figax}a and \ref{Figax}c, some parts of
the isotherms corresponds to a negative $ \beta $ which are not
physically acceptable. It should be noted that this also occurs in
the usual van der Waals fluid where the pressure can become
negative for certain values of $ T $. This oscillating part of the
isotherm indicates instability region  ($\frac{\partial
\beta}{\partial\chi}>0 $).  Indeed, to observe an acceptable
behavior, $ \beta $ should be a decreasing function of $\chi $. In
any place that such a principle is violated, the BH is unstable
and may undergo a phase transition in that region. In order to see
whether $ \beta $ is a decreasing/increasing function of its
conjugate quantity, we calculate its first order derivation with
respect to $\chi $
\begin{eqnarray}
\frac{d\beta}{d\chi}= \frac{l^{2}\left(Q^{2}\mathcal{B}^{3\omega -2}(3\omega -1)\chi^{\frac{2-3\omega}{3\omega}}+4\pi T \mathcal{B}^{3\omega +1}(3\omega +2)\chi^{-\frac{3\omega +1}{3\omega}}- (3\omega +1)\mathcal{B}^{3\omega}\chi^{-1}\right) - 9 \mathcal{B}^{3\omega +2}(\omega +1)\chi^{-\frac{3\omega +2}{3\omega}}}{8\pi l^{3\omega +5}},  \label{Eqdbeta}
\end{eqnarray}
If this expression is negative, the BH admits the mentioned
principle and that region is physically accessible, while
positivity of this expression means that a phase transition takes
place in that region. It is worthwhile to mention that places
where the signature of $\frac{d\beta}{d\chi} $ changes are where $
\beta $ has an extremum. To express such a possibility, we have
plotted diagrams in Fig. \ref{Fig2}.

\begin{figure}[!htb]
\centering \subfloat[$ Q=0.1 $ and $ \omega=-0.8 $]{
        \includegraphics[width=0.32\textwidth]{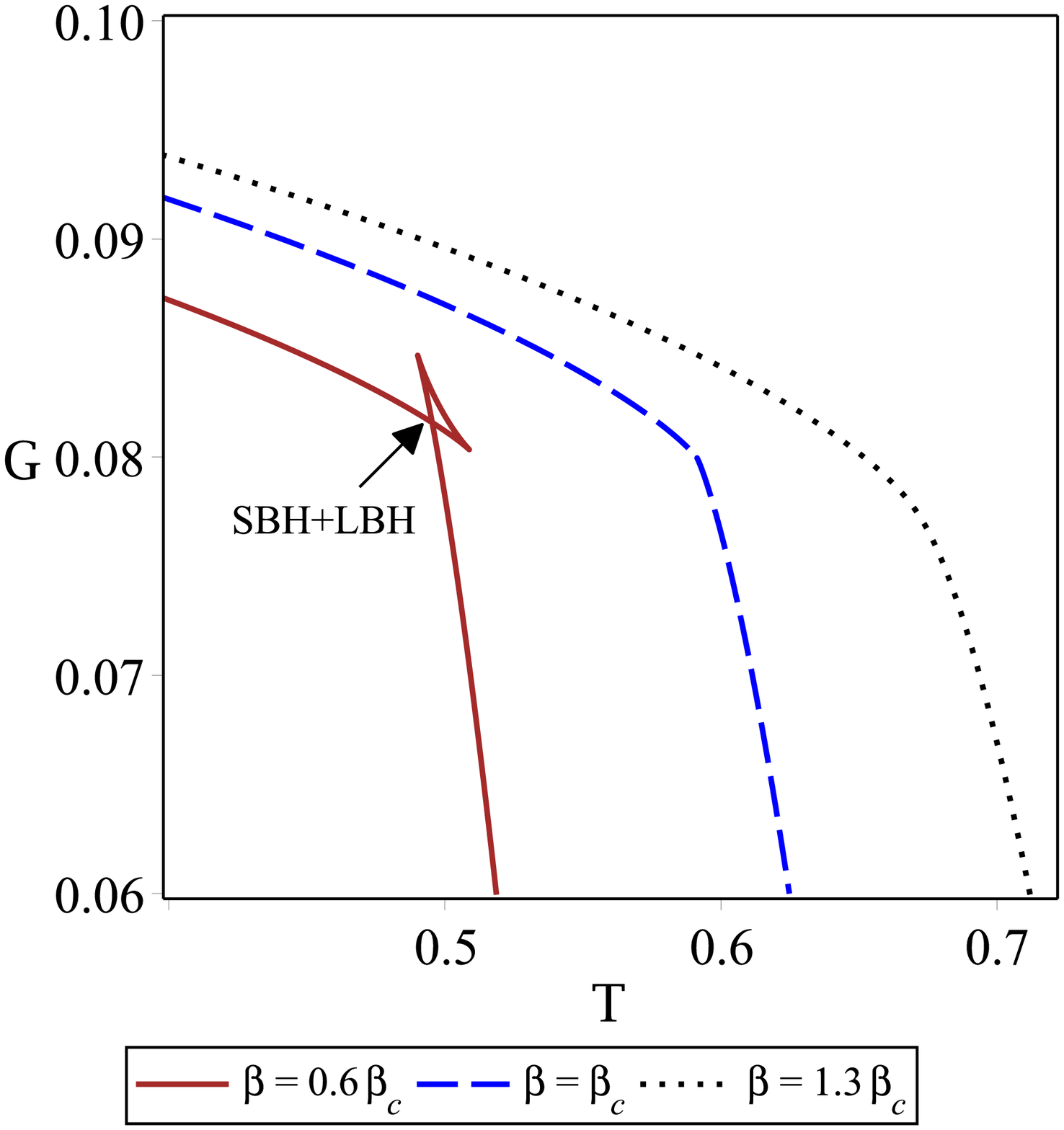}}
 \subfloat[$Q=0.1$ and $ \omega=-0.8 $]{
        \includegraphics[width=0.315\textwidth]{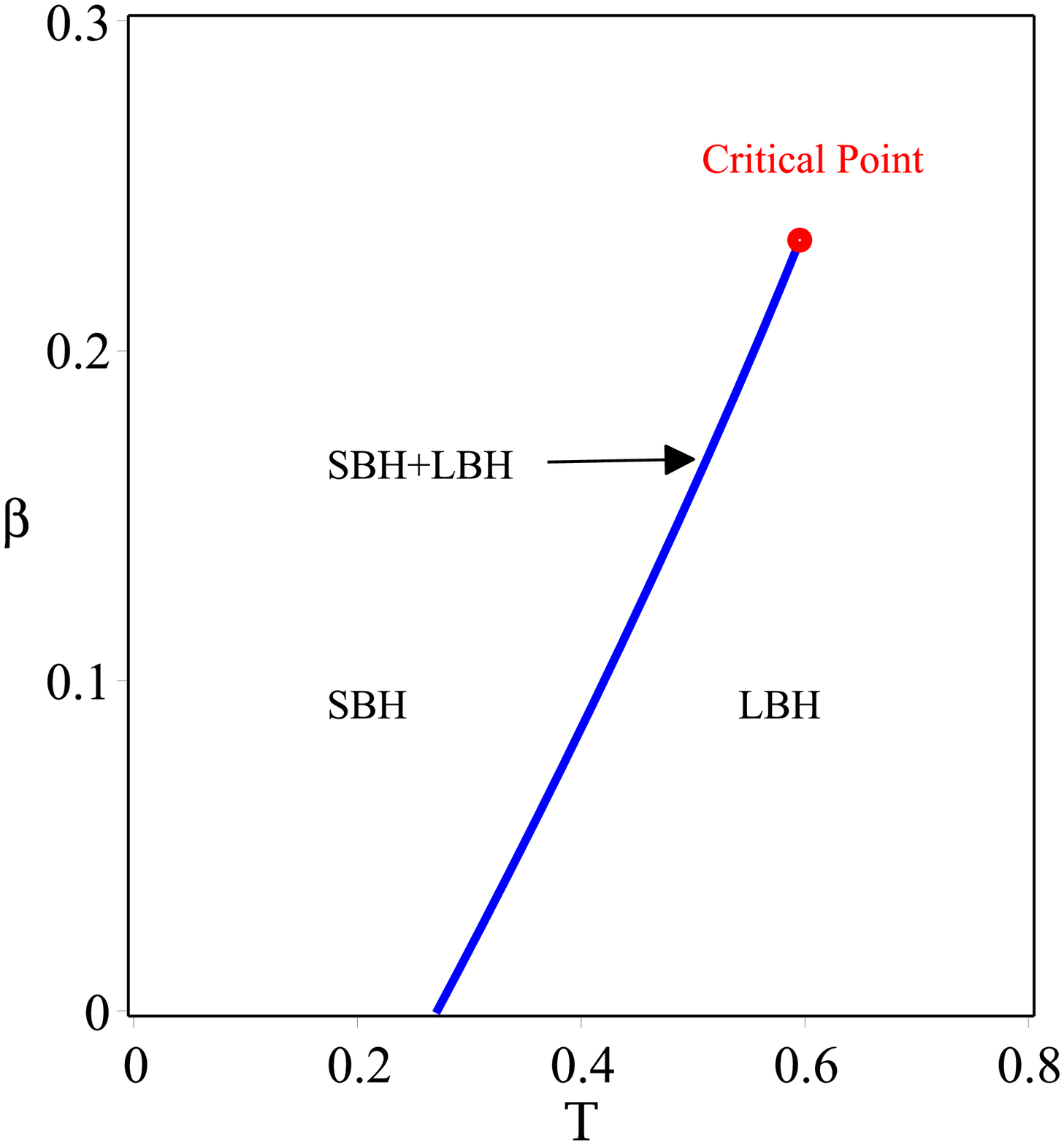}}\newline
\subfloat[$ Q=0.2 $ and $ \omega=-0.5 $]{
        \includegraphics[width=0.33\textwidth]{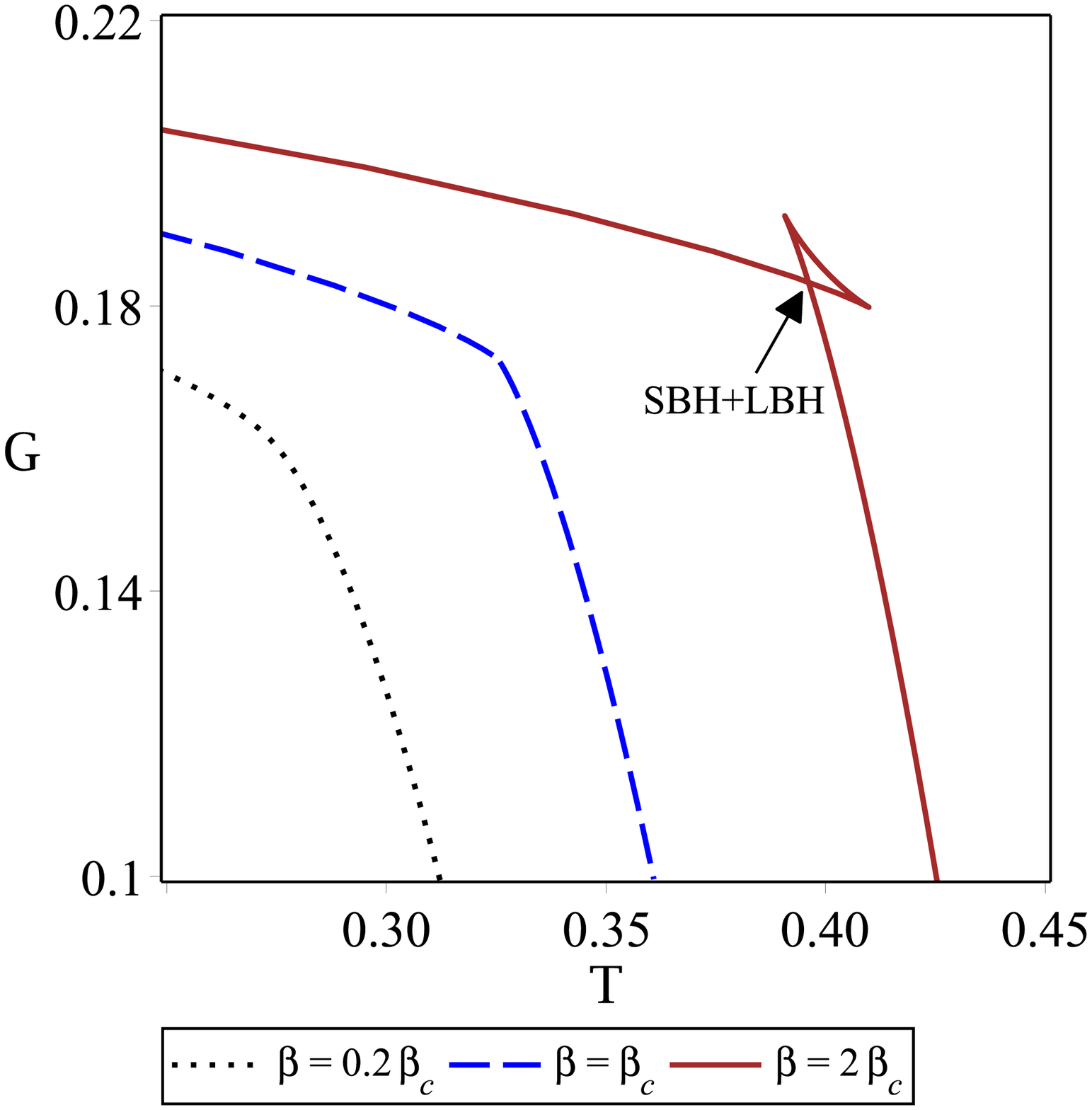}}
 \subfloat[ $Q=0.2$ and $ \omega=-0.5 $]{
        \includegraphics[width=0.315\textwidth]{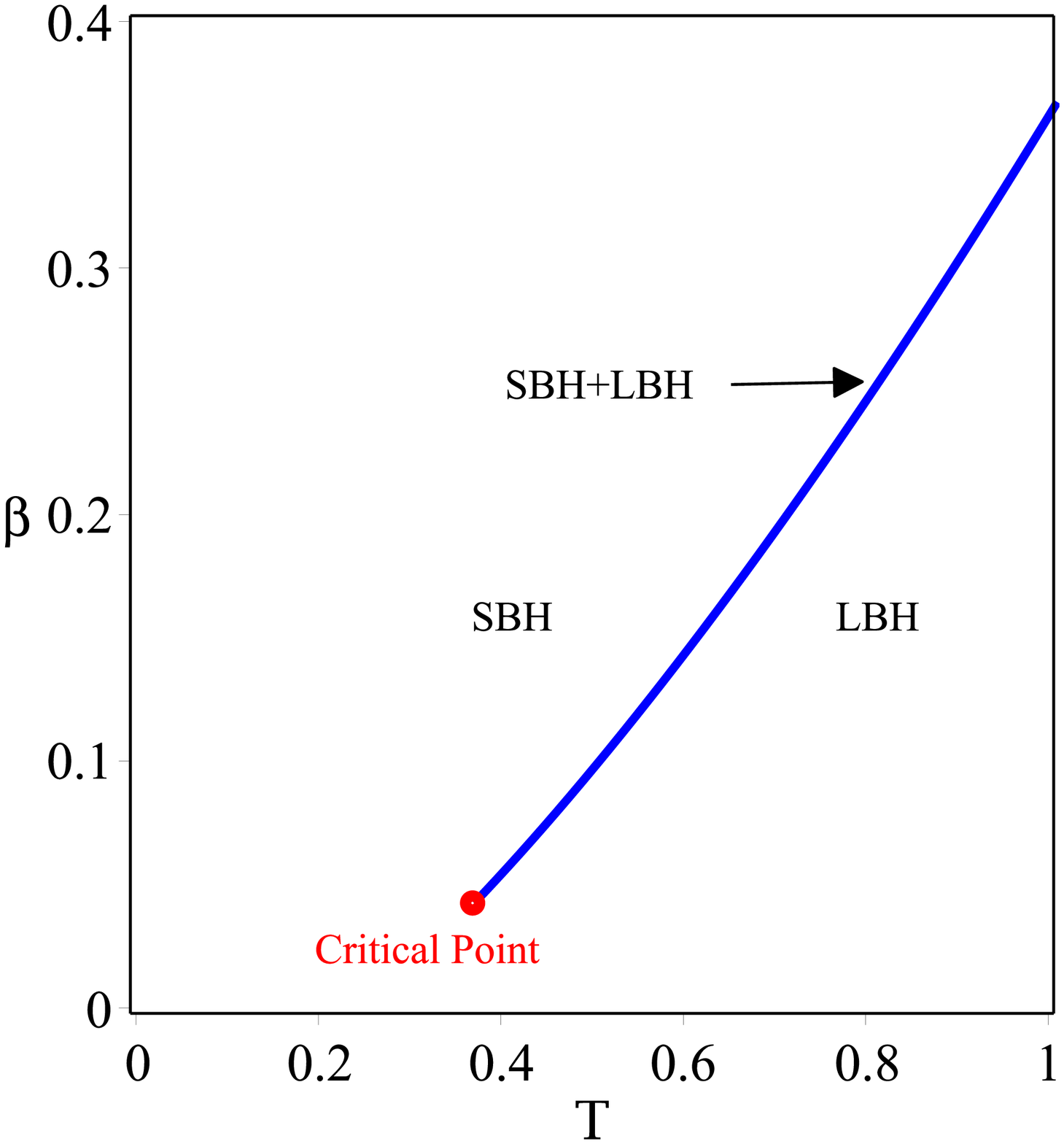}}\newline
\caption{Left panels: the Gibbs free energy diagram $ G - T$. Right
panels: Coexistence curve of small-large BH phase transition in
the $\beta - T  $ plane.} \label{Figax1}
\end{figure}

To get more information about the phase transition, we investigate $G-T$
diagram. The Gibbs free energy in the canonical ensemble can be calculated as
\begin{eqnarray}
G=\frac{3Q^{2}}{4r_{+}}+\frac{r_{+}}{4}-\frac{r_{+}^{3}}{4l^{2}}-\frac{2\pi \beta (3\omega +2) l^{3+3\omega}}{3\omega r_{+}^{3\omega}},
\label{EqGibbs}
\end{eqnarray}
where $r_{+} $ is related to $\chi $ which is a function of $T $
and $\beta $ through Eq. (\ref{beta3}). The phase transitions of a
system can be categorized by their orders which are characterized
by the discontinuity in n$^{th}$ derivatives of the Gibbs free
energy. For example, in a first-order phase transition, $G$ is a
continuous function but its first derivative (the entropy or
volume) changes abruptly whereas in the second-order one, both $G$
and its first derivative are continuous and the heat capacity (the
second derivative of $G$) is a discontinuous function. Formation
of the swallow-tail shape in $G-T$ diagram (continuous line of
Figs. \ref{Figax1}a and \ref{Figax1}c) represents a first-order
phase transition in the system. The phase transition point is
located at the cross point in the $G-T$ diagram, where small BH
(SBH) and large BH (LBH) exist simultaneously (see Figs.
\ref{Figax1}a and \ref{Figax1}c). Right panels of  Fig.
\ref{Figax1} display the coexistence line of small/large BH phase
transition. The critical point is located at the end of the
coexistence line which is indicated in Figs. \ref{Figax1}b and
\ref{Figax1}d. The first-order phase transition occurs when the
black hole crosses the coexistence line. Taking a look at Fig.
\ref{Figax1}a, we see that for $ \omega <-\frac{2}{3} $, the
first-order phase transition occurs for $ \beta < \beta_{c} $,
whereas for $ \omega >-\frac{2}{3} $, such a phase transition is
possible for $ \beta > \beta_{c} $ (see Fig. \ref{Figax1}c).

To calculate critical values, we use the properties of inflection
point which leads to
\begin{eqnarray}
\chi_{c} &=& \mathcal{B}^{3\omega}\left( \frac{\sqrt{2l}}{6}\sqrt{\frac{l(1+3\omega)+
\sqrt{108Q^{2}(1+\omega)(1-3\omega)+l^{2}(1+3\omega)^{2}}}{(1+\omega)}}\right)^{-3\omega},  \nonumber \\
&&  \nonumber \\
T_{c}&=&\frac{\mathcal{B}^{2}l^{2}(3\omega +1)\chi^{\frac{1}{3\omega}}-Q^{2}l^{2}(3\omega -1)\chi^{\frac{1}{\omega}}+9\mathcal{B}^{4}(\omega +1)\chi^{-\frac{1}{3\omega}}}
{4\pi (3\omega +2)\mathcal{B}^{3}l^{2}},  \nonumber \\
&&  \nonumber \\
\beta_{c}&=&
\frac{\mathcal{B}^{3\omega -1}\chi^{\frac{1-3\omega}{3\omega}}\left(3Q^{2}l^{2}-\mathcal{B}^{2}l^{2}\chi^{-\frac{2}{3\omega}}+3\mathcal{B}^{4}\chi^{-\frac{4}{3\omega}}\right) }{8\pi l^{3\omega +5}}.
\label{Eqaxc}
\end{eqnarray}

\begin{figure}[!htb]
\centering \subfloat[$ \omega =-0.8 $]{
        \includegraphics[width=0.3\textwidth]{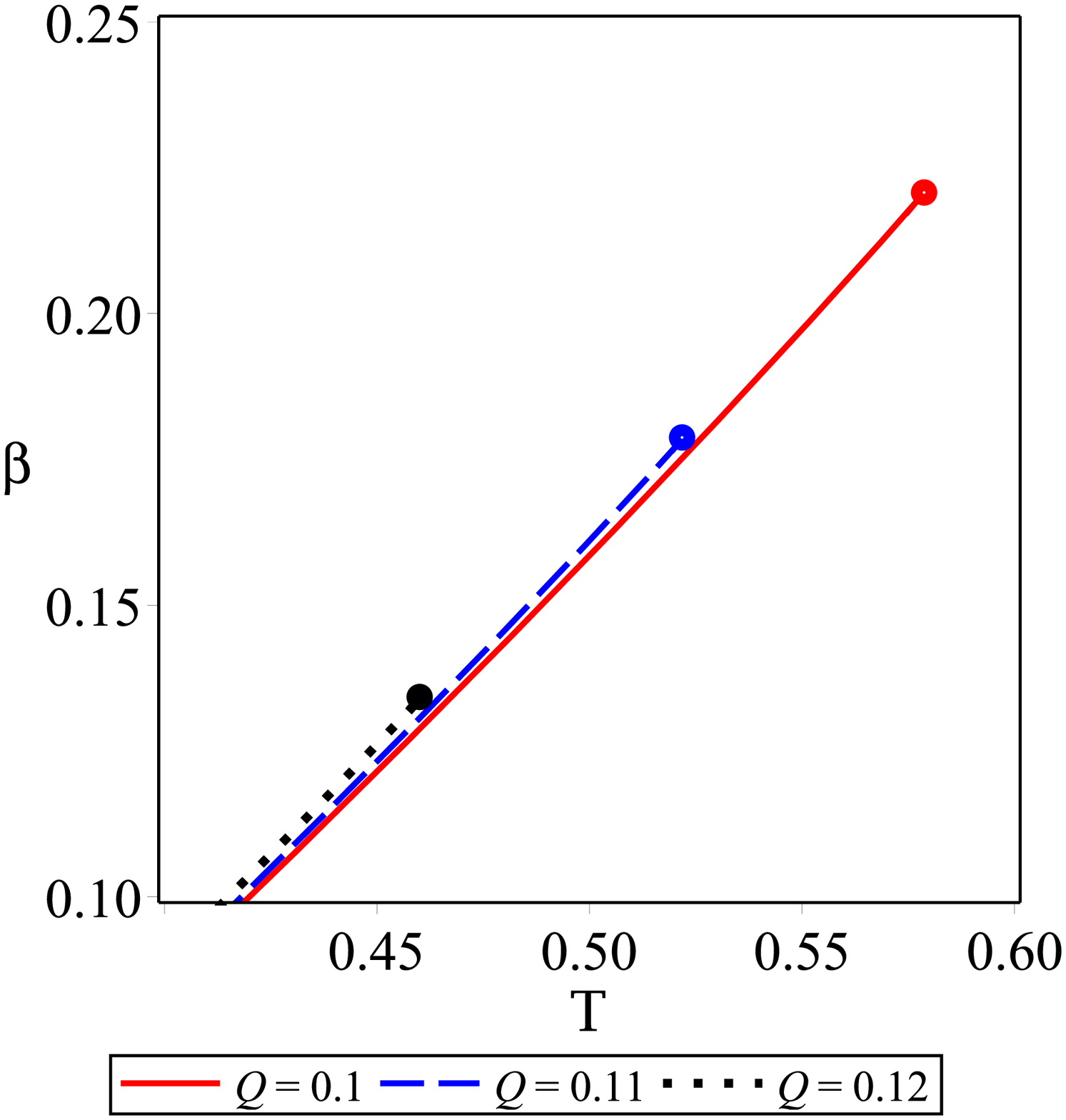}}
\subfloat[$Q=0.1 $]{
        \includegraphics[width=0.3\textwidth]{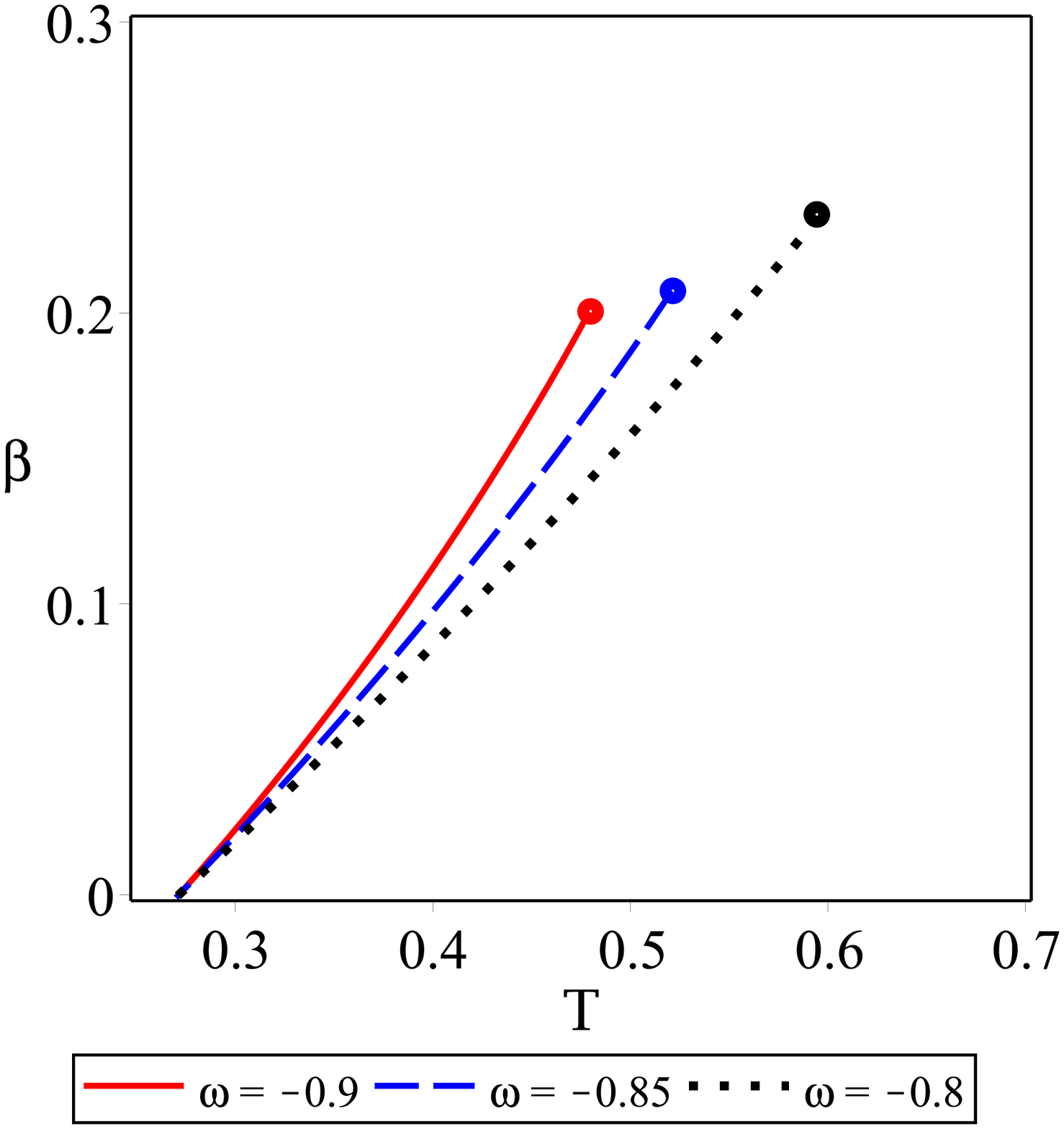}}\newline
\subfloat[$ \omega =-0.5 $]{
        \includegraphics[width=0.3\textwidth]{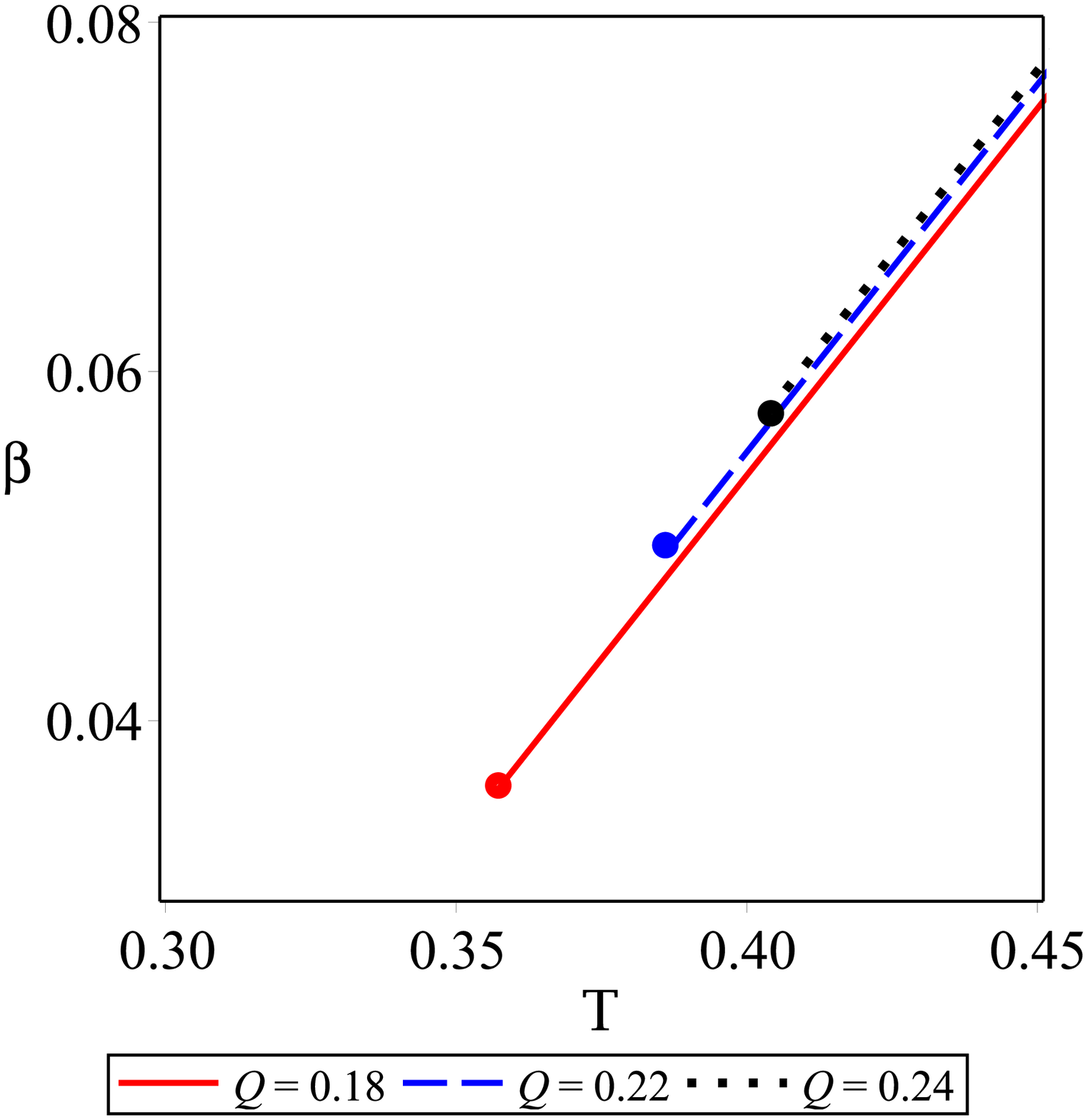}}
\subfloat[$Q=0.2 $]{
        \includegraphics[width=0.3\textwidth]{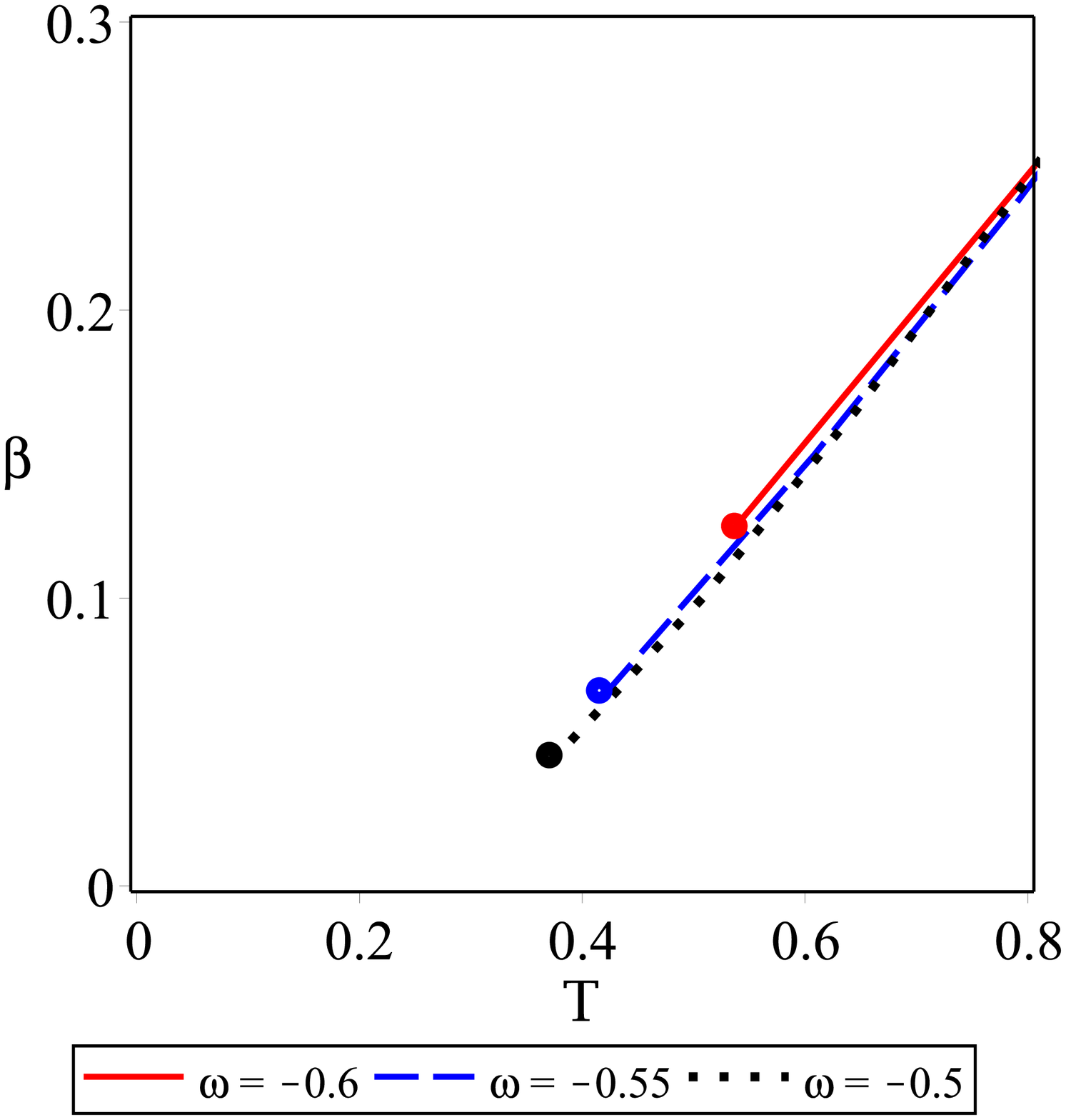}}\newline
\caption{$\beta-T$ diagram for $l=1$, different values of electric charge (left
panels) and different values of state parameter $\protect\omega $ (right
panels). Small circles in the endpoint of each line represent the critical points}.
\label{Fig4}
\end{figure}

To inspect the effects of electric charge and state parameter on
the critical values, we have depicted Fig \ref {Fig4}. As we see,
from Fig. \ref{Fig4}a, for $\omega <-\frac{2}{3} $, both $T_{c}$
and $ \beta_{c}$ are decreasing functions of $Q $ and vice versa
(see Fig. \ref{Fig4}c). Regarding the effect of $\omega$ on these
quantities, for $\omega <-\frac{2}{3} $, increasing the state
parameter from $-1 $ to $-\frac{2}{3} $ makes the increasing of $
T_{c} $ and $ \beta_{c} $ (see Fig. \ref{Fig4}b), whereas for
$\omega
>-\frac{2}{3} $, increasing this parameter leads to the decreasing
both quantities $ T_{c} $ and $ \beta_{c} $ (see Fig.
\ref{Fig4}d).

To calculate new relation for the parameter $ \beta $, we obtain the heat capacity as
\begin{eqnarray}
C_{Q}=T\left(\frac{\partial S}{\partial T} \right)_{Q,\beta} &=&
\frac{2\pi \mathcal{B}^{2}\chi^{-\frac{2}{3\omega}}
\left( 3\mathcal{B}^{4}\chi^{-\frac{4}{3\omega}}+\mathcal{B}^{2}l^{2}\chi^{-\frac{2}{3\omega}}-Q^{2}l^{2}+8\pi \beta \mathcal{B}^{1-3\omega}l^{3\omega +5}\chi^{\frac{3\omega -1}{3\omega}}
\right)}{ 3\mathcal{B}^{4}\chi^{-\frac{4}{3\omega}}-\mathcal{B}^{2}l^{2}\chi^{-\frac{2}{3\omega}}+3Q^{2}l^{2}-8\pi \beta (3\omega +2)\mathcal{B}^{1-3\omega}l^{3\omega +5}\chi^{\frac{3\omega -1}{3\omega}}
}.
\label{Eqheat}
\end{eqnarray}

Solving the denominator of the heat capacity with respect to
$\beta $, a new relation is determined as follows
\begin{equation}
\beta_{new}=
\frac{\mathcal{B}^{3\omega -1}\chi_{e}^{\frac{1-3\omega}{3\omega}}\left(3Q^{2}l^{2}-\mathcal{B}^{2}l^{2}\chi_{e}^{-\frac{2}{3\omega}}+3\mathcal{B}^{4}\chi_{e}^{-\frac{4}{3\omega}}\right) }{8\pi l^{3\omega +5}},
\label{EqNP}
\end{equation}
where $ \chi_{e} $ indicates $ \chi $ related to the extremum. The
resultant curve is displayed in Fig. \ref{Figax} by the dashed
lines. From Fig. \ref{Figax}a, we see that for $ \omega
<-\frac{2}{3} $, the function $ \beta_{new}$ has a maximum point
which coincides to $ \beta_{c}$, while vice versa happens for
$\omega >-\frac{2}{3} $ (see Fig. \ref{Figax}c). Inserting Eq.
(\ref{EqNP}) into the relation of temperature Eq. (\ref{TH1}), one
finds a new relation for the temperature which is independent of $
\beta $
\begin{equation}
T_{new}=\frac{\mathcal{B}^{2}l^{2}(3\omega +1)\chi_{e}^{\frac{1}{3\omega}}-Q^{2}l^{2}(3\omega -1)\chi_{e}^{\frac{1}{\omega}}+9\mathcal{B}^{4}(\omega +1)\chi_{e}^{-\frac{1}{3\omega}}}
{4\pi (3\omega +2)\mathcal{B}^{3}l^{2}},
\label{EqNT}
\end{equation}
the existence of extremum in the obtained relation is representing
the critical temperature (see dashed lines in Figs. \ref{Figax}b
and \ref{Figax}d). By deriving the new parameter $ \beta $ or
temperature with respect to $ \chi $, one can obtain $ \chi_{e} $
as
\begin{equation}
\frac{\partial P_{new}}{\partial \chi}=0~~~\Longrightarrow ~~~\chi_{e} =\mathcal{B}^{3\omega}\left( \frac{\sqrt{2l}}{6}\sqrt{\frac{l(1+3\omega)+
\sqrt{108Q^{2}(1+\omega)(1-3\omega)+l^{2}(1+3\omega)^{2}}}{(1+\omega)}}\right)^{-3\omega},
\label{EqNv}
\end{equation}
which is the same as $\chi_{c}$. Inserting Eq. (\ref{EqNv}) into
Eqs. (\ref{EqNT}) and (\ref{EqNP}), one can find that $ T_{new} $
and $ \beta_{new} $ are, respectively, the same as $T_{c}$ and $
\beta_{c} $ in Eq. (\ref{Eqaxc}).

\subsection*{Behavior near the critical point}\label{Crit-Exp}

Let us now compute the critical exponents for the BH system. We
start with the behavior of the entropy and rewrite it in terms of
$T $ and $\chi $ as
\begin{equation}
S(T,\chi)=\pi \mathcal{B}^{2}\chi^{-\frac{2}{3\omega}},  \label{EqNS}
\end{equation}
which is independent of temperature. So, we find that
\begin{equation}
C_{\chi}=T\frac{\partial S}{\partial T}\bigg|_{\chi}=0,
\end{equation}
and hence $\alpha =0 $.

Expanding the the equation of state around the critical point
\begin{equation}
\varphi = \frac{\chi}{\chi_{c}} -1,~~~t=\frac{T}{T_{c}}-1.  \label{EqNereduced}
\end{equation}
and defining $ \varrho=\frac{\beta}{\beta_{c}} $, Eq.
(\ref{beta3}) is rewritten as
\begin{equation}
\varrho = \mathcal{B}_{1}+B_{1}t+\mathcal{B}_{2}\varphi t+
\mathcal{B}_{3}\varphi+\mathcal{B}_{4}\varphi^{2}+\mathcal{B}_{5}\varphi^{3}+O(t\varphi^{2},\varphi^{4}),
 \label{Eqnuprime}
\end{equation}
 where
\begin{eqnarray}
\mathcal{B}_{2}&=&-\frac{1}{3\omega}B_{1}(2+3\omega),  \nonumber \\
&&  \nonumber \\
\mathcal{B}_{3}&=&-\frac{1}{3\omega}\left( 3\mathcal{B}_{1}\omega +2B_{1}-B_{2}+B_{3}+3B_{4}\right) ,  \nonumber \\
&&  \nonumber \\
\mathcal{B}_{4}&=&\frac{1}{18\omega^{2}}\left( 9\omega (2B_{1}-B_{2}+B_{3}+3B_{4})+3B_{1}+8B_{4}+\mathcal{B}_{1}+18\mathcal{B}_{1}\omega^{2}\right),  \nonumber \\
&&  \nonumber \\
\mathcal{B}_{5}&=& -\frac{1}{162\omega^{3}}\left( 99\omega^{2} (2B_{1}-B_{2}+B_{3}+3B_{4})+18\omega (\mathcal{B}_{1}+3B_{1}+8B_{4})+8B_{1}-B_{2}+B_{3}+27B_{4}+162\mathcal{B}_{1}\omega^{3} \right)
, \label{EqaTc2}
\end{eqnarray}
which is the same as Eq. (\ref{EqaTc1}), and the only
difference between these two equations is $ B_{i} $ coefficients
given by
\begin{eqnarray}
\mathcal{B}_{1}&=& B_{1}+B_{2}+B_{3}+B_{4},  ~~~B_{1}=\frac{T_{c}\mathcal{B}^{3\omega +2}\chi_{c}^{-\frac{3\omega +2}{3\omega}}}{2\beta_{c}l^{3\omega +3}},~~~
B_{2}=\frac{Q^{2}\mathcal{B}^{3\omega -1}\chi_{c}^{\frac{1-3\omega}{3\omega}}}{8\pi \beta_{c}l^{3\omega +3}}
,\nonumber \\
&&  \nonumber \\
B_{3}&=& -\frac{\mathcal{B}^{3\omega +1}\chi_{c}^{-\frac{1+3\omega}{3\omega}}}{8\pi \beta_{c}l^{3\omega +3}}
, ~~~
B_{4}= -\frac{3\mathcal{B}^{3\omega +3}\chi_{c}^{-\frac{1+\omega}{\omega}}}{8\pi \beta_{c}l^{3\omega +5}}
. \label{EqaTc21}
\end{eqnarray}

Our numerical analysis showed that the coefficients $
\mathcal{B}_{3} $ and $ \mathcal{B}_{4} $ are very small and can
be considered zero as in the previous case. The obtained critical
coefficients $ \lambda $, $ \gamma $ and $ \delta $ are the same as
those presented in the previous subsection and we do not write
them here to avoid repetition. So the obtained critical
exponents in this approach coincide with those obtained for van
der Waals fluid, similar to the previous case.

\begin{figure*}[!htb]
\centering
\includegraphics[width=0.45\linewidth]{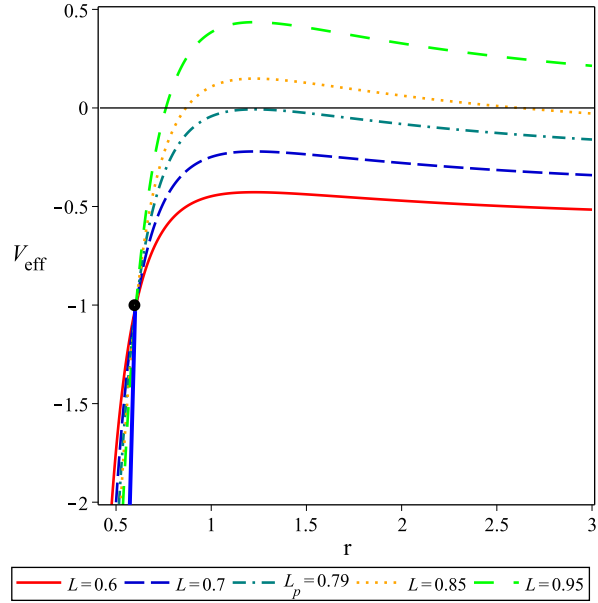}
\caption{Effective potential $ V_{eff} $ as a function of $ r $
for $ E=l=1 $, $ M=0.5 $ $, \beta=0.05 $,  $ Q=0.1 $, $ \omega =-0.8 $ and various $ L $. The blue
solid line is related to the place of the horizon where $
V_{eff}=-1 $.} \label{FigVef}
\end{figure*}
\section{ PHOTON SPHERE AND SHADOW}\label{PS-S}

The image of a supermassive BH in the galaxy $M87$, a dark part
which is surrounded by a bright ring, was direct support of the
Einstein's general relativity and the existence of the BH in our
universe \cite{Akiyama}. The image of the BH gives us the
information regarding its jets and matter accretion. The BH shadow
is one of the useful tools for a better understanding of the
fundamental properties of the BH and comparing alternative
theories with general relativity. The gravitational field near the
black hole's event horizon is so strong that can affect light
paths and causes spherical light rings. The shadow of a BH is
caused by gravitational light deflection.

It is worthwhile to mention that in preliminary studies in the
context of BH shadow, the BH was assumed to be eternal, i.e, the
spacetime was assumed to be time independent. So, a static or
stationary observer could see a time-independent shadow. But,
modern observational results have shown that our universe is
expanding with acceleration. This reveals the fact that shadow
depends on time. Although for the BH candidates at the center of
the Milky Way galaxy and at the centers of nearby galaxies the
effect of the cosmological expansion is negligible, for galaxies
at a larger distance the influence on the diameter of the shadow
is significant \cite{Perlick}. One method to explain the amazing
accelerating expansion is to introduce dark energy which makes up
about $70$ percent of the universe. The cosmological constant and
quintessence are two well-known candidates of dark energy
scenarios. Recently, the role of the cosmological constant in
gravitational lensing has been the subject of focused studies
\cite{Sh1,Sh2,Sh3,Sh4,Sh5}. The BH shadow arises as a result of
gravitational lensing in a strong gravity regime. So, one can
inspect the effect of the cosmological constant on the shadow of
black holes \cite{Sh6,Sh7,Sh8}.
\begin{figure}[!htb]
\centering \subfloat[RN BH]{
        \includegraphics[width=0.325\textwidth]{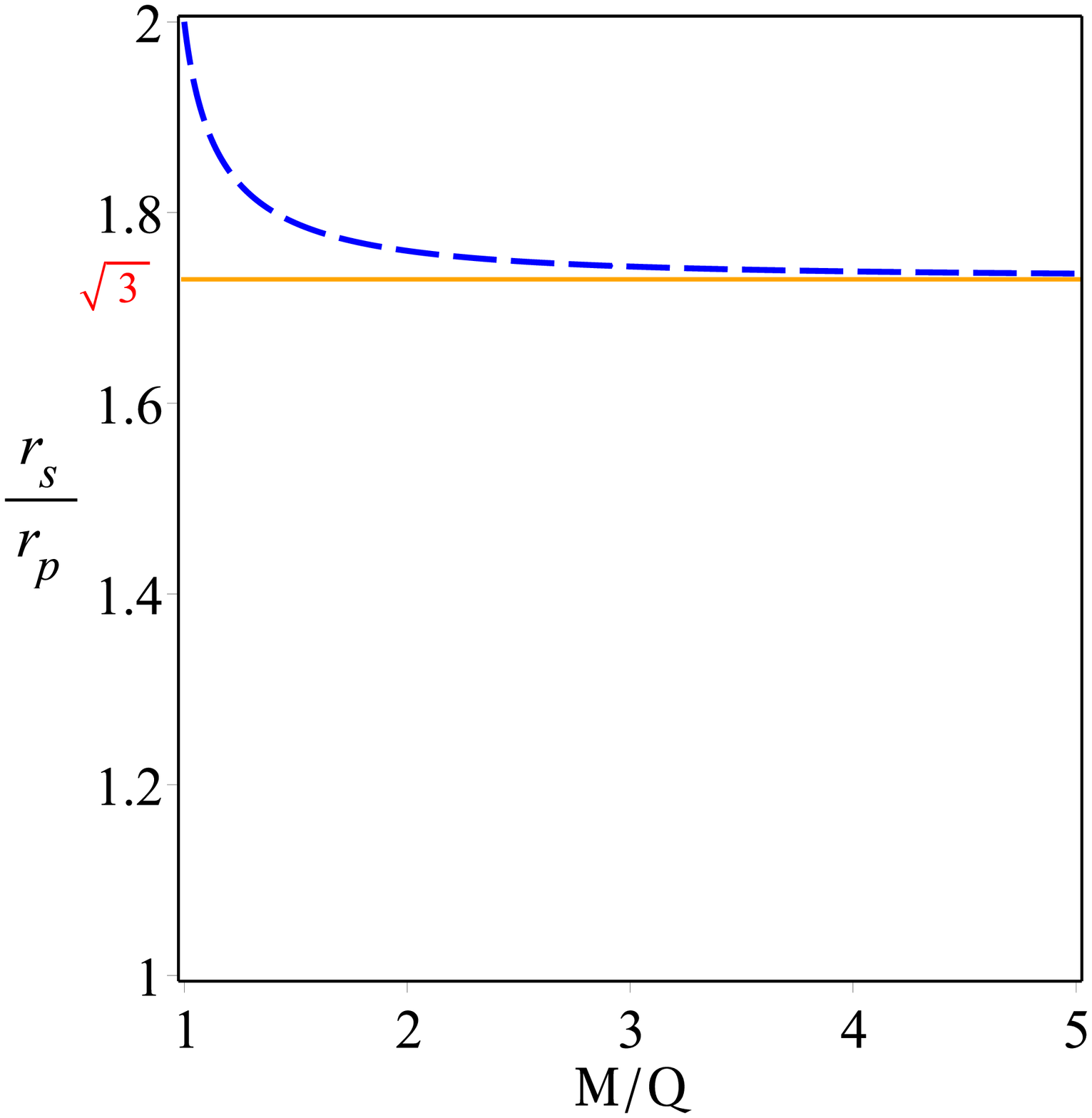}}
 \subfloat[RN AdS BH]{
        \includegraphics[width=0.315\textwidth]{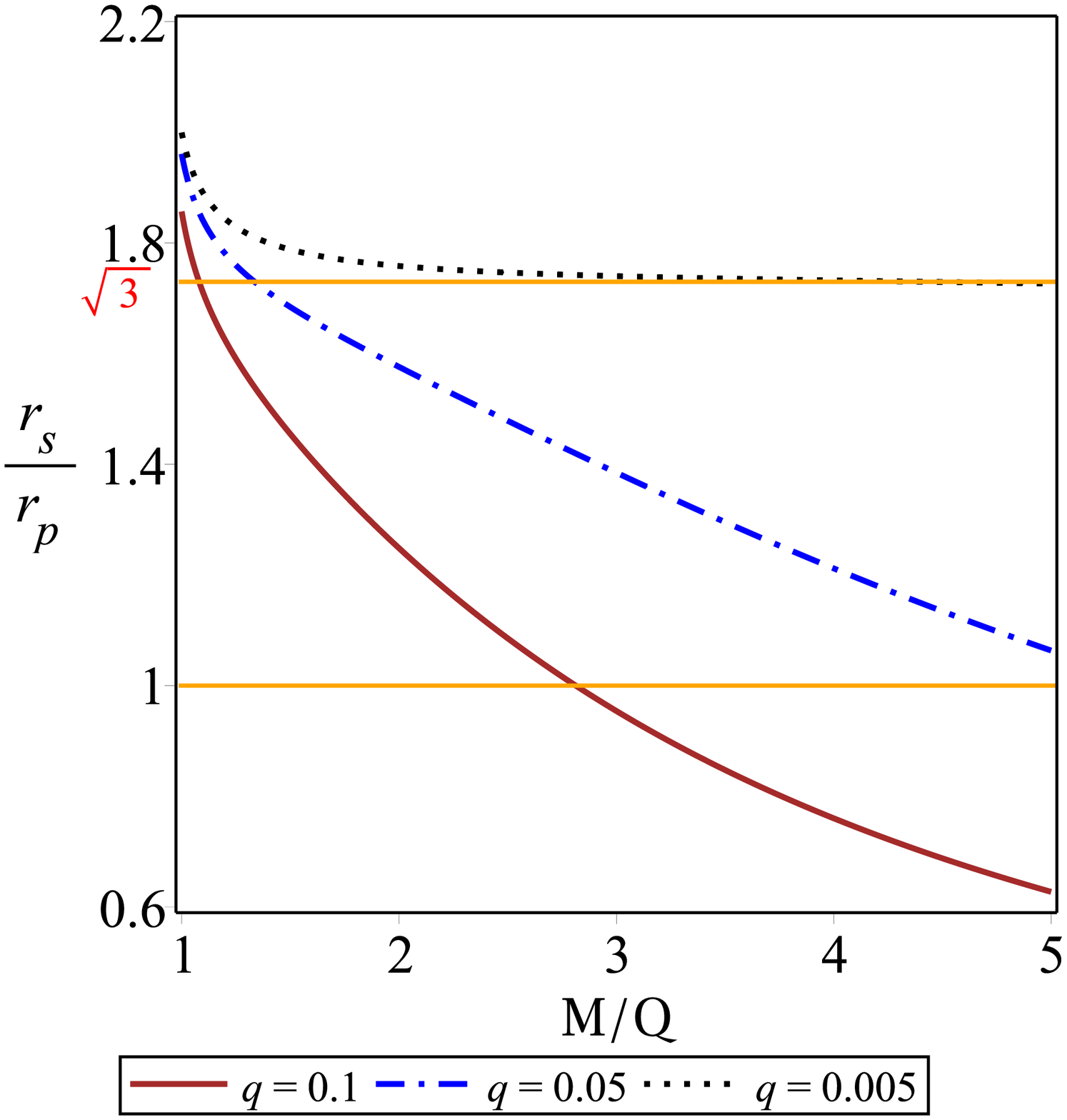}}\newline
\subfloat[ RNQ BH for $ \omega=-1/3 $]{
        \includegraphics[width=0.33\textwidth]{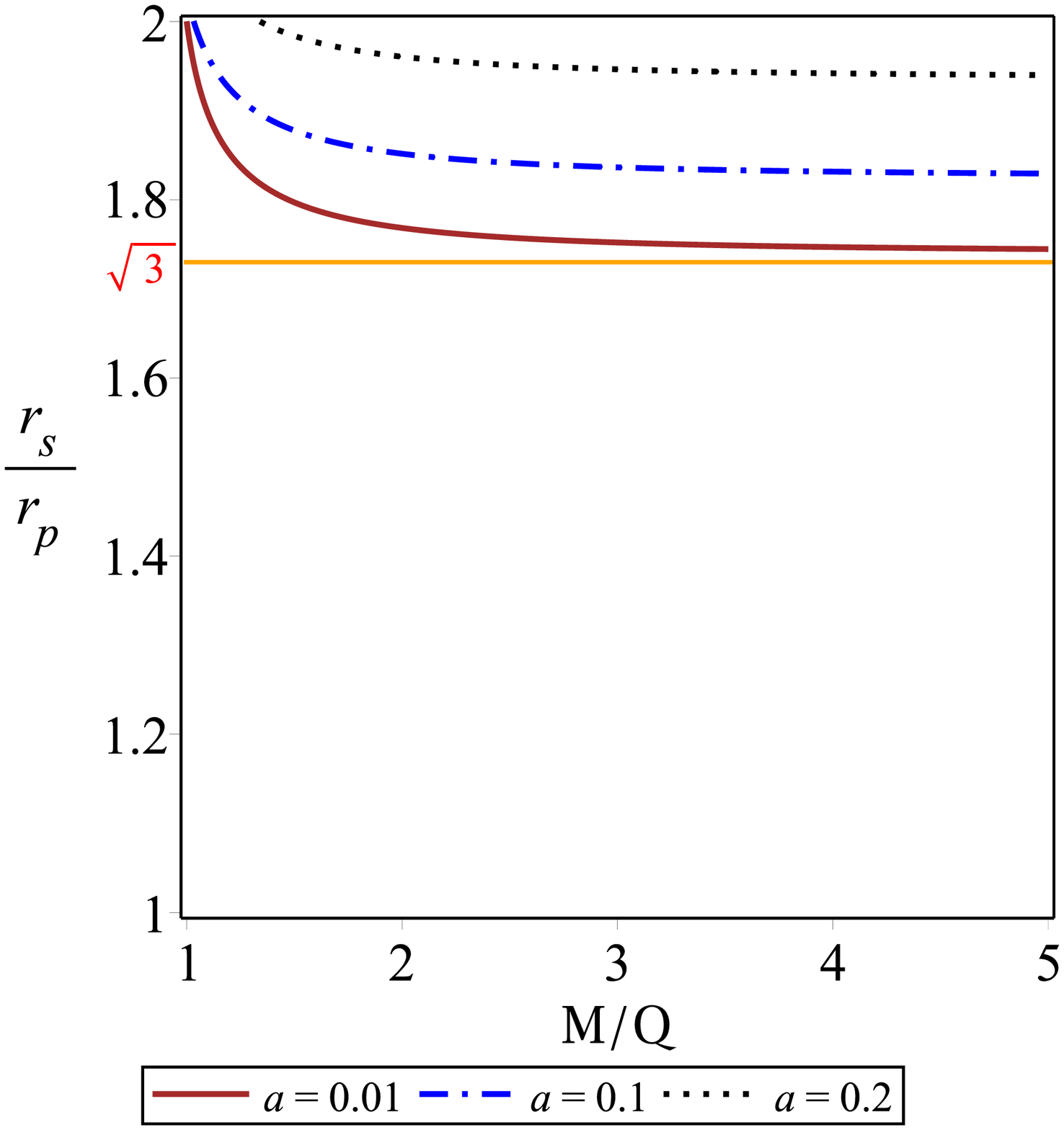}}
 \subfloat[ RNQ BH for $ \omega=-1 $]{
        \includegraphics[width=0.315\textwidth]{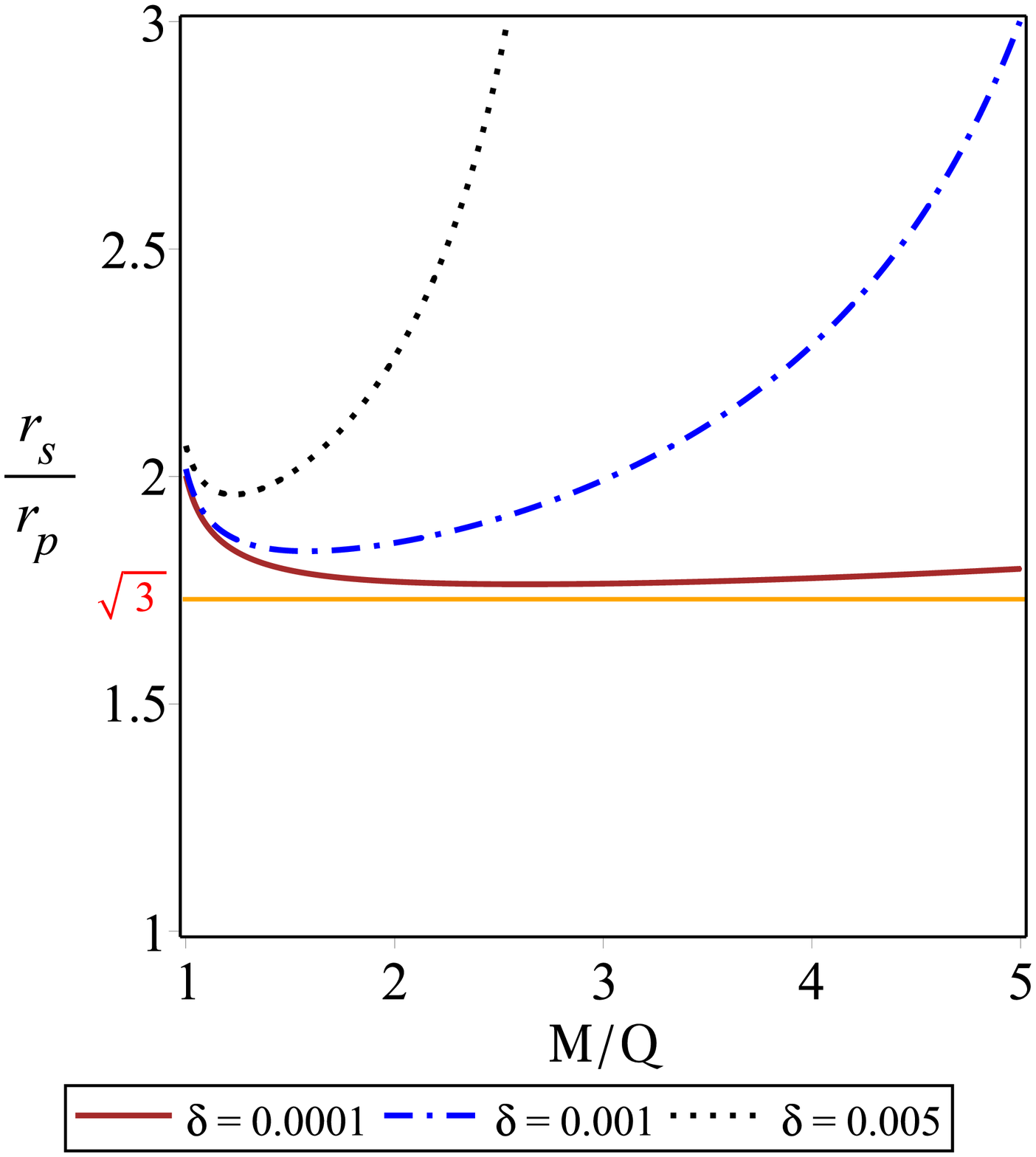}}\newline
\caption{The dependce of $\frac{r_{s}}{r_{p}} $ on the ratio $ \frac{M}{Q} $. In up right panel, we consider $ q=\frac{Q}{l} $. In down right panel, we set $ \delta =a Q^{2} $ where $ a $ is the normalization factor. } \label{FigRsp}
\end{figure}
It should be noted that although the expansion of the universe was
based on a positive cosmological constant, some pieces of evidence
show that it can be associated with a negative cosmological
constant.  As we know, an interesting approach to examine the
accelerated cosmic expansion and study properties of dark energy
is through observational Hubble constant data which has gained
significant attention in recent years \cite{Sh9,Sh10,Sh11}. The
Hubble constant, $H(z)$, is measured as a function of cosmological
redshift. The investigation of $H(z)$ behavior at low redshift
data showed that the dark energy density has a negative minimum
for certain redshift ranges which can be simply modeled through a
negative cosmological constant \cite{Dutta12}. The other reason to
consider a negative cosmological constant is the concept of
stability of  the accelerating universe. In Ref. \cite{Maedaa},
authors analyzed the possibility of de Sitter expanding spacetime
with a constant internal space and demonstrated that de Sitter
solution would be stable just in the presence of the negative
cosmological constant. The other interesting reason is through
supernova data. Although there is strong observational evidence
from high-redshift supernova that the expansion of the Universe is
accelerating due to a positive cosmological constant, the
supernova data themselves derive a negative mass density in the
Universe \cite{Riess,Perlmutter}. Several galaxy cluster
observations appear to have inferred the presence of a negative
mass in cluster environments. It was shown that a negative mass
density can be equivalent to a negative cosmological constant
\cite{Farnes}. In fact, the introduction of negative masses can
lead to an AdS space. This would correspond to one of the most
researched areas of string theory, the AdS/CFT correspondence.

Now, we would like to investigate how BH parameters affect the
shadow radius of the corresponding black hole. To do so, we employ
the Hamilton-Jacobi method for a photon in the BH spacetime. The
Hamilton-Jacobi equation is expressed as \cite{Carter,Decanini}
\begin{equation}
\frac{\partial S}{\partial \sigma}+H=0,
\end{equation}
where $S $ and $\sigma $ are the Jacobi action and affine parameter along
the geodesics, respectively. The Hamiltonian of the photon moving in the
static spherically symmetric spacetime is
\begin{equation}
H= \frac{1}{2}g^{\mu\nu}\frac{\partial S}{\partial x^{\mu}}\frac{\partial S}{%
\partial x^{\nu}}=0.  \label{EqHamiltonian}
\end{equation}
\begin{figure}[!htb]
\centering \subfloat[$ \sigma =0.001 $ and $ \omega=-1/3 $]{
        \includegraphics[width=0.32\textwidth]{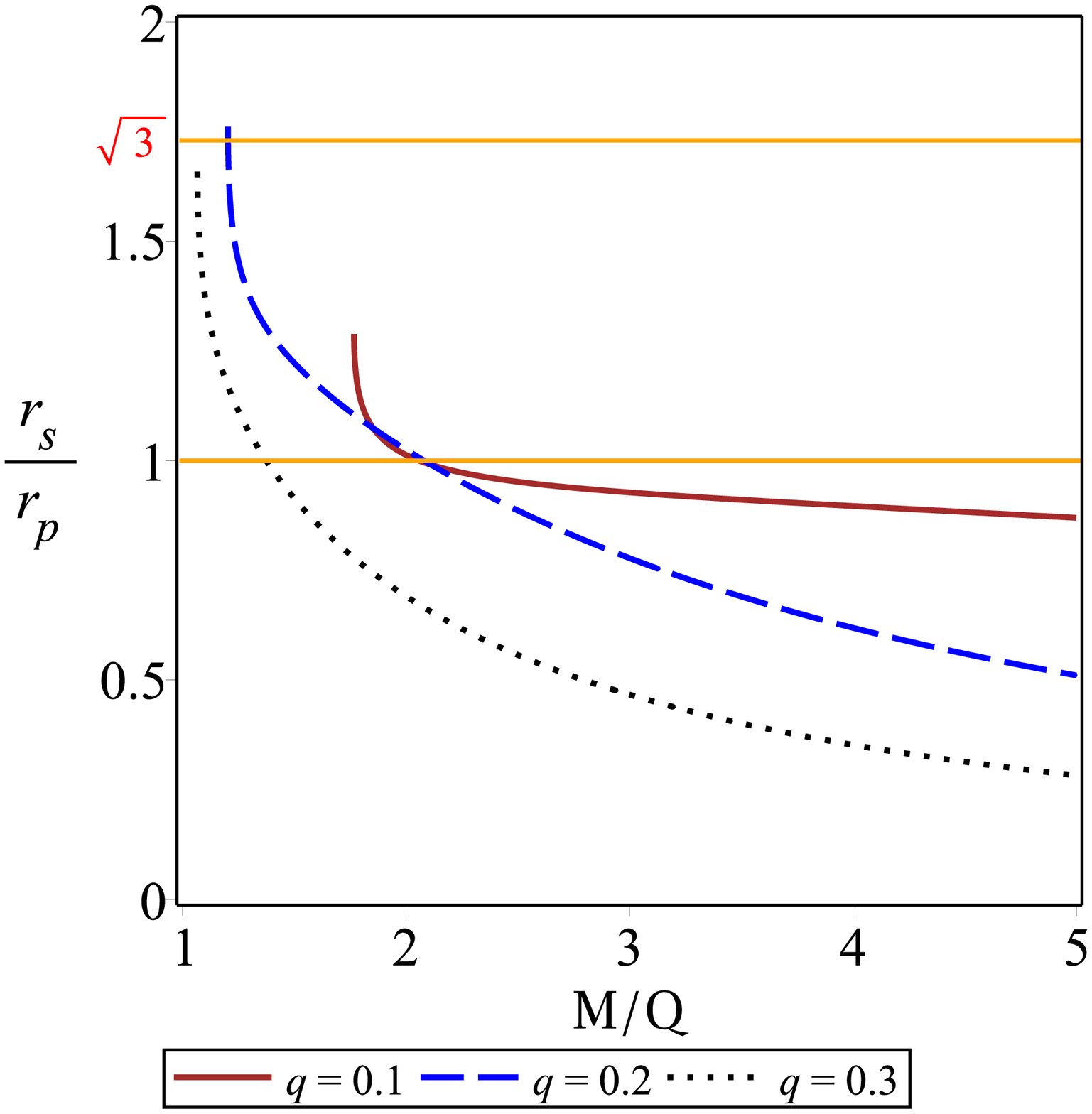}}
 \subfloat[$q=0.2$ and $ \omega=-1/3 $]{
        \includegraphics[width=0.315\textwidth]{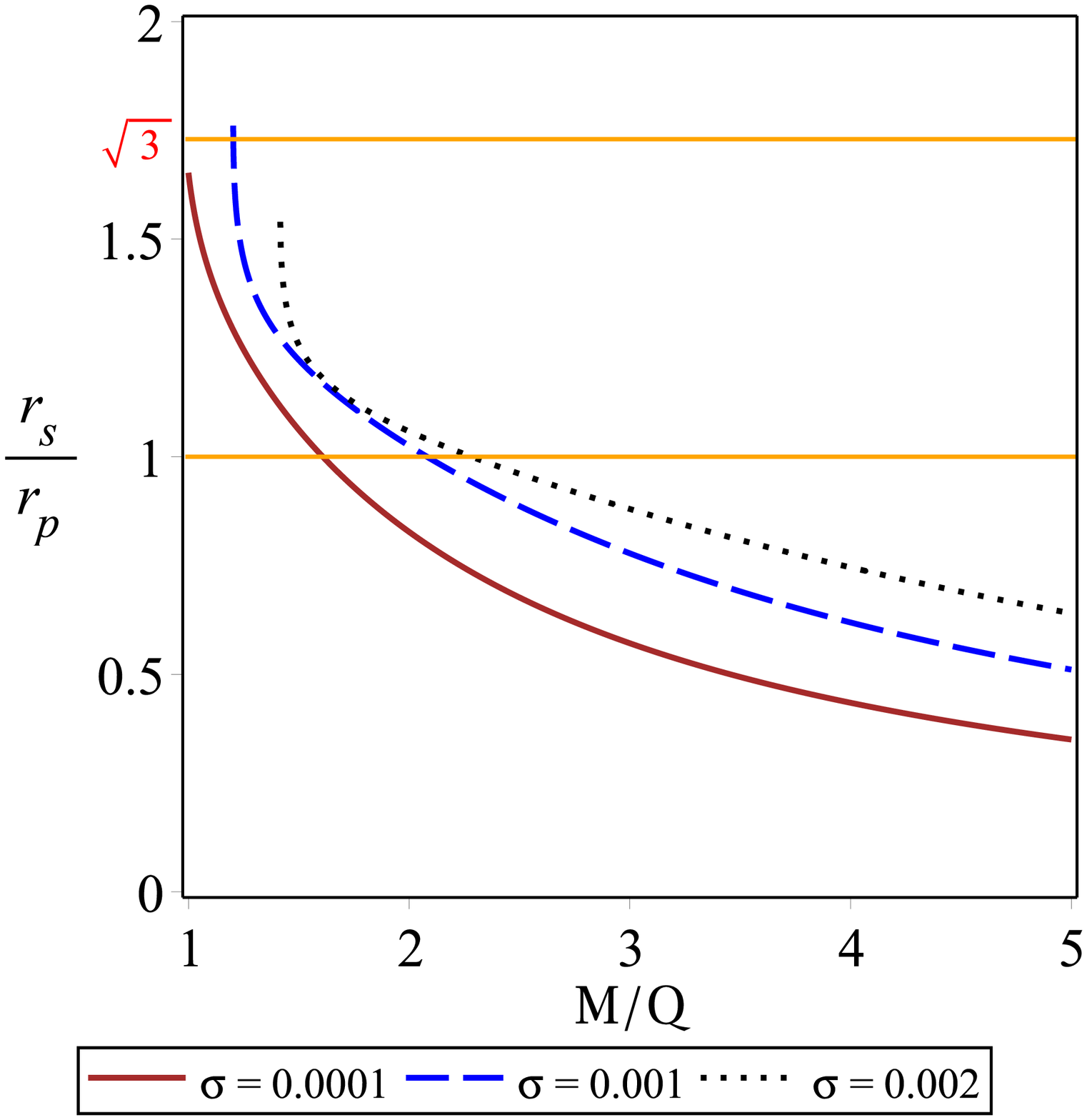}}\newline
\subfloat[$ \sigma =0.001 $ and $ \omega=-1 $]{
        \includegraphics[width=0.33\textwidth]{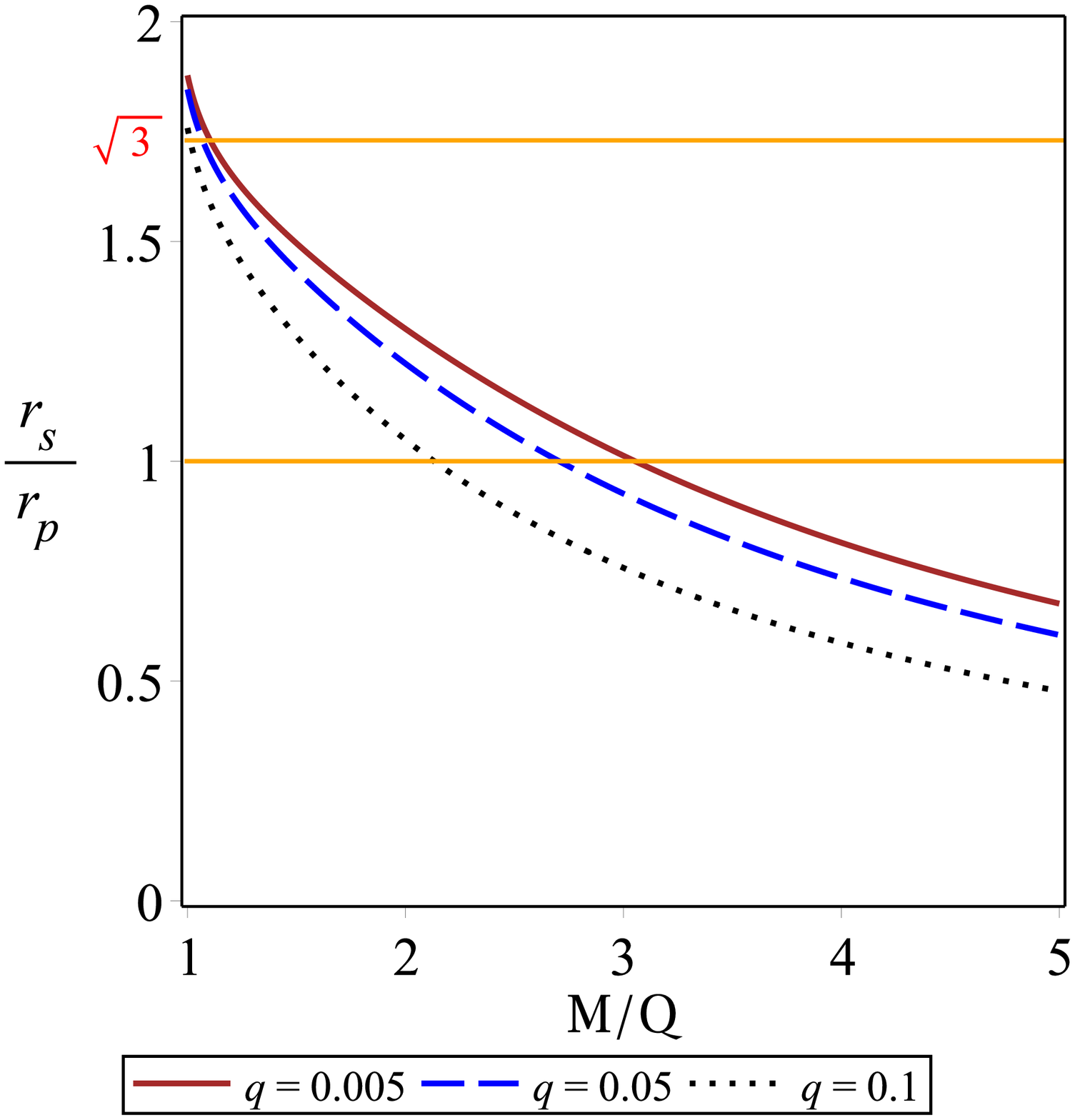}}
 \subfloat[ $q=0.1$ and $ \omega=-1 $]{
        \includegraphics[width=0.315\textwidth]{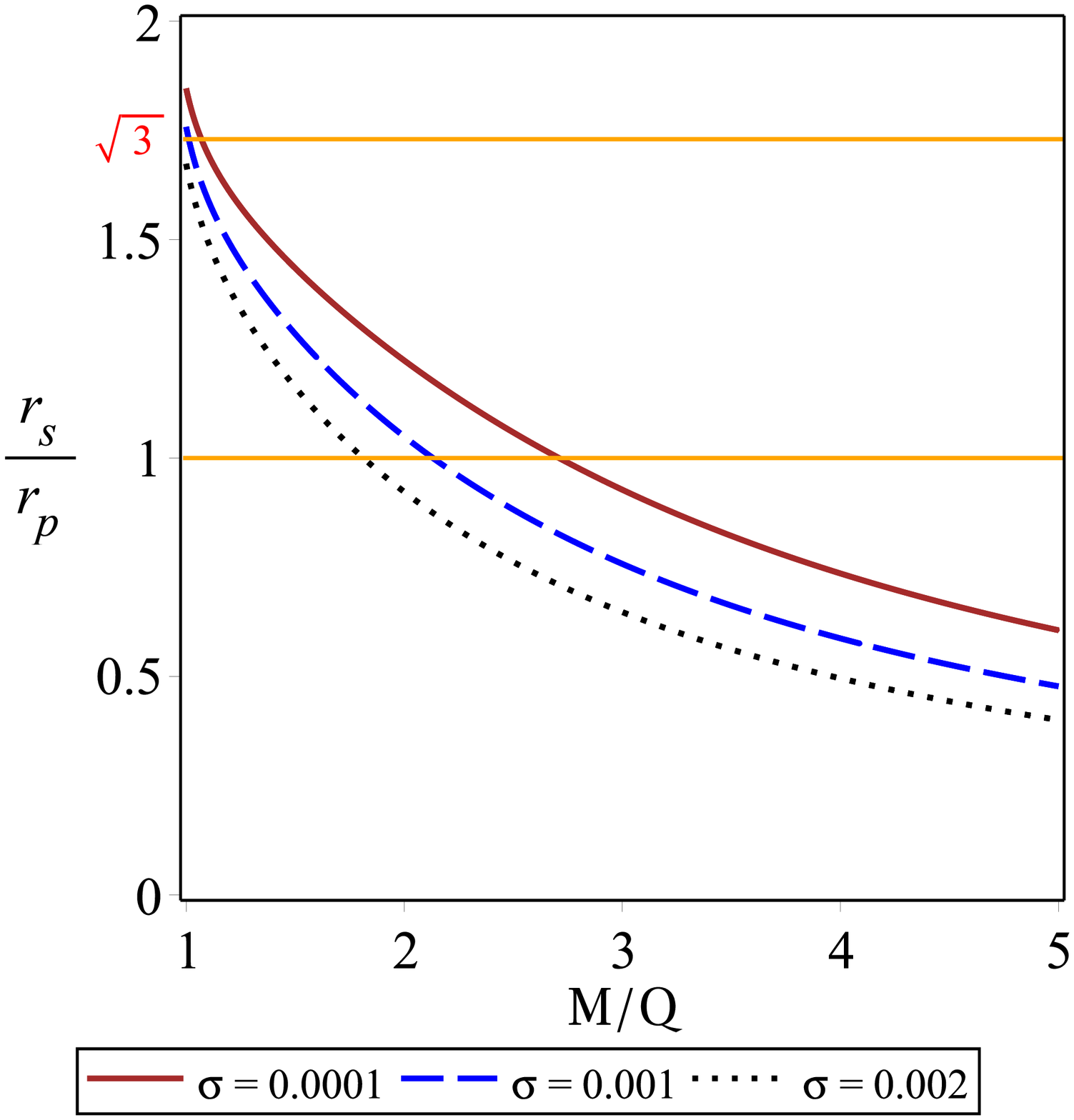}}\newline
\caption{The dependce of $\frac{r_{s}}{r_{p}} $ on the ratio $ \frac{M}{Q} $. We have set $ q=\frac{Q}{l} $ and $ \sigma=\beta Q^{2} $.} \label{FigRsp1}
\end{figure}
Due to the spherically symmetric property of the black hole, one can
consider a photon motion on the equatorial plane with $\theta=\frac{\pi}{2}$%
. So, Eq. (\ref{EqHamiltonian}) reduces to
\begin{equation}
\frac{1}{2}\left[-\frac{1}{f(r)}\left(\frac{\partial H}{\partial\dot{t}}
\right)^{2}+f(r)\left(\frac{\partial H}{\partial\dot{r}} \right)^{2}+\frac{1%
}{r^{2}}\left(\frac{\partial H}{\partial\dot{\phi}} \right)^{2} \right]=0.
\label{EqNHa}
\end{equation}

Regarding the fact that the Hamiltonian does not depend explicitly
on the coordinates $t$ and $\phi$, one can define
\begin{equation}
\frac{\partial H}{\partial\dot{t}}=-E ~~~~and~~~~ \frac{\partial H}{\partial%
\dot{\phi}}=L ,  \label{Eqenergy}
\end{equation}
where constants $E$ and $L$ are, respectively, the energy and angular
momentum of the photon. Using the Hamiltonian formalism, the equations of
motion are obtained as
\begin{equation}
\dot{t}=\frac{dt}{d\sigma}=-\frac{1}{f(r)}\left(\frac{\partial H}{\partial%
\dot{t}} \right)~~~,~~~\dot{r}=\frac{dr}{d\sigma}=-f(r)\left(\frac{\partial H%
}{\partial\dot{r}} \right)~~~,~~~\dot{\phi}=\frac{d\phi}{d\sigma}=\frac{1}{%
r^{2}}\left(\frac{\partial H}{\partial\dot{\phi}} \right).  \label{Eqmotion}
\end{equation}

The effective potential of the photon is obtained as
\begin{equation}
\dot{r}^{2}+V_{eff}(r)=0~~~\Longrightarrow ~~~ V_{eff}(r)=f(r)\left[ \frac{%
L^{2}}{r^{2}}-\frac{E^{2}}{f(r)}\right] .  \label{Eqpotential}
\end{equation}

Figure \ref{FigVef} depicts the behavior of the photon's effective
potential for $E = 1$ with various $  L$. As we see, there exists
a peak of the effective potential which increases with increasing
$L$. Due to the constraint $ \dot{r}^{2} \geq 0 $, we expect that
the effective potential satisfies $ V_{eff}\leq 0 $. So, an
ingoing photon from infinity with the negative effective potential
falls into the BH inevitably, whereas it bounces back if $ V_{eff}
> 0 $. An interesting occurrence is related to the critical
angular momentum $L = L_{p}$ ($ V_{max~eff} = 0$). In this case,
the ingoing photon loses both its radial velocity and acceleration
at $r = r_{max}$ completely. But for the sake of its non-vanishing
transverse velocity, it can circle the black hole. So, the case of
$r = r_{max}$ is called the photon orbit and it is denoted as $r =
r_{p}$. From what was expressed, one can find that the photon
orbits are circular and unstable associated to the maximum value
of the effective potential. In order to obtain such a maximum
value, we use the following conditions, simultaneously
\begin{equation}
V_{eff}(r)\Bigg\vert_{r=r_{p}}=0,~~~~~\frac{\partial V_{eff}(r)}{\partial r}%
\Bigg\vert_{r=r_{p}}=0,  ~~~~~~\frac{\partial^{2} V_{eff}(r)}{\partial r^{2}}\Bigg\vert_{r=r_{p}}<0 ,
\label{Eqcondition}
\end{equation}
where the first two conditions, determining the critical angular
momentum of the photon sphere ($ L_{p} $) and the photon sphere
radius ($ r_{p} $), respectively, result in the following equation
\begin{equation}
\frac{4Q^{2}}{r_{p}^{5}}-\frac{6M}{r_{p}^{4}} +\frac{2}{r_{p}^{3}}-\frac{8\pi\beta (\omega +1)l^{3\omega +3}}{\omega r_{p}^{3\omega +4}}=0.
\label{Eqrpn1}
\end{equation}

Besides, the third condition ensures that the photon orbits are
unstable. The orbit equation for the photon is obtained in the
following form
\begin{equation}
\frac{dr}{d\phi}=\frac{\dot{r}}{\dot{\phi}}=\frac{r^{2}f(r)}{L}\left(\frac{%
\partial H}{\partial\dot{r}} \right).  \label{Eqorbit}
\end{equation}

The turning point of the photon orbit is expressed by the following
constraint
\begin{equation}
\frac{dr}{d\phi}\Bigg\vert_{r=R}=0.  \label{EqTpoint}
\end{equation}

Using Eqs. (\ref{EqNHa}) and (\ref{EqTpoint}), one gets
\begin{equation}
\frac{dr}{d\phi}=\pm r\sqrt{f(r)\left[\frac{r^{2}f(R)}{R^{2}f(r)} -1\right] }%
.  \label{EqNorbit}
\end{equation}

Considering a light ray sending from a static observer placed at $r_{0} $
and transmitting into the past with an angle $\Theta$ with respect to the
radial direction, one can write \cite{M.Zhang,Belhaj}
\begin{equation}
\cot \Theta =\frac{\sqrt{g_{rr}}}{g_{\phi\phi}}\frac{dr}{d\phi}\Bigg\vert%
_{r=r_{0}}.  \label{Eqangle}
\end{equation}

Hence, the shadow radius of the BH can be obtained as
\begin{equation}
r_{s}=r_{0}\sin \Theta =R\sqrt{\frac{f(r_{0})}{f(R)}}\Bigg\vert_{R=r_{p}},
\label{Eqshadow}
\end{equation}
where $r_{0} $ is the position of the observer.

 \begin{table*}[htb!]
\centering \caption{The event horizon, photon sphere radius and
shadow radius for the variation of the total mass, the electric
charge,  the state parameter and the parameter $\beta$  for $l
=1$.} \label{table2}
\begin{tabular}{c c c c c c}
 \footnotesize $M$  \hspace{0.3cm} & \hspace{0.3cm}$0.2$ \hspace{0.3cm} &
\hspace{0.3cm} $0.25$\hspace{0.3cm} & \hspace{0.3cm} $0.3$\hspace{0.3cm} & \hspace{0.3cm}$0.35$\hspace{0.3cm} & \hspace{0.3cm} \\ \hline\hline
$ r_e (Q=0.1 $, $\beta=0.01$ and  $\omega=-0.8 $)  & $ 0.24 $ & $0.32$ & $0.39$&$0.46$ \\
$ r_p (Q=0.1 $, $\beta=0.01$ and  $\omega=-0.8 $) & $ 0.39 $ & $0.55$ & $0.7$&$0.85$\\
$ r_s (Q=0.1 $, $\beta=0.01$ and  $\omega=-0.8 $) & $ 0.56 $ & $0.68$ & $0.75$&$0.8$\\\hline
\\
\footnotesize $Q$  \hspace{0.3cm} & \hspace{0.3cm}$0.01$ \hspace{0.3cm} &
\hspace{0.3cm} $0.05$\hspace{0.3cm} & \hspace{0.3cm} $0.1$\hspace{0.3cm} & \hspace{0.3cm}$0.15$\hspace{0.3cm} & \hspace{0.3cm} \\ \hline\hline
$ r_e (\beta=0.01 $, $M=0.3$ and  $\omega=-0.8 $)  & $ 0.42 $ & $0.41$ & $0.399$&$0.37$ \\
$ r_p (\beta=0.01 $, $M=0.3$ and  $\omega=-0.8 $) & $ 0.730 $ & $0.721$ & $0.708$&$0.671$\\
$ r_s (\beta=0.01 $, $M=0.3$ and  $\omega=-0.8 $) & $ 0.762 $ & $0.761$ & $0.754$&$0.743$\\\hline
\\
\footnotesize $\beta$  \hspace{0.3cm} & \hspace{0.3cm}$0.01$ \hspace{0.3cm} &
\hspace{0.3cm} $0.02$\hspace{0.3cm} & \hspace{0.3cm} $0.04$\hspace{0.3cm} & \hspace{0.3cm}$0.06$\hspace{0.3cm} & \hspace{0.3cm} \\ \hline\hline
$ r_e (Q=0.1 $, $M=0.3$ and  $\omega=-0.8 $)  & $ 0.398 $ & $0.392$ & $0.37$&$0.36$ \\
$ r_p (Q=0.1 $, $M=0.3$ and  $\omega=-0.8 $) & $ 0.70 $ & $0.69$ & $0.67$&$0.65$\\
$ r_s (Q=0.1 $, $M=0.3$ and  $\omega=-0.8 $) & $ 0.75 $ & $0.72$ & $0.68$&$0.64$\\\hline
\\
 \footnotesize $\omega$  \hspace{0.3cm} & \hspace{0.3cm}$-1$ \hspace{0.3cm} &
\hspace{0.3cm} $-0.9$\hspace{0.3cm} & \hspace{0.3cm} $-0.8$\hspace{0.3cm} & \hspace{0.3cm}$-0.7$\hspace{0.3cm} & \hspace{0.3cm}  \\ \hline\hline
$ r_e (Q=0.1 $, $M=0.3$ and  $\beta=0.01 $)  & $ 0.403 $ & $0.402$ & $0.399$&$0.395$ \\
$ r_p (Q=0.1 $, $M=0.3$ and  $\beta=0.01 $) & $ 0.722 $ & $0.716$ & $0.708$&$0.696$\\
$ r_s (Q=0.1 $, $M=0.3$ and  $\beta=0.01 $) & $ 0.764 $ & $0.760$ & $0.754$&$0.747$\\\hline
\end{tabular}
\end{table*}

Since Eqs. (\ref{Eqrpn1}) and (\ref{Eqshadow}) are complicated to
solve analytically, we employ numerical methods to obtain the
radius of the photon sphere and shadow. In this regard, several
values of the event horizon ($ r_{e} $), photon sphere radius ($
r_{p} $) and shadow radius ($ r_{s} $) are presented in Table
\ref{table2}. We find that for intermediate values of the total
mass, the photon sphere radius would be larger than the shadow
radii which is physically not acceptable. To have a physical
behavior, we consider the small mass for our investigation. We
note that the size of shadow shrinks with increasing the electric
charge and parameter $ \beta $. Regarding the effect of state
parameter, we find that increasing this parameter from $ -1 $ to $
-2/3 $ results in a decrease of the photon sphere and shadow
radius.
\begin{figure}[!htb]
\centering \subfloat[]{
        \includegraphics[width=0.325\textwidth]{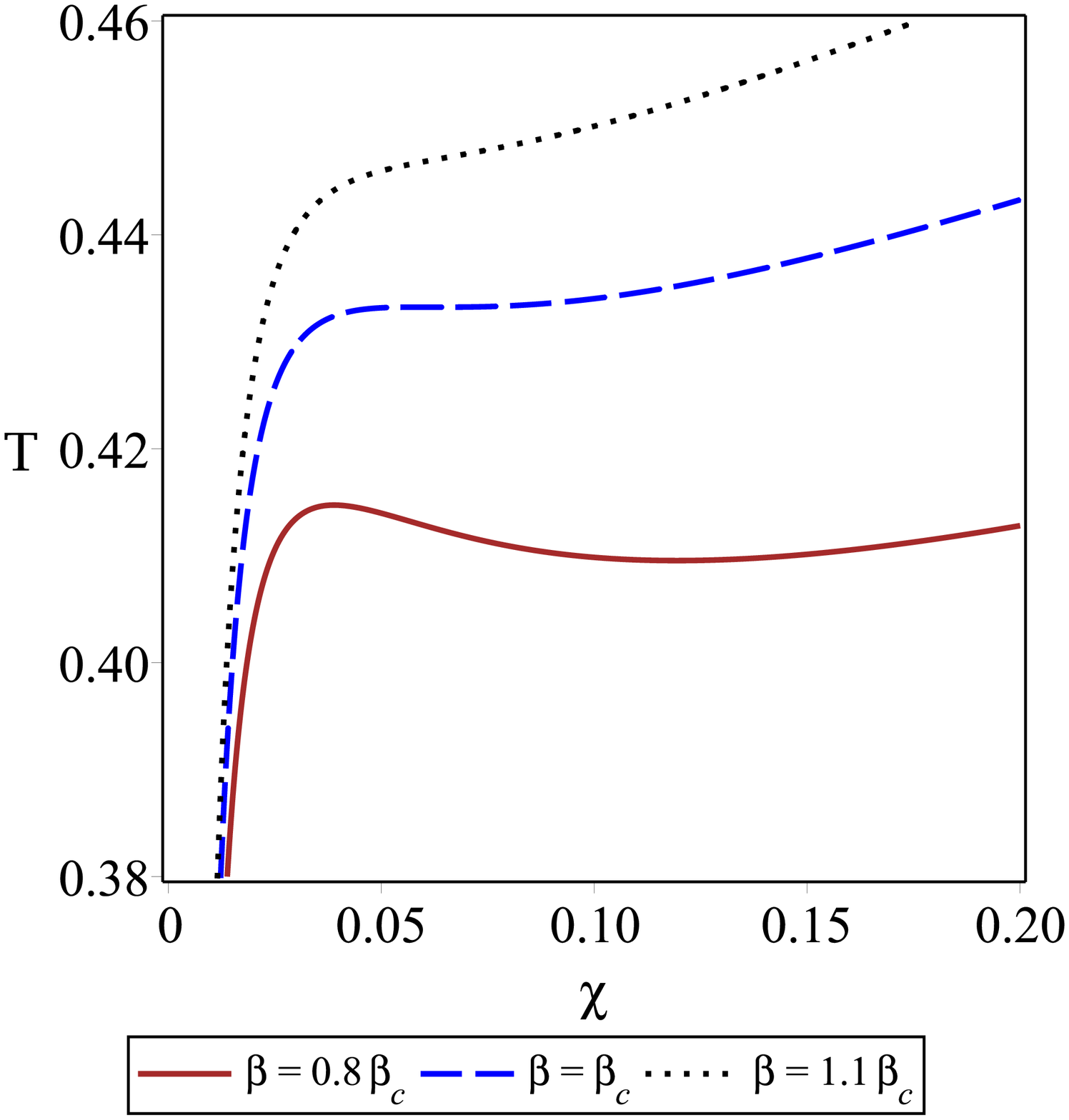}}
 \subfloat[]{
        \includegraphics[width=0.315\textwidth]{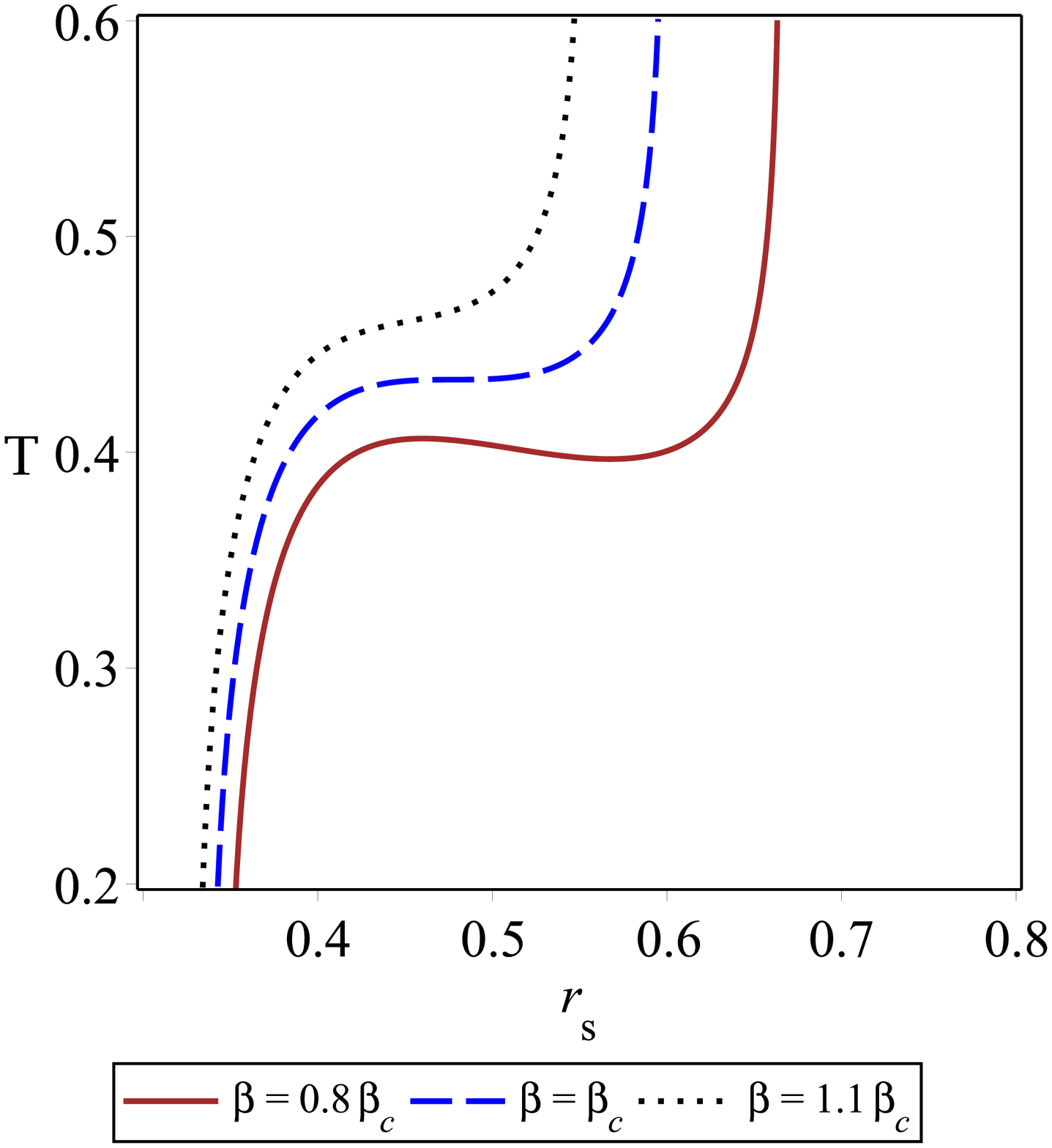}}\newline
\subfloat[]{
        \includegraphics[width=0.33\textwidth]{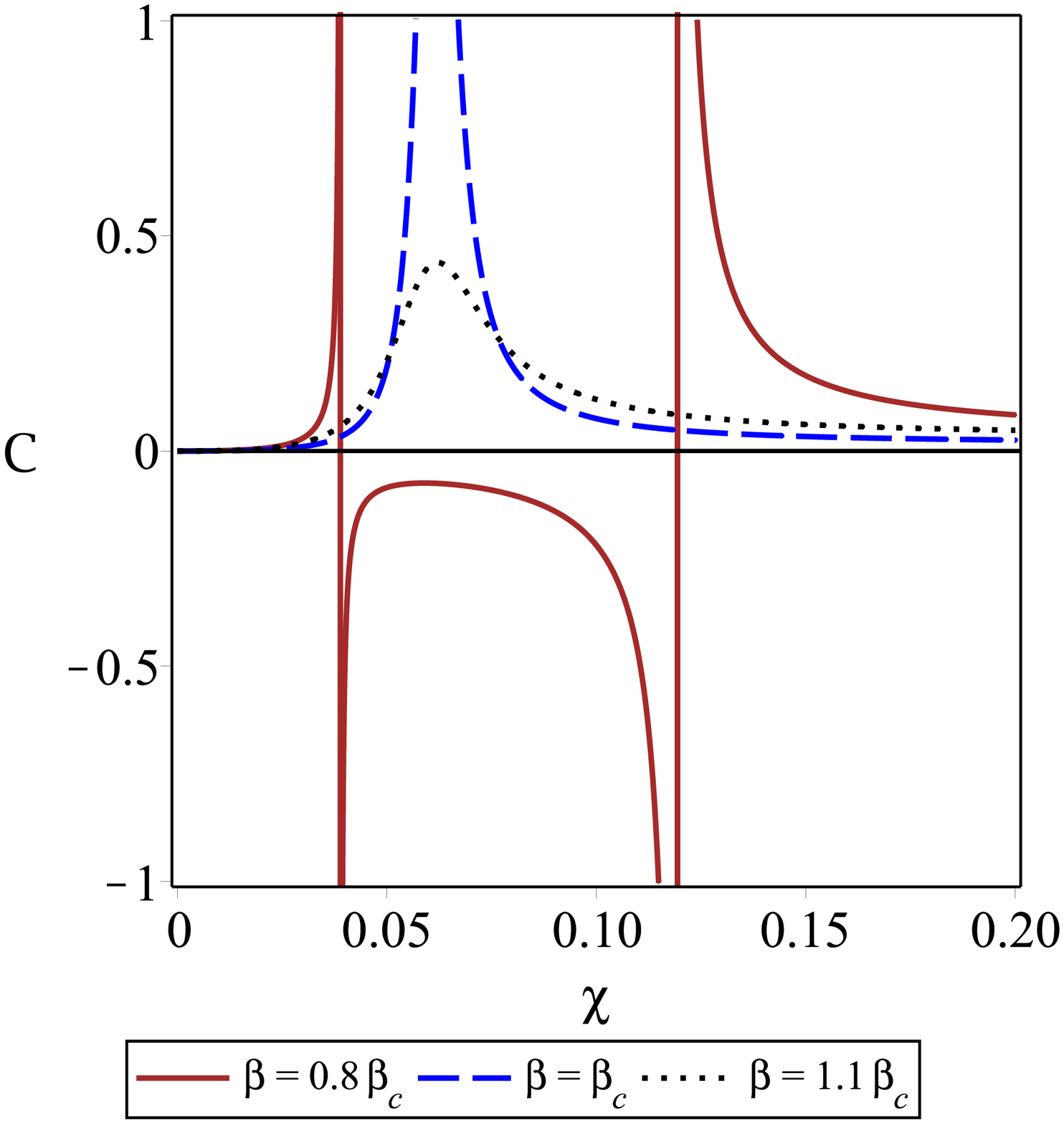}}
 \subfloat[]{
        \includegraphics[width=0.33\textwidth]{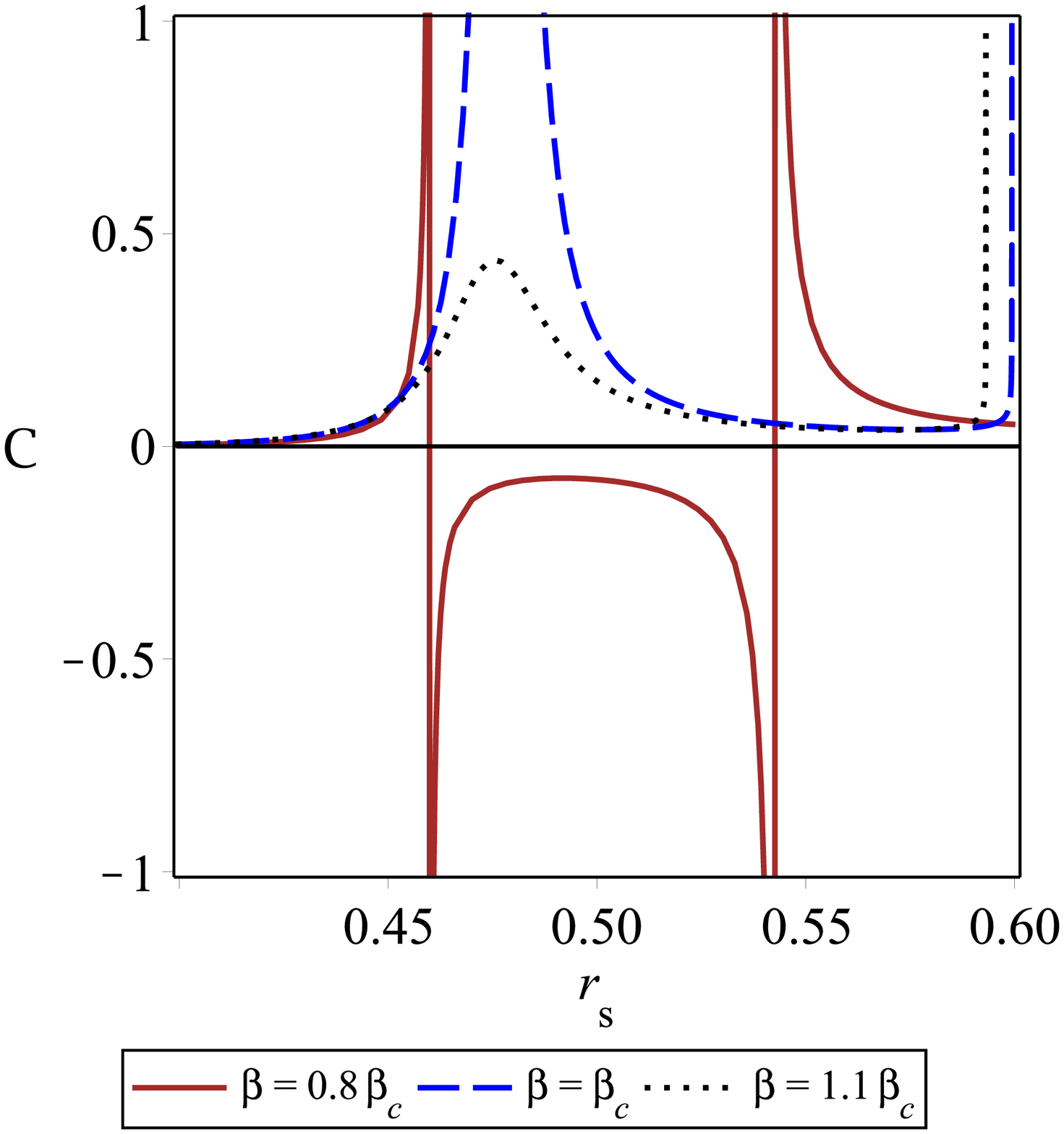}}\newline
\caption{Hawking temperature with respect to $ \chi $ (up-left panel) and $ r_{s} $ (up-right panel); The heat capacity
versus $ \chi $ (down-left panel) and $ r_{s} $ (down-right panel) for $l=1$,  $Q=0.1$, $ \omega =-1 $ and various $ \beta $.} \label{Fig7}
\end{figure}

As it was already mentioned, no acceptable optical behavior is
observed for large values of total mass. This shows that a
restriction between the total mass and electric charge should be
imposed to observe an acceptable physical result. To have a better
understanding of this issue, we examine the ratio of shadow radius
and photon sphere ($\frac{r_{s}}{r_{p}}$) for two limited states,
$ \omega =-\frac{1}{3} $ and $ \omega =-1 $. For more precise
study, we first inspect this ratio for three BH solutions as the
Reissner-Nordstr\"{o}m (RN) BH, RN - AdS BH and  RN BH in the
presence of quintessence. The qualitative behavior of the ratio
$\frac{r_{s}}{r_{p}}$ with respect to $\frac{M}{Q}$ is displayed
in Fig. \ref{FigRsp}. From this figure, one can find that the
proposed inequality relationship ($ \frac{r_{s}}{r_{p}} \geq
\sqrt{3}$) in Ref. \cite{HDLyu} is satisfied for charged BH (see
Fig. \ref{FigRsp}a) and charged one in the quintessence (see Figs.
\ref{FigRsp}c and \ref{FigRsp}d), whereas for RN - AdS BH, as we
see from Fig. \ref{FigRsp}b, this ratio is fulfill just for very
small values of $q=\frac{Q}{l}$ (for small electric charge or
large AdS radius). Figure \ref{FigRsp} also shows that the ratio
$\frac{r_{s}}{r_{p}}$ monotonically decreases as the ratio
$\frac{M}{Q}$ increases. This means that the increase of
difference between the values of total mass and electric charge
will make the ratio $\frac{r_{s}}{r_{p}}$ decrease. Only for RN
black hole in the presence of quintessence with $\omega=-1$,  the
ratio $\frac{r_{s}}{r_{p}}$ increases by increasing the ratio
$\frac{M}{Q}$, however for very small values of $ \delta = a Q^{2}
$ (for small electric charge and normalization factor), the ratio
$\frac{r_{s}}{r_{p}}$ is a decreasing function of $ \frac{M}{Q} $.
\begin{figure}[!htb]
\centering \subfloat[]{
        \includegraphics[width=0.32\textwidth]{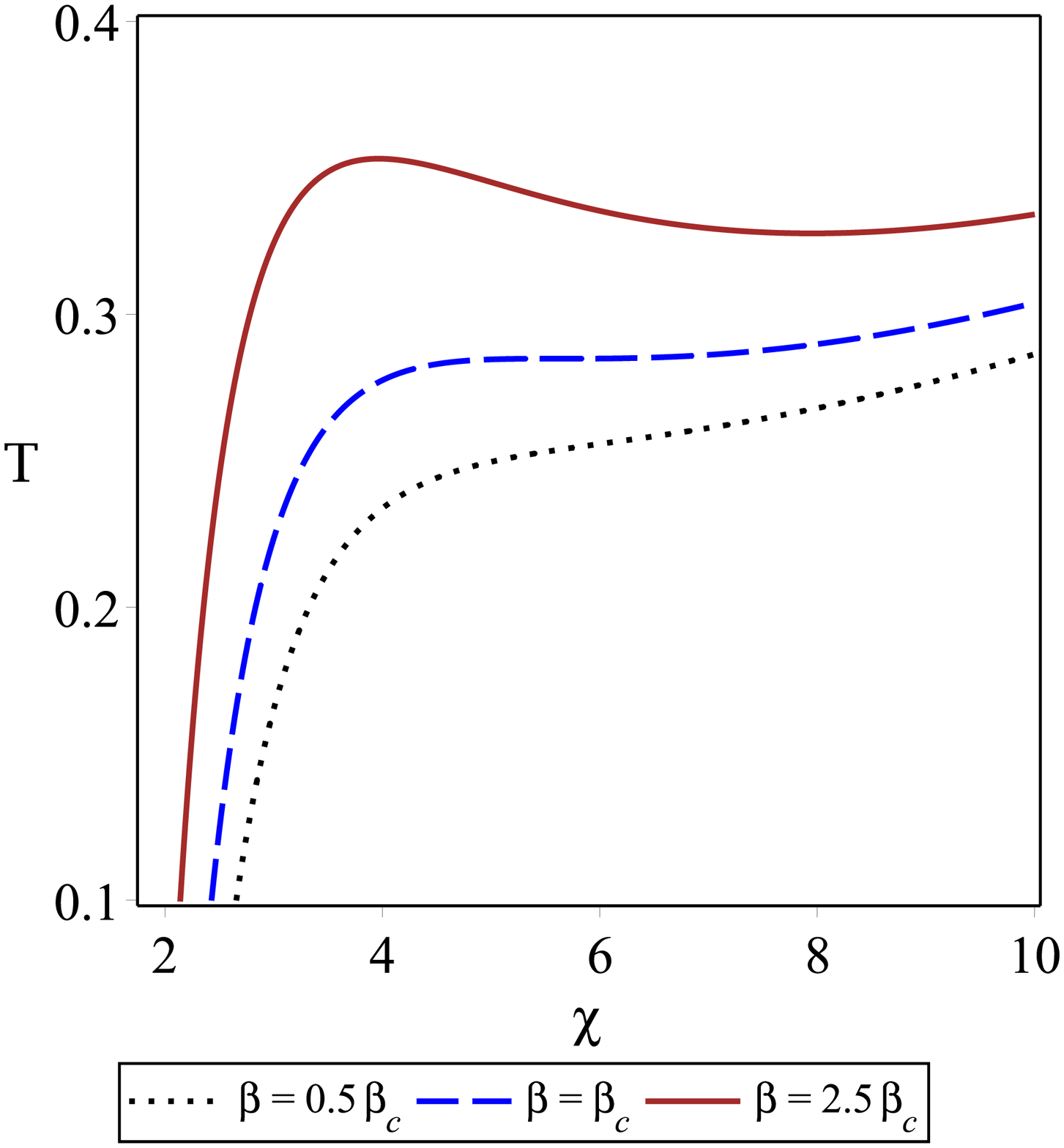}}
 \subfloat[]{
        \includegraphics[width=0.315\textwidth]{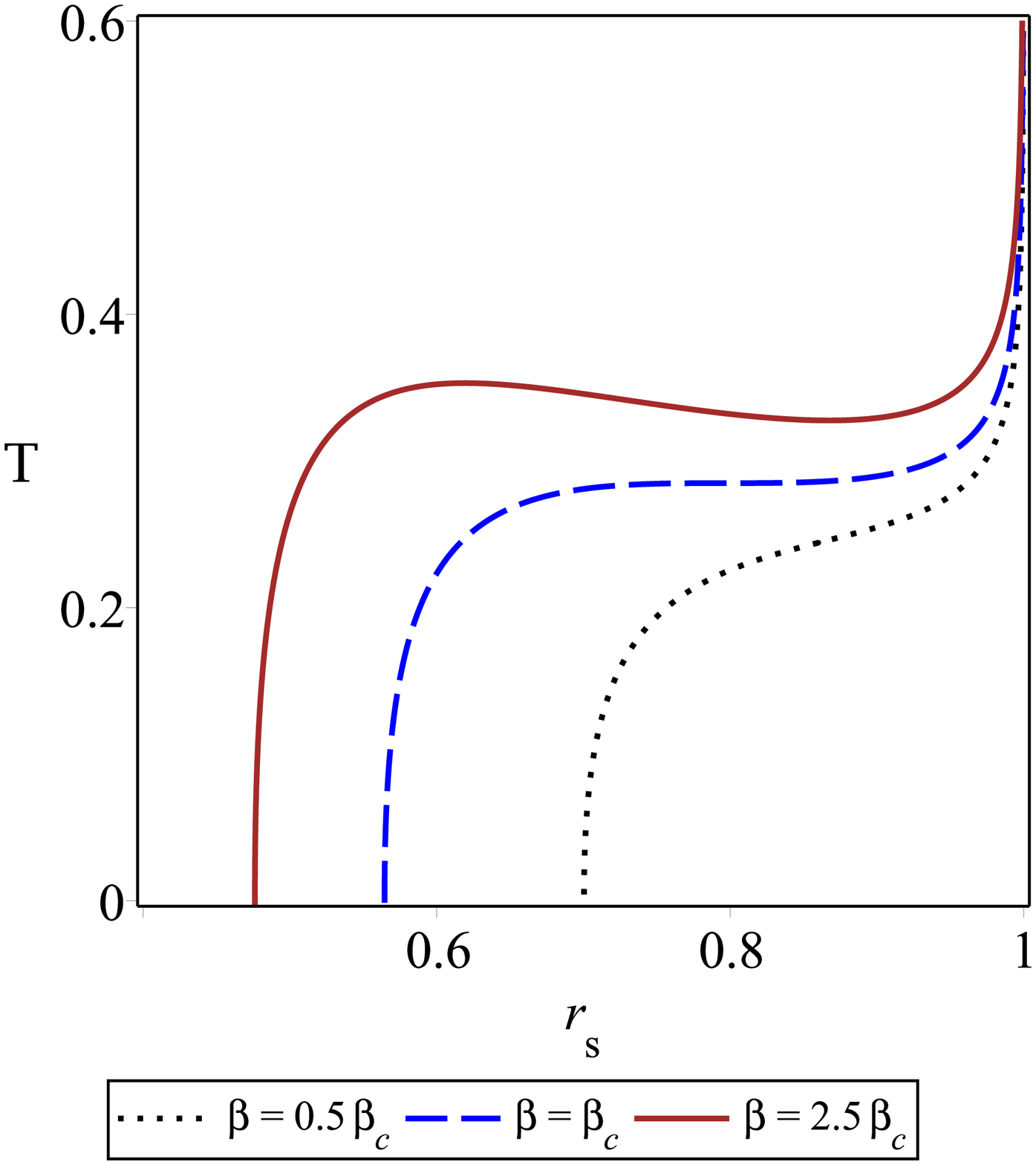}}\newline
\subfloat[]{
        \includegraphics[width=0.33\textwidth]{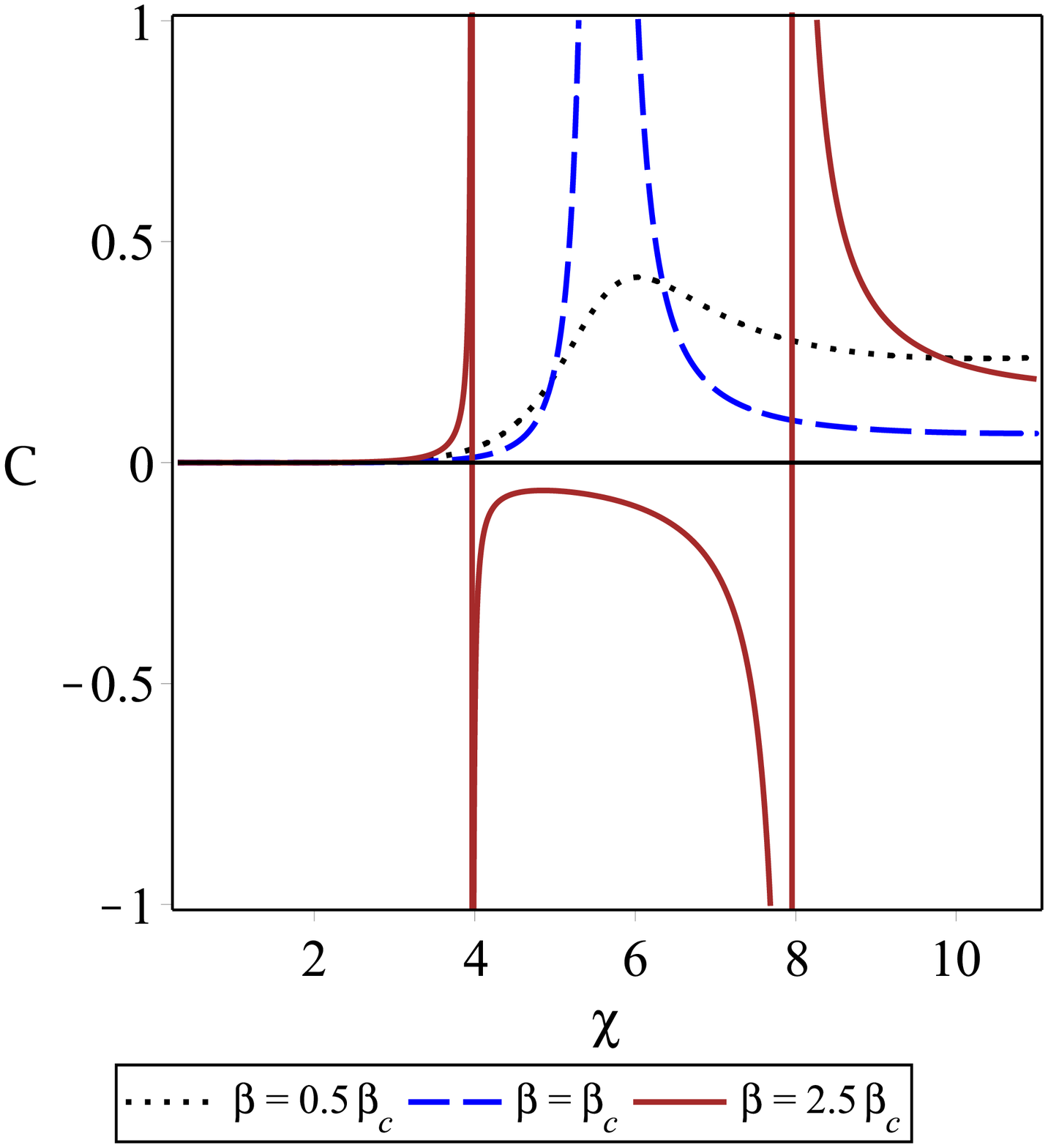}}
 \subfloat[]{
        \includegraphics[width=0.33\textwidth]{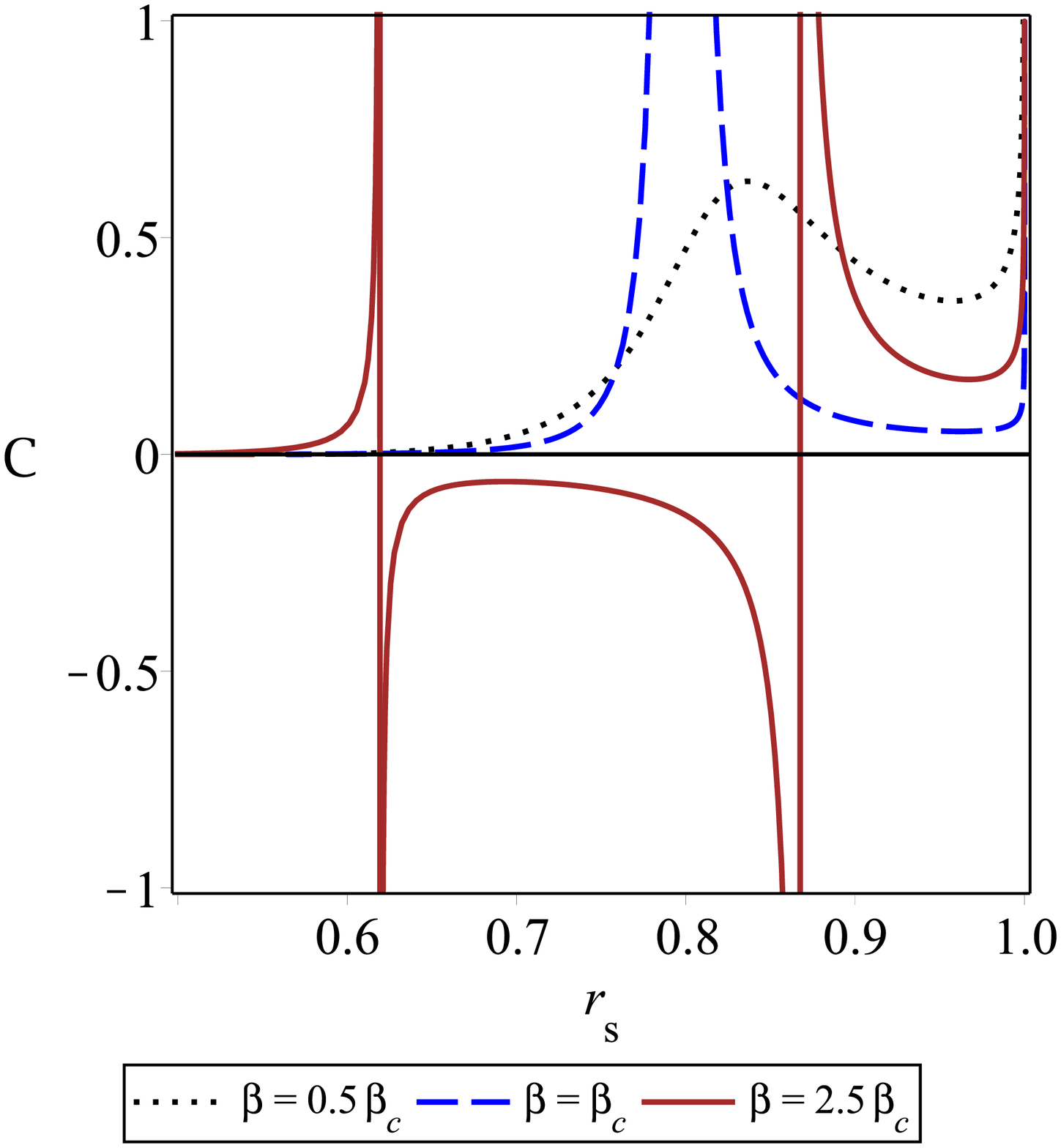}}\newline
\caption{Hawking temperature with respect to $ \chi $ (up-left panel) and $ r_{s} $ (up-right panel); The heat capacity
versus $ \chi $ (down-left panel) and $ r_{s} $ (down-right panel) for $l=1$,  $Q=0.2$, $ \omega =-1/3 $ and various $ \beta $.} \label{Fig8}
\end{figure}
Now, we would like to investigate qualitative behavior of the
ratio $\frac{r_{s}}{r_{p}}$ for our solution. From Eq.
(\ref{Eqrpn1}), the radius of photon sphere can be obtained as

\begin{eqnarray}
r_{p}&=&\frac{3M+\sqrt{9M^{2}+8Q^{2}-64\pi \beta Q^{2}l^{2}}}{2(8\pi \beta l^{2}+1)} ~~~~\omega =-\frac{1}{3}, \label{rpa}\\
r_{p}&=&\frac{3}{2}M+\frac{1}{2}\sqrt{9M^{2}-8Q^{2}}~~~~~~~~~~~~~~~~~~\omega =-1.
\label{rpb}
\end{eqnarray}

Inserting $ r_{p} $ into Eq. (\ref{Eqshadow}), one can obtain the
radius of shadow. The ratio $\frac{r_{s}}{r_{p}}$ with respect to
$ \frac{M}{Q} $ is depicted in Fig. \ref{FigRsp1}. Taking a look
at this figure, one can find that the inequality relationship $
\frac{r_{s}}{r_{p}} \geq \sqrt{3}$ is not fulfill for RN-AdS BH in
the presence of quintessence. Moreover for large $ \frac{M}{Q} $,
the radius of the photon sphere would be larger than the shadow
radii which is not physically acceptable.
\begin{figure}[!htb]
\centering \subfloat[$ Q=0.2 $ and $\omega=-1/3 $]{
        \includegraphics[width=0.32\textwidth]{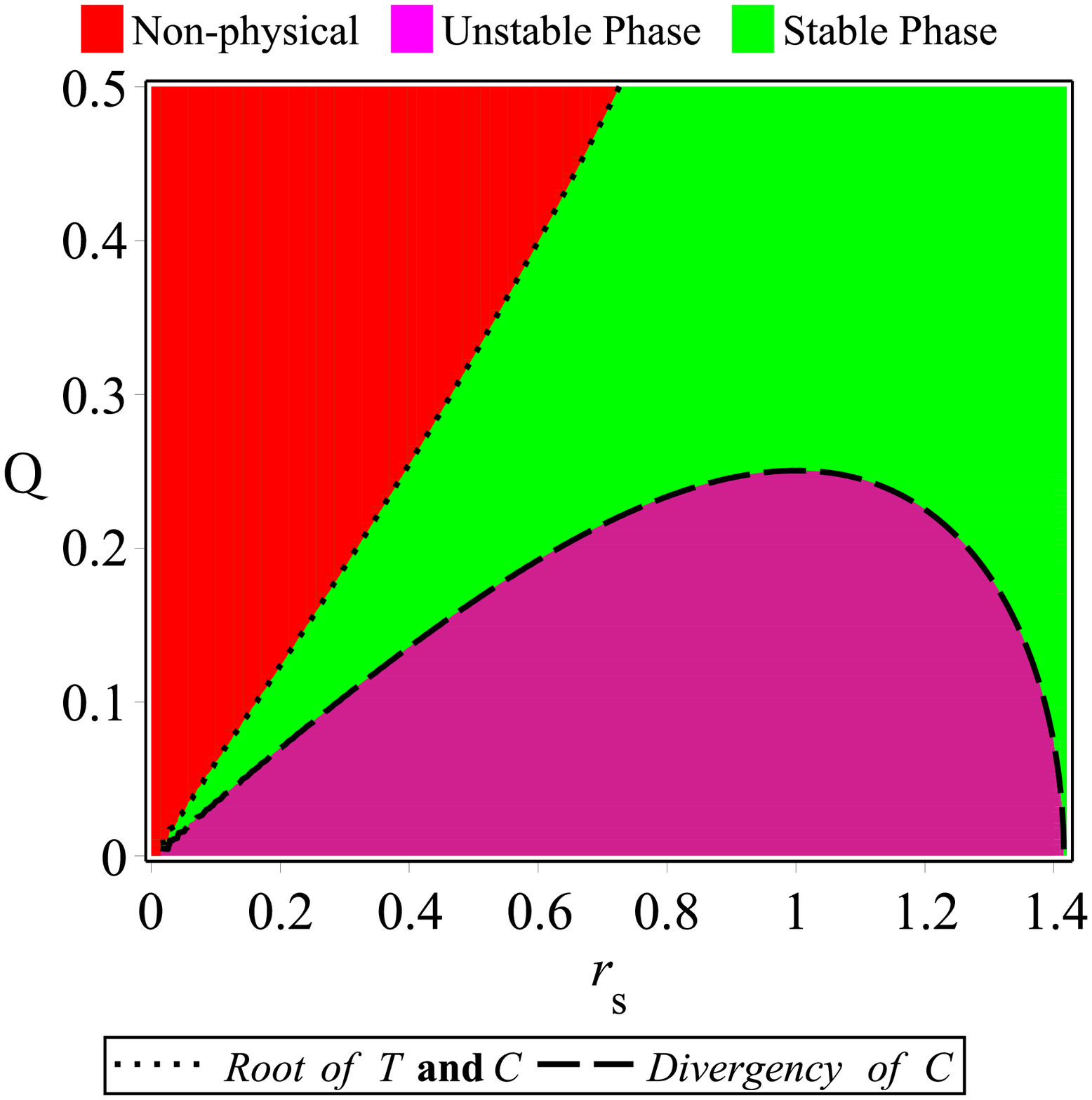}}
 \subfloat[$Q=0.2$ and $\omega=-1/3 $]{
        \includegraphics[width=0.32\textwidth]{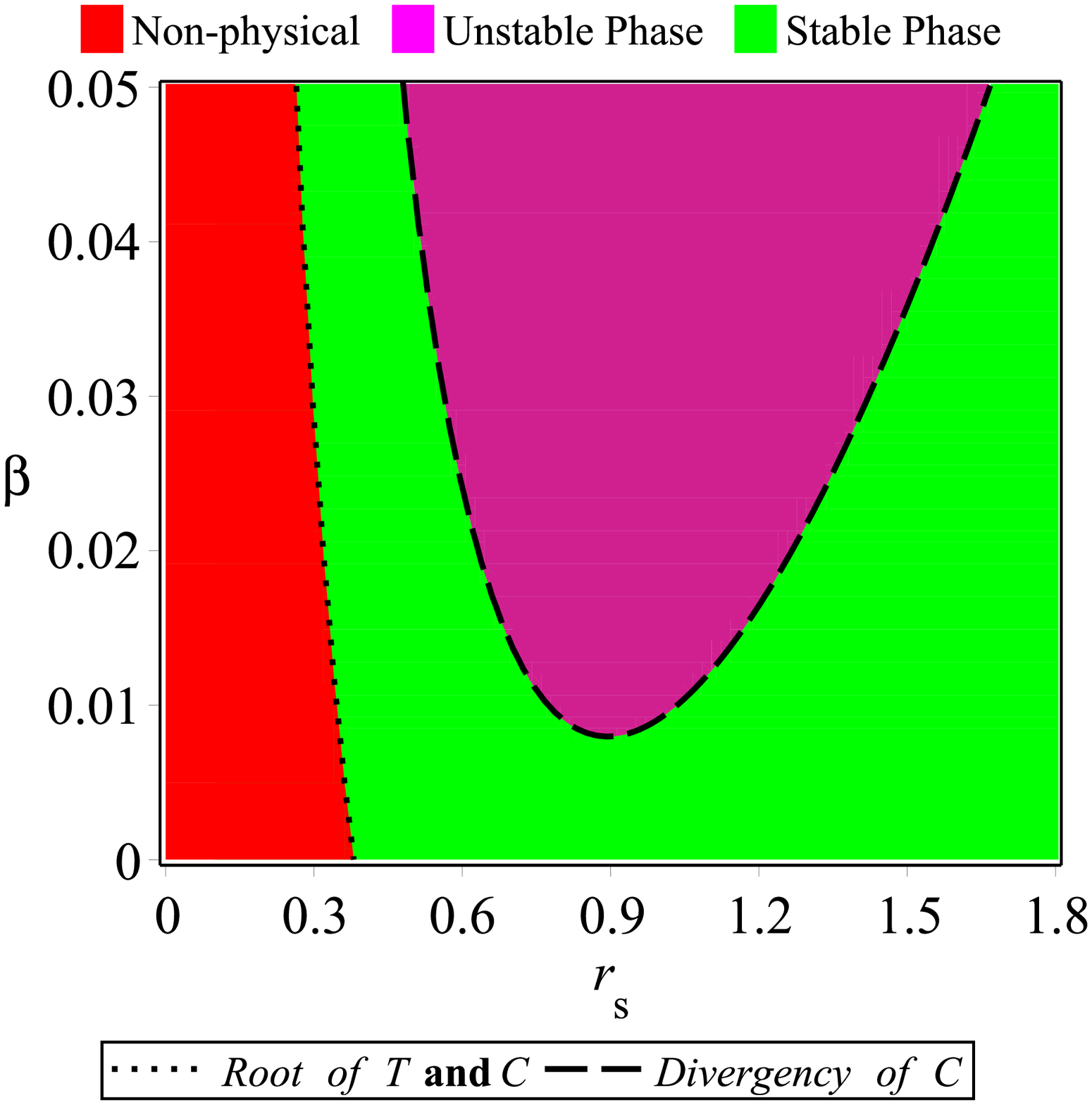}}\newline
\subfloat[$ Q=0.1 $ and $ \omega=-1 $]{
        \includegraphics[width=0.32\textwidth]{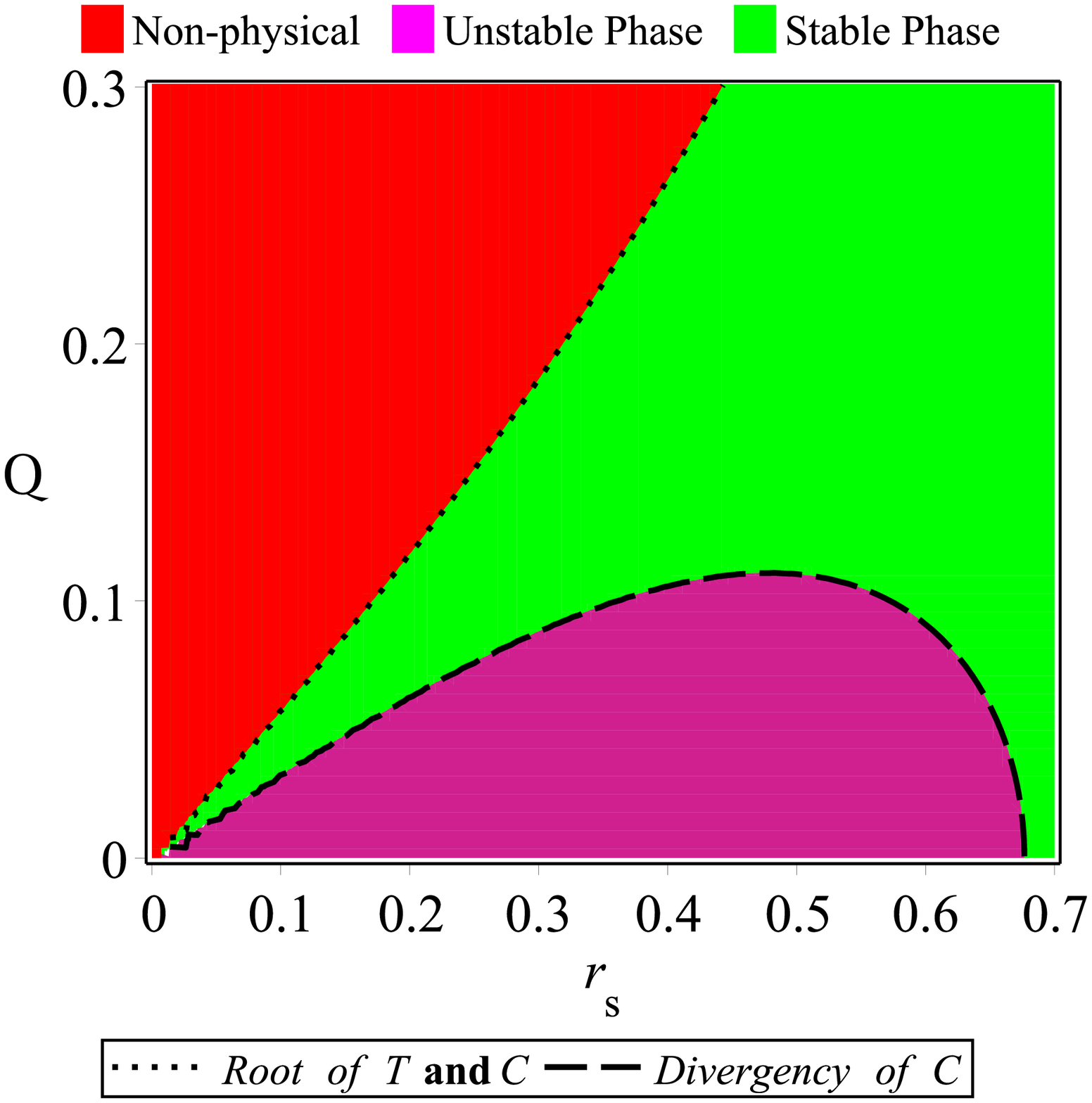}}
 \subfloat[ $Q=0.1$ and $ \omega=-1 $]{
        \includegraphics[width=0.32\textwidth]{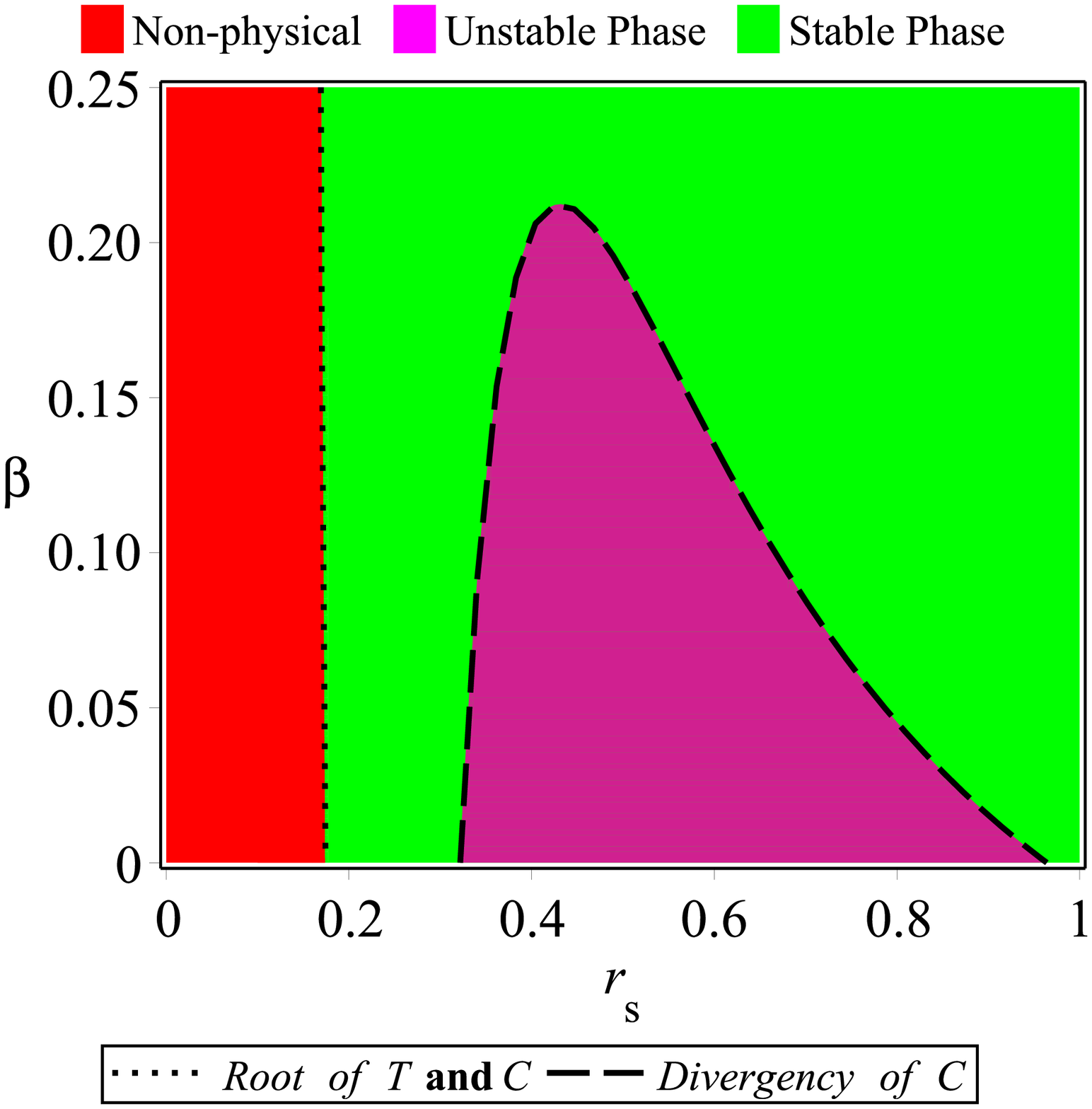}}\newline
\caption{Thermally stable and/or unstable regions of the black holes for $l=1$.} \label{Fig9}
\end{figure}

\subsection{Relations between shadow radius and phase transitions} \label{shadow-critical}

Now, we are interested in examining the relations between the
shadow radius and phase transitions. According to
\cite{M.Zhang,Belhaj,Wei2,Li2}, there is a close connection
between BH shadows and the BH thermodynamics. The heat capacity is
one of the interesting thermodynamic quantities which provides the
information related to the thermal stability and phase transition
of a thermodynamic system. The sign of heat capacity determines
thermal stability/instability of black holes. The positivity
(negativity) of this quantity indicates a BH is thermally stable
(unstable). Besides, the discontinuities in heat capacity could be
interpreted as the possible phase transition points. Indeed, the
phase transition points are where heat capacity diverges.
According to Eq. (\ref{Eqheat}), heat capacity can be written as
\begin{equation*}
C=T\left(\frac{\partial S}{\partial r_{+}}\frac{\partial
r_{+}}{\partial T} \right).
\end{equation*}
By using the fact that $ \frac{\partial S}{\partial r_{+}}>0 $,
the sign of $ C $ is directly inducted from $ \frac{\partial
T}{\partial r_{+}} $ which can be rewritten as
\begin{equation*}
\frac{\partial T}{\partial r_{+}}=\frac{\partial T}{\partial r_{s}}\frac{\partial r_{s}}{\partial r_{+}}.
\end{equation*}

\begin{figure}[!htb]
\centering \subfloat[$ Q=0.2 $ and $ \omega=-1/3 $]{
        \includegraphics[width=0.32\textwidth]{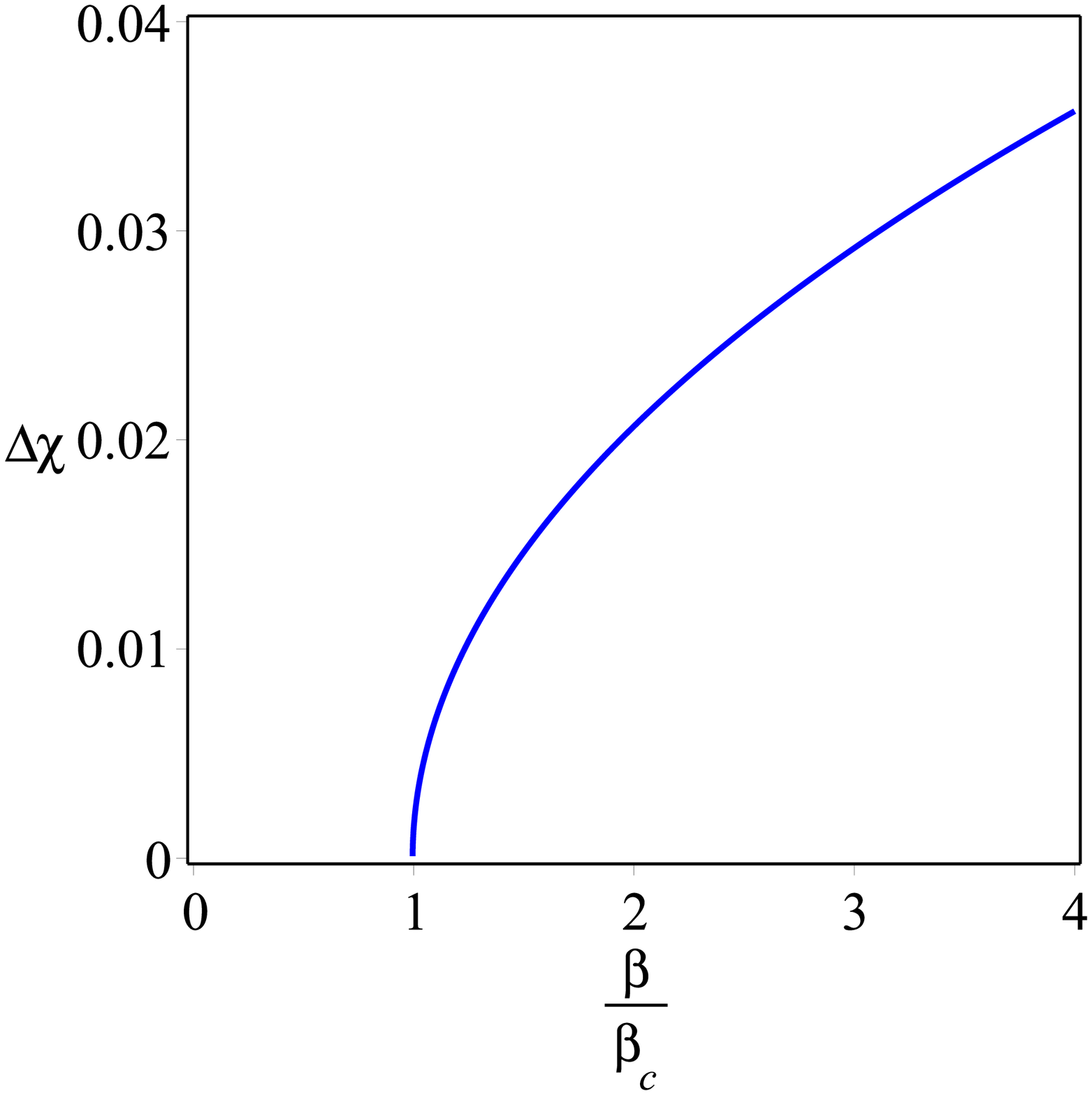}}
 \subfloat[$Q=0.2$ and $ \omega=-1/3 $]{
        \includegraphics[width=0.325\textwidth]{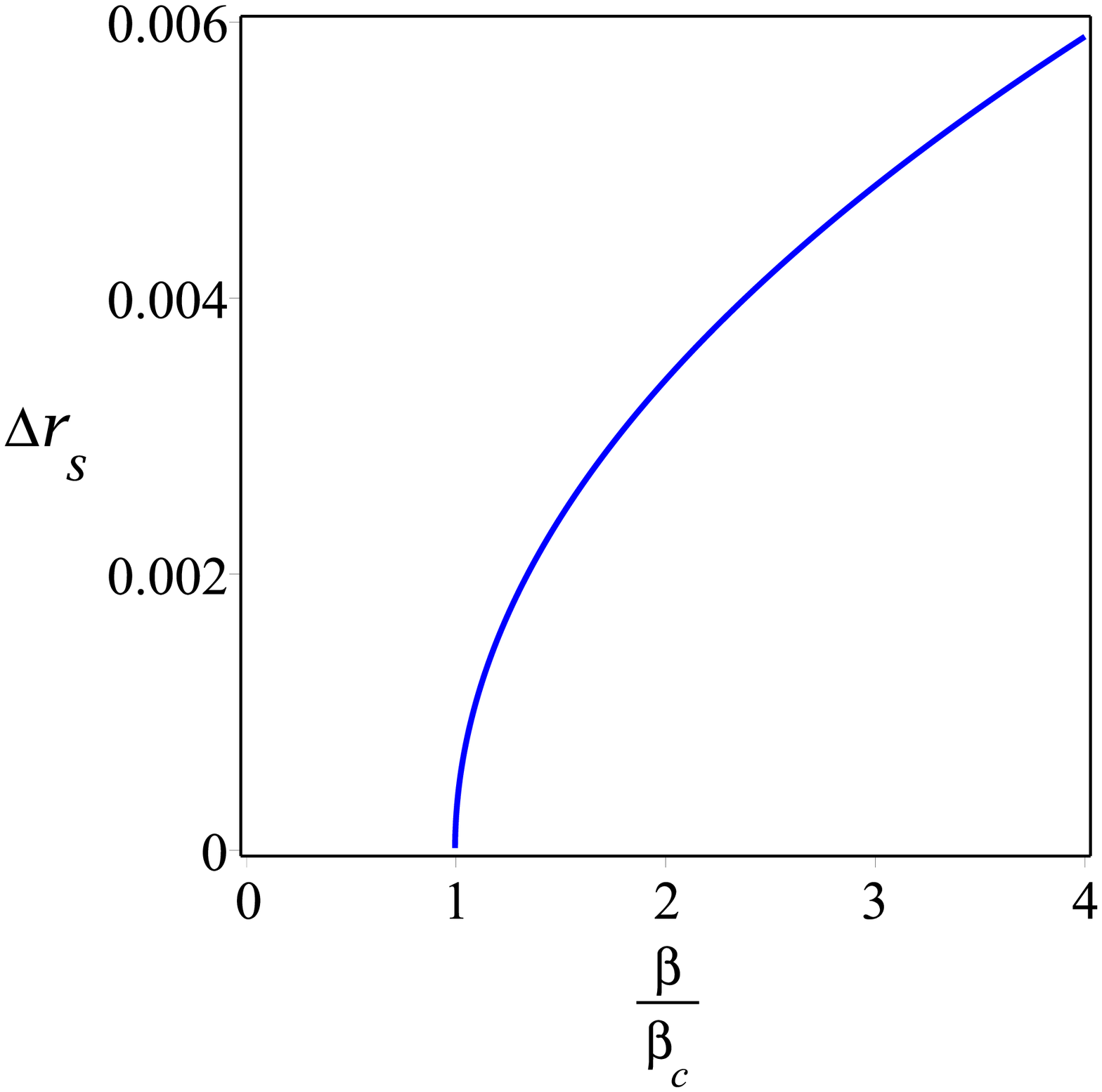}}\newline
\subfloat[$ Q=0.1 $ and $ \omega=-1 $]{
        \includegraphics[width=0.33\textwidth]{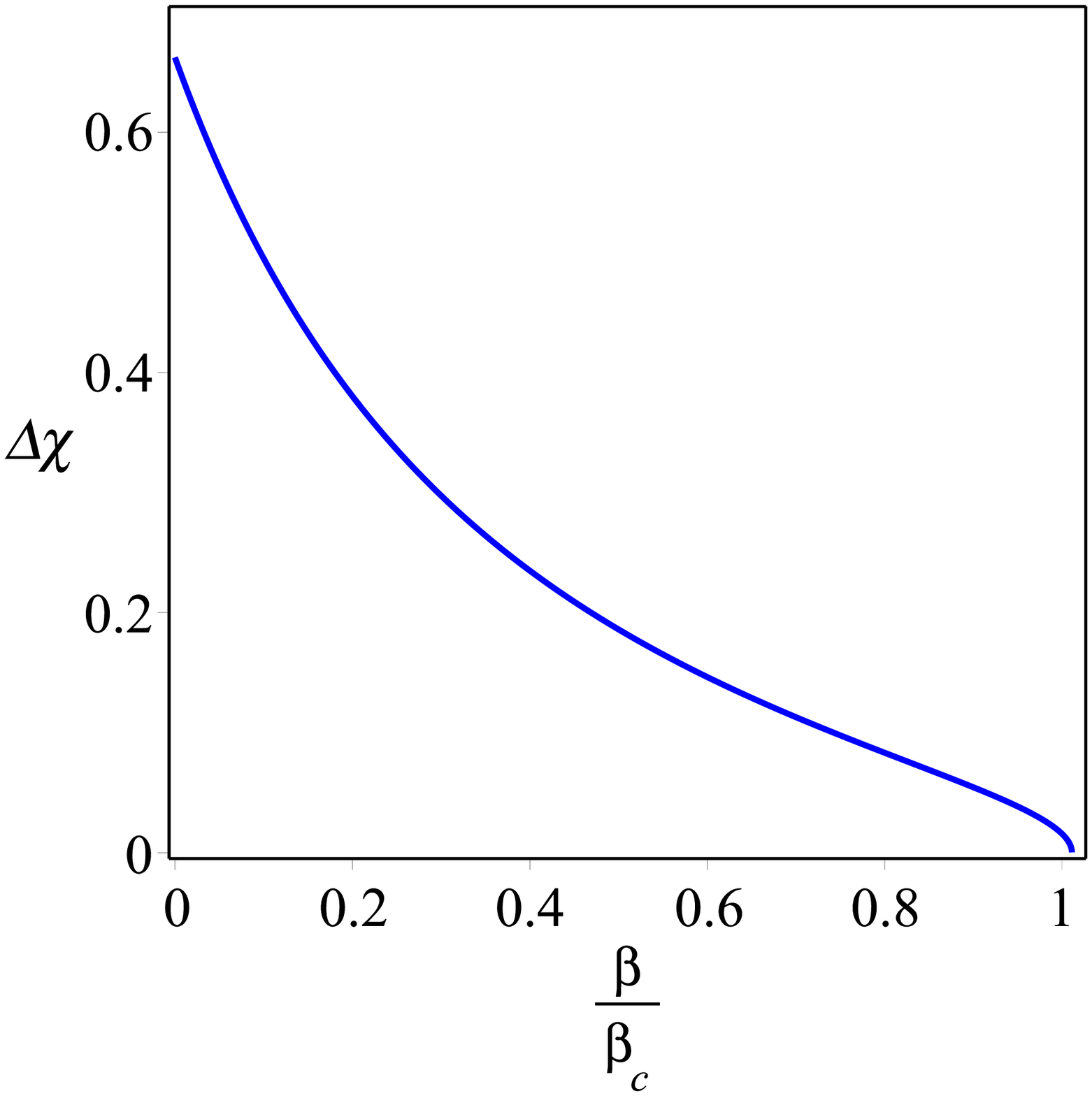}}
 \subfloat[ $Q=0.1$ and $ \omega=-1 $]{
        \includegraphics[width=0.31\textwidth]{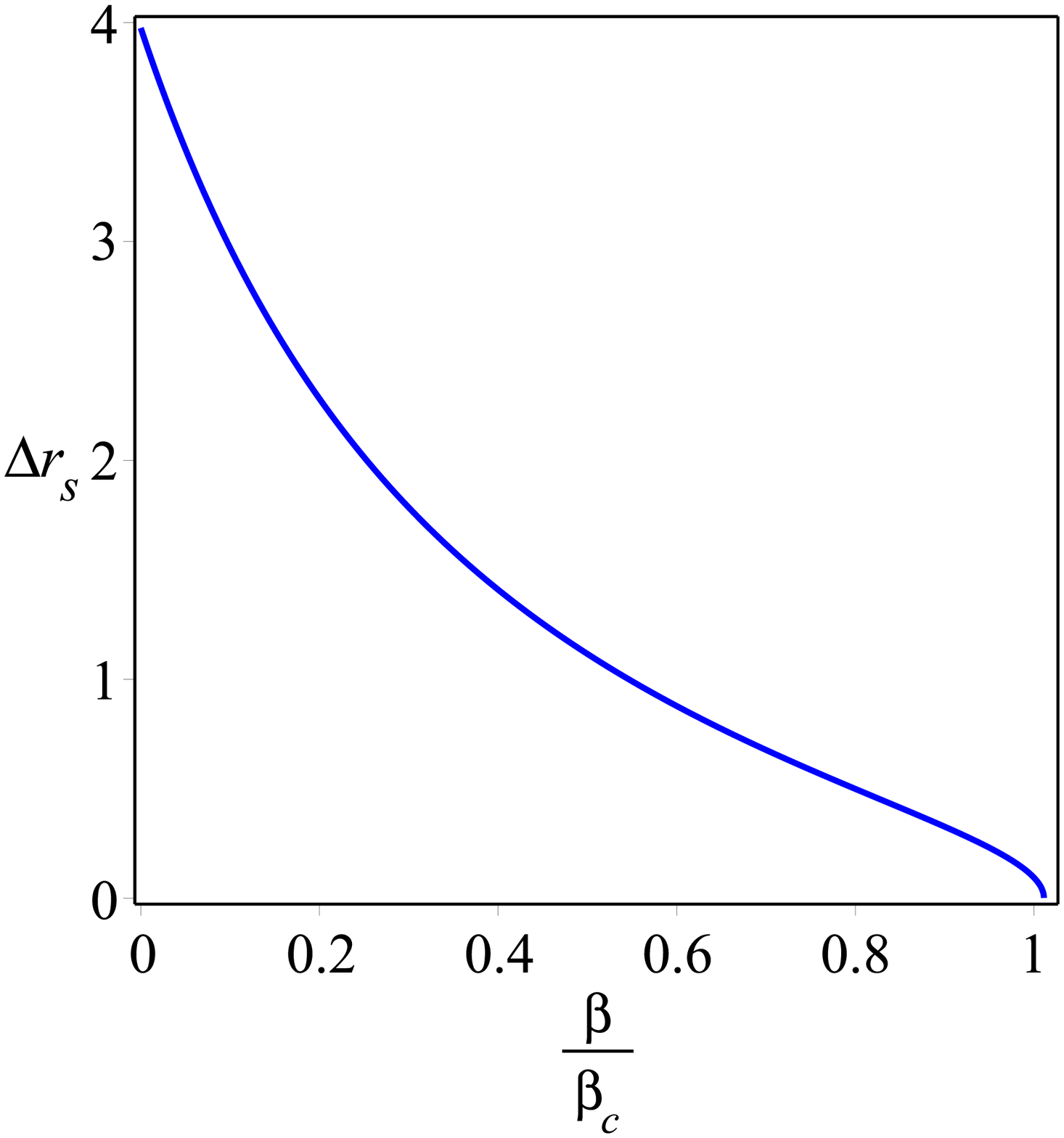}}\newline
\caption{Behavior of $ \Delta \chi $ and $ \Delta r_{s} $ as a
function of the $\frac{\beta}{\beta_{_{c}}}$ for $l=1$.}
\label{Fig10}
\end{figure}

By satisfying the constraint $ \frac{\partial r_{s}}{\partial
r_{+}}>0 $ \cite{M.Zhang,Belhaj}, one can draw a conclusion that
the sign of $ C $ is controlled by $ \frac{\partial T}{\partial
r_{s}} $.  To investigate the link between phase transition and
shadow of the black hole, we consider the temperature expression
Eq. (\ref{TH1}) and the heat capacity Eq. (\ref{Eqheat}), and
examine the behavior of temperature and heat capacity with respect
to $ r_{s} $. Since the thermodynamic behavior of the system is
different for $ \omega
>-\frac{2}{3} $ and $ \omega <-\frac{2}{3} $, we consider $ \omega
$ toward  $ -1 $ and $ -\frac{1}{3} $
 as two limited states and solve the problem analytically.
The isobar curves on the $T - r_{s}$ and $C - r_{s}$ are displayed
in Figs. \ref{Fig7}b and \ref{Fig7}d for $ \omega=-1 $. As we see,
these curves exhibit similar behaviors as $T - \chi$ and $C -
\chi$ curves in Figs. \ref{Fig7}a and \ref{Fig7}c. For
$\beta>\beta_{c}$,  the temperature is only a monotone increasing
function of $  r_{s} $ without any extremum (see dotted line in
Fig. \ref{Fig7}b). The heat capacity is also a continuous function
for variable $  r_{s} $  (see dotted line in Fig. \ref{Fig7}d).
For the case $\beta<\beta_{c}$,  a non-monotonic behavior appears
for temperature with one local maximum and one minimum which
corresponds to the first-order phase transition. According to the
definition of heat capacity, these extrema coincide with
divergence points of $ C $. Evidently, a change of signature
occurs at these points. In other words, it changes from positive
to negative at the first divergency and then it becomes positive
again at the second one. For $ \omega=-1/3 $, $T - r_{s}$ and $C -
r_{s}$ diagrams are plotted in Fig. \ref{Fig8}. Evidently, its
behavior is opposite to $\omega=-1$ case, meaning that divergency
is appeared for $\beta>\beta_{c}$, whereas a monotonic behavior is
observed for $\beta<\beta_{c}$. So, three black holes are
thermodynamically competing. Small black holes which are located
between root and smaller divergency of the heat capacity are in a
stable phase. For an intermediate range of the shadow radius, the
black holes are thermodynamically unstable. The region after
larger divergency is related to large black holes which are
thermally stable. Figure \ref{Fig9} displays these three phases
and the effect of BH parameters on these regions. As we see, the
parameter $ \beta $ (normalization factor) and state parameter
have significant effects on the stability/instability of the black
hole. In other words, a stable BH may exit in its stable state
when it is surrounded by quintessence. This reveals the fact that
it is logical to consider the normalization factor as a variable
quantity.

For $\beta=\beta_{c}  $,  the small BH and the large one merge
into one squeezing out the unstable black hole. This can be found
as a deflection point in the $T - r_{s}$ plot which forms critical
point of the second-order phase transition.  Such behavior of
temperature is very similar to van der Waals liquid-gas system
which goes under a second-order phase transition at $ T=T_{c} $.
One can analyze the behavior of the shadow radius before and after
the second-order small-large BH phase transition. To do so, we
have depicted the changes of the shadow radius ($ \Delta
r_{s}=r_{s}^{L}- r_{s}^{S} $) as a function of the reduced $ \beta
$  ($ \beta/\beta_{c} $) in left panels of Fig. \ref{Fig10}. We
see that $ \Delta r_{s}$ and $ \Delta \chi$ have similar behaviors
and they are monotonically decreasing functions of the reduced
parameter $ \beta $. They approach to zero at $ \beta=\beta_{c} $,
where the first-order phase transition becomes a second-order one.

\section{conclusion}

In this paper, we have studied the analogy of charged AdS black
holes surrounded by quintessence with van der Waals fluid system
with a new viewpoint, in which we kept the cosmological constant
as a constant parameter and instead allow the normalization factor
to vary as a thermodynamic quantity. The obtained results showed
that for $ \omega<-2/3 $, the system has similar thermodynamic
behavior as the van der Waals fluid system such that it goes under
a first order phase transition for $T<T_{c}$ and
$\beta<\beta_{c}$, and undergoes a second-order phase transition
at $T=T_{c}  $ and $\beta=\beta_{c} $. For $ \omega>-2/3 $, an
opposite behavior was observed, meaning that a first order phase
transition takes place for $T>T_{c}  $ and $\beta>\beta_{c}  $. We
also derived all the critical exponents of the system and found
that they are exactly coincident with the van der Waals fluid
system.

It is worthwhile to mention that although one can investigate
phase transition of a BH with help of $ P-V $ and $ Q^{2}-\Psi $
planes, it is more logical to take the normalization factor, which
indicates the intensity of the quintessence field as a variable
quantity.  As we know, the cosmological constant does not change
and basically has a constant value. In contrast, the quintessence
field is a dynamic parameter that changes over time. Regarding the
consideration the square of the electric charge as a thermodynamic
quantity, although one can employ this method to explore phase
transition, it should be noted that according to
\cite{Dehyadegari,Yazdikarimi} a first-order phase transition
takes place for temperatures above its critical values which is a
little different from the behavior of van der Waals fluid and
other ordinary phase transitions in everyday systems. In addition,
since the electromagnetic repulsion in compressing an electrically
charged mass is dramatically greater than the gravitational
attraction, it is not expected that black holes with a significant
electric charge will be formed in nature. So, the electric charge
of a BH cannot change over the time so much.

Finally, we investigated the photon sphere and the shadow observed
by a distant observer. We found that the shadow size shrinks with
increasing the electric charge, state parameter $\omega $ and
parameter $ \beta $ (normalization factor). We also explored the
connection between shadow radius and phase transition and found
that for $ \omega>-2/3 $ ($ \omega<-2/3 $) there exists a
non-monotonic behavior of the shadow radius for $ \beta
>\beta_{c} $ ($ \beta <\beta_{c} $)  which corresponds to a
first-order phase transition. We have shown such a phase
transition becomes a second-order one at $ \beta_{c} $. Studying
thermal stability of the system in this point of view, we noticed
that the normalization factor and state parameter have a
significant influence on the stability/instability of the black
hole. This revealed the fact that a stable BH may exit in its
stable state if it is surrounded by the quintessence.

\section{Acknowledgements}
We would like to thank the anonymous referees for their
constructive comments. SHH also thank Shiraz University Research
Council.

\end{document}